%% file: main-journal_Final.tex
\documentclass[journal]{IEEEtran}
\usepackage{amsmath,amssymb,amsfonts,amsthm,array,multirow}
\usepackage{algorithm}
\usepackage{algpseudocode}
\usepackage{subcaption}
\usepackage[utf8]{inputenc}
\usepackage{booktabs}
\usepackage{color}
\usepackage{setspace}

\usepackage[noadjust]{cite}

\usepackage{graphicx}

\usepackage{array}
\newcolumntype{L}[1]{>{\raggedright\let\newline\\\arraybackslash\hspace{0pt}}m{#1}}
\newcolumntype{C}[1]{>{\centering\let\newline\\\arraybackslash\hspace{0pt}}m{#1}}
\newcolumntype{R}[1]{>{\raggedleft\let\newline\\\arraybackslash\hspace{0pt}}m{#1}}

\algnewcommand{\IIf}[1]{\State\algorithmicif\ #1\ \algorithmicthen}
\algnewcommand{\IElse}[1]{\State\algorithmicelse\  #1 \ }
\algnewcommand{\EndIIf}{\unskip\ \algorithmicend\ \algorithmicif}

\input{defs.tex}

\begin{document}

\title{High-Contrast Reflection Tomography with Total-Variation Constraints}

\author{Ajinkya~Kadu, Hassan~Mansour, Petros T. Boufounos
\thanks{A preliminary version of this work appeared in~\cite{KMBL_2019}.}
\thanks{A. Kadu is with the Mathematical Institute of Utrecht University in The Netherlands. The work was partially completed while A. Kadu was with MERL.}
\thanks{H. Mansour and P. T. Boufounos are with Mitsubishi Electric Research Laboratories (MERL), 201 Broadway, Cambridge, MA, 02139.}
}

\maketitle
%

\begin{abstract}
Inverse scattering is the process of estimating the spatial distribution of the scattering potential of an object by measuring the scattered wavefields around it. In this paper, we consider reflection tomography of high contrast objects that commonly occurs in ground-penetrating radar, exploration geophysics, terahertz imaging, ultrasound, and electron microscopy. Unlike conventional transmission tomography, the reflection regime is severely ill-posed since the measured wavefields contain far less spatial frequency information of the target object. We propose a constrained incremental frequency inversion framework that requires no side information from a background model of the object. Our framework solves a sequence of regularized least-squares subproblems that ensure consistency with the measured scattered wavefield while imposing total-variation and non-negativity constraints. We propose a proximal Quasi-Newton method to solve the resulting subproblem and devise an automatic parameter selection routine to determine the constraint of each subproblem. We validate the performance of our approach on synthetic low-resolution phantoms and with a mismatched forward model test on a high-resolution phantom.
\end{abstract}


\section{Introduction}
\label{sec:intro}

Inverse scattering addresses the problem of reconstructing an image of the scattering potential of an object by probing it with electromagnetic or acoustic waves of finite bandwidth. An incident wavefield propagating inside the object induces multiple scattering of the waves that are generally measured on the boundary of the material. The scattered waves carry information about the spatial distribution of the scattering potential of the material, which has led to applications in numerous fields, such as, non-destructive testing \cite{laurens2005non}, optical tomography \cite{arridge1999optical}, geophysical imaging \cite{sirgue2010thematic,virieux2009overview}, ground-penetrating radar \cite{witten1994ground}, medical imaging \cite{yuan2007three,haynes2010large}, and electron microscopy \cite{humphry2012ptychographic,zewail2010four,barton1988photoelectron}.

\begin{figure}[t]
	\centering
	\begin{tabular}{ccc}
	\fbox{\includegraphics[width=0.26\columnwidth]{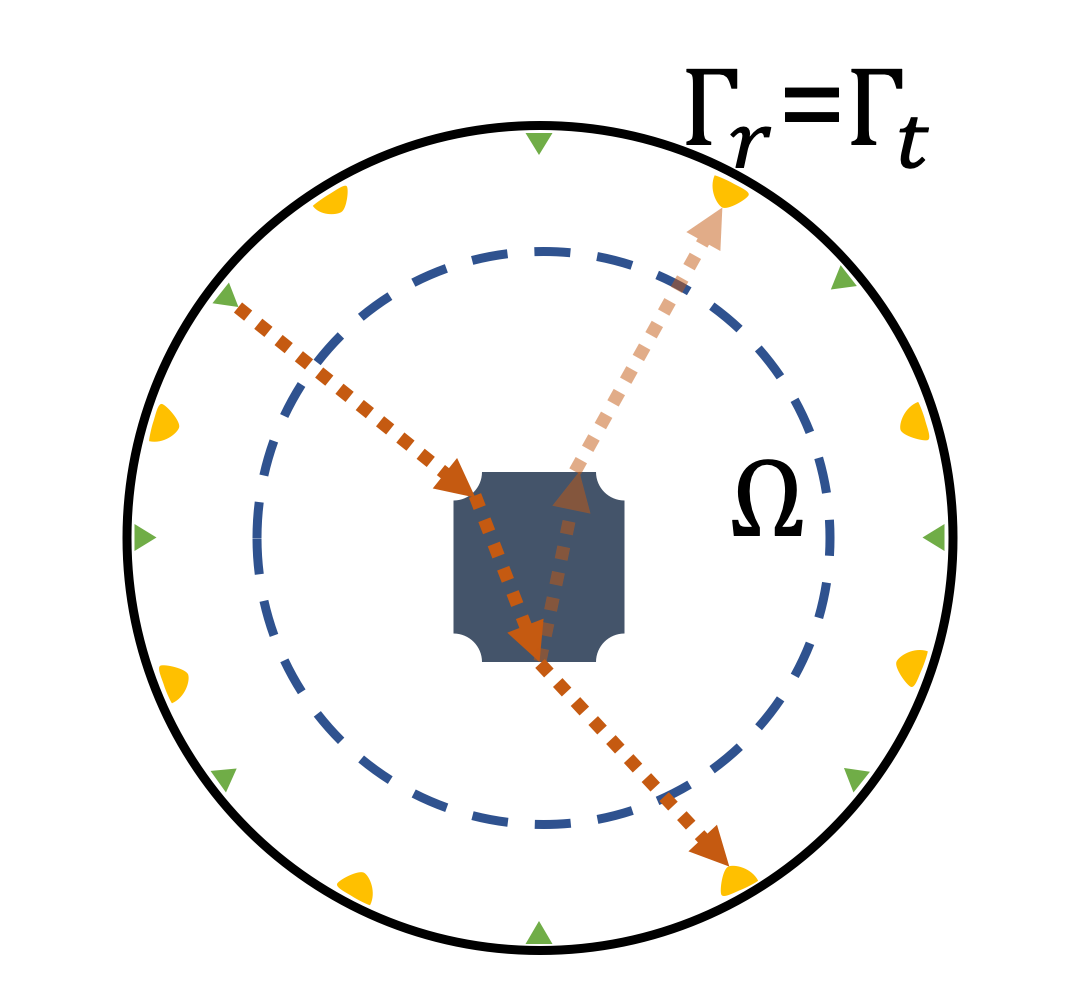}} & \fbox{\includegraphics[width=0.26\columnwidth]{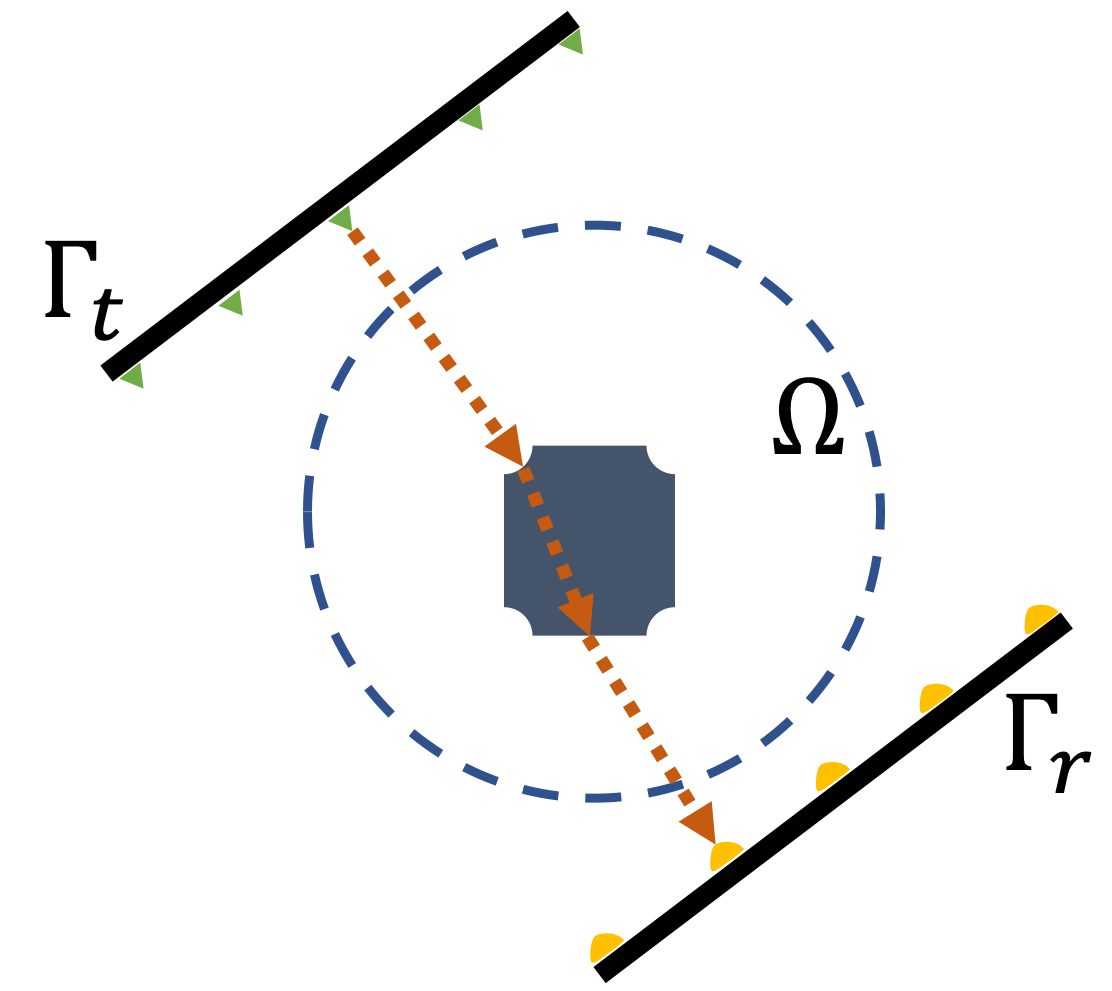}} &  \fbox{\includegraphics[width=0.26\columnwidth]{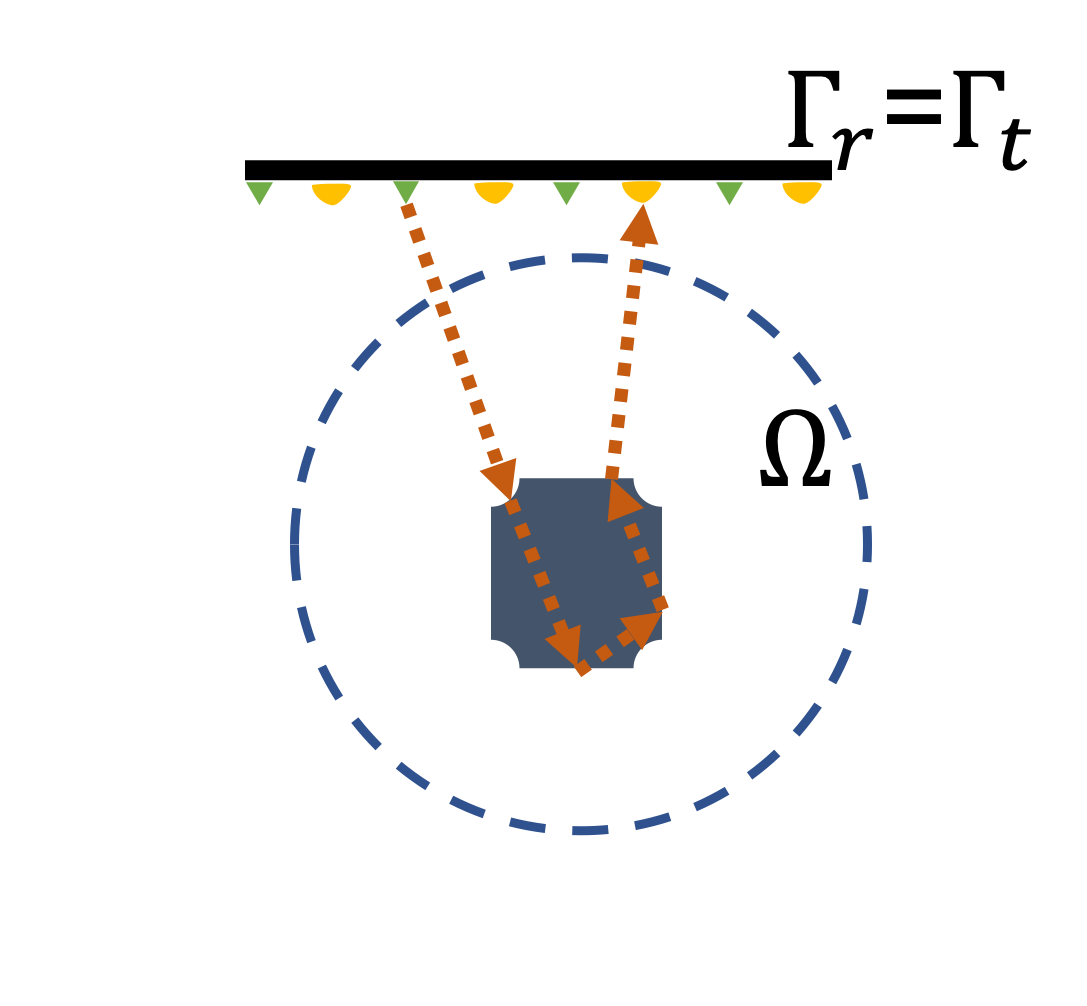}} \\
	(a)  & (b)  & (c)
	\end{tabular}
	\caption{Three acquisition scenarios in inverse scattering, (a) full-view, (b) Transmission, and (c) Reflection. $\Omega$ is the domain of interest, $\Gamma_t$ is a transmission domain, and $ \Gamma_r$ is a receiver domain. A single experiment consists of - a set of transmitters in $\Gamma_t$ sending a wavefield into $\Omega$, and scattered wavefield being measured by set of receivers in $\Gamma_r$.}
	\vspace{-4mm}
	\label{fig:acq-setup}
\end{figure}

A scattering experiment consists of a transmission domain $\Gamma_t \subset \mathbb{R}^d$, an object domain $\Omega \subset \mathbb{R}^d$ and a receiver domain $\Gamma_r \subset \mathbb{R}^d$, where $d  (= 2, 3)$ is the dimension of the scene. A set of transmitters located in $\Gamma_t$ sends incident waves into the scene that interact with an object in $\Omega$. This interaction leads to scattering of the incident waves. The scattered waves are then measured at the set of receivers located in $\Gamma_r$. Based on the location of transmission and receiver domain with respect to the object, we can classify the acquisition scheme into three different types: \textit{(i)} \textit{full-view}, where $\Gamma_t$ and $\Gamma_r$ surround the domain $\Omega$; \textit{(ii)} \textit{transmission} mode, where $\Gamma_t$ and $\Gamma_r$ are located on opposite sides of the object; and \textit{(iii)} \textit{reflection} mode, where $\Gamma_r$ and $\Gamma_t$ are co-located. Figure~\ref{fig:acq-setup} illustrates these acquisition schemes. The \textit{full-view} mode provides the most information about the spatial distribution of the object. The \textit{transmission} mode offers less information than that of \textit{full-view}, but it reduces the cost of the experiment due to the requirement of fewer transmitters and receivers. Tomographic imaging in this acquisition mode, known as \textit{transmission tomography}, has found applications in many areas, for example, X-ray tomography in medicine and non-destructive testing.

The \textit{reflection} mode generally arises due to a limitation in the ability to access different sides of the material, as in the case of underground imaging. 
We focus our presentation on the reflection tomography scenario where the problem is severely ill-posed. The ill-posedness arises due to restricted measurements and the limited availability of low spatial-frequency content in the measured wavefields. We discuss this further in Section~\ref{sec:TransvsRef}. The underground imaging setup often appears in ground-penetrating radar, seismic imaging, and ultrasound imaging.

The spatial scattering potential of a material can be described by its contrast level. The contrast indicates the power of interaction of an object material with a probing wave. A low-contrast material is semi-transparent, meaning that the interaction of the waves with it induces weak scattering. A high-contrast material strongly interacts with waves inducing \textit{multiple} scattering events. In this paper, we classify objects according to their contrast level, with a contrast below 1 being \textit{low}, a contrast ranging from 1 to 10 being medium-contrast, and a contrast above 10 to be \textit{high}. In general, the contrast varies with the frequency of the wave, but here we assume it to be independent of frequency. We also restrict the study to lossless objects. Extension of the analysis to lossy objects would require that the scattering potential be complex-valued with a non-negative real component. However, such an extension is out of the scope of this paper.

\vspace{-0.3cm}
\subsection{Related Work}
\vspace{-0.1cm}
Numerous techniques have been proposed for solving the inverse scattering problem in the reflection regime. Earlier approaches linearize the scattering model iteratively using straight-ray theory, the Born approximation, and the Rytov approximation~\cite{spencer1962general,born2013principles,devaney1981inverse,baysal1983reverse}. However, such linear models fail to account for the complex interaction between the wavefield and the material properties that result in multiple scattering. As a result, these methods require an accurate initial target model to enable the inversion and generally suffer from poor reconstruction quality, especially when the material is inhomogeneous or contains highly scattering objects. Recently, the nonlinear interaction between the wavefield and the object has been incorporated into the inversion process using the wave equation~(for example,~\cite{colton2012inverse,bao2005inverse,bao2015inverse,borges2017high,estatico2015multifrequency}). The inverse problem that deals with the wave-equation based scattering model is known as full-waveform inversion (FWI)~\cite{tarantola1984inversion,pratt1999seismic,virieux2009overview}. FWI has been applied in multiple domains and across all modes of acquisition. A considerable amount of research has focused on full-view tomography and transmission tomography with FWI. Since we are mainly interested in reflection tomography, we do not address the literature for other modes of tomography and nonlinear inverse problems, see for example~\cite{ISLP:1996,IPP:2001}. We note, however, that contrary to other modes of tomography, measurements in  reflection tomography are dominated by the high spatial frequencies of the contrast map. Moreover, contrary to other nonlinear inverse problems, the reflection tomography problem suffers from the fact that for every new frequency included in the measurements, a significantly larger number of unknowns are added to the estimation problem.

Reflection tomography with FWI has been heavily investigated in the geophysical community. Since the problem is nonlinear and nonconvex, the convergence of the inversion depends heavily on the initial model \cite{mulder2008exploring,symes2008migration}. Various approaches have been proposed to mitigate the effect of an initial model on the reconstructed solution \cite{warner2016adaptive,van2013mitigating,biondi2012tomographic,van2015penalty,bharadwaj2016full}. While these methods work well in the low contrast regime, they require regularization and additional constraints in the high contrast regime to deliver good reconstruction~\cite{esser2016constrained, esser2018total,asnaashari2013regularized,ambrosanio2015compressive}. Our work is also inspired by the TV-regularization strategies proposed in~\cite{esser2018total}. However, since the total variation parameter may be unknown~\cite{esser2018total}, we develop a framework to estimate this parameter from the noise level in the data.

The sequential workflow has a long history in the geophysical imaging literature. It was introduced in~\cite{Bunks1995} under the name multiscale full-waveform inversion. We work with a regularized version of this multiscale approach. Since we add one frequency at the time, as opposed to a frequency batch in \cite{Bunks1995}, our approach is more robust against local minima (more discussions in Section~\ref{sec:seqWorkflow}).

\subsection{Contributions and Outline}
\vspace{-0.1cm}
We develop an inversion framework for high-contrast limited-angle reflection tomography. Our contributions are three-fold:
\begin{itemize}
	\item \textit{Formulation}: We adopt a regularized sequential approach based on incremental frequency inclusion. We keep the low frequencies in the cost function to avoid potential local minima. For a total of $k$ frequencies in the data, we solve $k$ constrained nonlinear least-squares problems sequentially.
	\item \textit{Regularization and Optimization}: We introduce a combination of non-negative and total-variation regularization for the contrast function. Note that both the regularizers are non-differentiable. To solve the regularized nonlinear least-squares problem, we propose a proximal Quasi-Newton (prox-QN) method that is computed using a primal-dual method.
	\item \textit{Parameter Estimation}: We develop a strategy for estimating the total-variation constraint parameter from the noise-level in the data.
\end{itemize}

We introduce the forward and the inverse scattering problem in Section~\ref{sec:problem}. Here, we also discuss the challenges of reflection tomography. In Section~\ref{sec:proposed}, we present the details of our sequential approach and describe the optimization framework as well as the regularization strategies. We validate the proposed method on numerical phantoms and compare it with other methods in Section~\ref{sec:results}, and conclude the paper in Section~\ref{sec:conc}.

\section{Inverse Scattering Problem}\label{sec:problem}
We begin by presenting the scattering model that describes the relationship between the wavefield and the contrast function. Next, we formulate the discrete inverse problem to reconstruct the contrast function from the set of measured scattered wavefields. Finally, we discuss some challenges in estimating the contrast of an object in the reflection regime.

\subsection{Forward problem}
The \textit{forward} scattering problem constructs a mapping from a contrast function (determined by materials in the object) to the scattered waves measured at receivers. A wave-equation governs this mapping in the frequency-domain. For simplicity, we restrict our discussion to scalar waves, but the map can naturally be constructed for the other types of scattering problems with some modifications (see, for example, \cite{born2013principles}).

Consider the setup shown in Figure~\ref{fig:acq-setup} where an object is located in a bounded domain $\Omega \subset \mathbb{R}^d$, where $d = 2,3$ denotes the dimension. The object has a spatial distribution of permittivity given by $\epsilon(\mathbf{r})$, where $\mathbf{r}$ denotes the spatial co-ordinate. The object lies in a homogeneous background (free space) of permittivity $\epsilon_b$. We define the contrast function (or the relative permittivity) of the object as the difference of the permittivity of the object from the background, \ie, $f(\mathbf{r}) = \epsilon(\mathbf{r}) - \epsilon_b$. In this section, we consider the single frequency setting and drop the frequency index from the equations. The variables are expressed as scalar functions of the position $\mathbf{r}$. We will later use vector notation to represent the variables over the entire domain $\Omega$. 

We illuminate the target object using the waves generated from a source function $q: \Gamma_t \mapsto \mathbb{C}$. Subsequently, the scattered wavefield is measured inside the receiver domain $\Gamma_r$. The total wavefield $u: \Omega \mapsto \mathbb{C}$ in the object domain $\Omega$ is related to the contrast function $f$ by the \textit{Lippmann-Schwinger} integral equation
\begin{equation}
	u(\mathbf{r}) = u_\text{in}(\mathbf{r}) +  k^2 \! \! \int_{\mathbf{r}' \in \Omega}  g(\mathbf{r} - \mathbf{r}') u(\mathbf{r}') f(\mathbf{r}') \mathrm{d} \mathbf{r}' \quad \forall \mathbf{r} \in \Omega,
	\label{eq:LippSch}
\end{equation}
where $g: \mathbb{C}^d \mapsto \mathbb{C}$ is the Green function, $u_\text{in}: \mathbb{C}^d \mapsto \mathbb{C}$ is an input wavefield, $k = 2 \pi \omega/ c$ represents the wavenumber in vacuum, $\omega$ is the frequency and $c$ denotes the speed of light in vacuum. We assume that $f$ is real, or in other words, the object is lossless. The input wavefield in~\eqref{eq:LippSch} depends on the source function $q$ as
\begin{equation}
	u_\text{in}(\mathbf{r}) = k^2 \int_{\mathbf{r}' \in \Gamma_t} \, g(\mathbf{r} - \mathbf{r}')q(\mathbf{r}') \, \mathrm{d} \mathbf{r}' \qquad \forall \, \mathbf{r} \in \mathbb{R}^d.
	\label{eq:IntRepIncWav}
\end{equation}
Finally, the scattered wavefield measured in the receiver domain, $y: \Gamma_r \mapsto \mathbb{R}$ is given by
\begin{equation}
	y(\mathbf{r}) = \int_{\Omega}  g(\mathbf{r} - \mathbf{r}') f(\mathbf{r}') u(\mathbf{r}') \;\mathrm{d}\mathbf{r}' , \qquad \forall \, \mathbf{r} \in \Gamma_r.
	\label{eq:DataEqn}
\end{equation}
We provide detailed derivation of equations \eqref{eq:LippSch}, \eqref{eq:IntRepIncWav}, and \eqref{eq:DataEqn}  in Appendix~A. The \textit{forward} problem finds the measurements $y$ from the known source function $q$, the contrast function $f$, and the Green function $g$.  In essence, it consists of solving equation \eqref{eq:IntRepIncWav}, the \textit{Lippmann-Schwinger} equation \eqref{eq:LippSch}, and finally, the data equation \eqref{eq:DataEqn}. Generally, we pre-compute the input wavefield $u_{\text{in}}$ for each wavenumber $k$, since it is independent of the contrast function.

In the discrete setting, the scattering equation~\eqref{eq:LippSch} and data equation~\eqref{eq:DataEqn} reduce to the following system of linear equations for each transmitter illumination and the wavenumber:
\begin{equation}
	\begin{split}
		\mathbf{u} &= \mathbf{v} +  \mathbf{G} \diag \left( \mathbf{f} \right) \mathbf{u}, \\
		\mathbf{y} &=  \mathbf{H} \diag (\mathbf{f}) \mathbf{u},
	\end{split}
	\label{eq:forwardProb}
\end{equation}
where $\mathbf{u} \in \mathbb{C}^{N}$ and $\mathbf{v} \in \mathbb{C}^N$ are the total and input wavefields, respectively, $N$ denotes the number of grid points used to discretize the domain $\Omega$, $\mathbf{f} \in \mathbb{R}^N$ denotes the discretized contrast function, while $\mathbf{G} \in \mathbb{C}^{N \times N}$ and $\mathbf{H} \in \mathbb{C}^{n_r \times N}$ are the Green functions of the domain and receivers, respectively. Let $n_r$ be the number of receivers that discretizes the receiver domain $\Gamma$, then $\mathbf{y} \in \mathbb{C}^{n_r}$ is the noise-free scattered wavefield measured at the receivers. The critical step in the forward problem involves estimating the wavefield $\mathbf{u}$ by inverting the matrix $\mathbf{A} :=  \mathbf{I} - \mathbf{G} \diag \left( \mathbf{f} \right) $, where $\mathbf{I}$ denotes the identity operator. As the discretization dimension $N$ increases, explicitly forming the matrix $\mathbf{A}$ and computing its inverse become prohibitively expensive. Therefore, a functional form of $\mathbf{A}$ along with the conjugate-gradient method (CG) are often used to perform the inversion. We note here that the convergence of CG depends on the conditioning of the operator $\mathbf{A}$, which becomes ill-conditioned for large wavenumber and high-contrast media, \ie, for large values of $\| \mathbf{f} \|_\infty$.

\subsection{Inverse problem}
An \textit{inverse scattering} problem is defined as the estimation of the contrast function given the measurement of the scattered wavefield at $n_r$ receivers for each input wavefield generated from $n_t$ transmitters. We use $\mathcal{J} = \{1, \dots, n_f \}$ and $\mathcal{I} = \{1, \dots, n_t \}$ to denote the index sets for frequencies and transmitters respectively, $n_f$ to represent the number of frequencies, and $n_t$ to represent the number of transmitters. Let $\mathbf{y}_{ijl}$ be the measured signal at frequency $j$ and receiver $l \in \{1,\dots n_r\}$ and illuminated by transmitter $i$. Also,  $\mathbf{H}_{jl}$ denotes the forward operator mapping associated with frequency $j$ and receiver $l$. Assuming that the measurements are contaminated by white Gaussian noise, we can formulate the discrete \textit{inverse} problem as a constrained least-squares problem:
\begin{equation}
	\begin{split}
		\underset{\mathbf{f}}{\min} \quad & \sum_{l=1}^{n_r}\sum_{j \in \mathcal{J}, i \in \mathcal{I}}  \tfrac{1}{2} \| \mathbf{y}_{ijl} - \mathbf{H}_{jl} \diag( \mathbf{f}) \mathbf{u}_{ij} \|^2, \\[1ex]
		\mbox{subject to} \quad & \left( \mathbf{I} - \mathbf{G}_j \diag \left( \mathbf{f} \right) \right) \mathbf{u}_{ij} = \mathbf{v}_{ij} \quad \forall i,j
	\end{split}
	\label{eq:PDE-constr-LS}
\end{equation}

We assume that the $\mathcal{J}$ is ordered according to the frequencies (in an increasing order).  For the rest of this paper, $\| \cdot \|$ denotes the Euclidean norm (if there is no subscript). In general, problem \eqref{eq:PDE-constr-LS} is ill-posed and admits multiple solutions. Therefore, spatial regularization in the form of a penalty function $\mathcal{R}(\mathbf{f})$ is often added to make the solution space smaller.

Let us introduce, for each frequency $j \in \mathcal{J}$, a data matrix $\mathbf{Y}_j \in \mathbb{C}^{n_r \times n_t}$, a wavefield matrix $\mathbf{U}_j \in \mathbb{C}^{N \times n_t}$ and the input wavefield matrix $\mathbf{V}_{j} \in \mathbb{C}^{N \times n_t}$. Hence, the cost function and the constraint for each frequency takes the form
\begin{align*}
		\mathcal{D}_{j} \! \left( \mathbf{f},\mathbf{U}_j \right) &= \tfrac{1}{2} \| \mathbf{Y}_j - \mathbf{H}_j \diag( \mathbf{f}) \mathbf{U}_j\|_F^2, \\[1ex]
	\mathbf{C}_j \! \left( \mathbf{f},\mathbf{U}_j \right) &= \left( \mathbf{I} - \mathbf{G}_j \diag \left( \mathbf{f} \right) \right) \mathbf{U}_j - \mathbf{V}_{j},
\end{align*}
where $\| \cdot \|_F$ denotes the Frobenius norm. It is possible to eliminate the wavefields $\mathbf{U}$ by satisfying the constraints, $ \ie, \mathbf{U}_j^\star = \left( \mathbf{I} - \mathbf{G}_j \diag(\mathbf{f}) \right)^{-1} \mathbf{V}_j $. Such reduced cost-function at frequency $j$ is given by
\begin{align}
	\mathcal{F}_j (\mathbf{f}) \triangleq \bigg\lbrace \mathcal{D}_j(\mathbf{f}, \mathbf{U}_j) \quad \mbox{subject to} \quad \mathbf{C}_j(\mathbf{f}, \mathbf{U}_j) = \mathbf{0} \, \bigg\rbrace .
\label{eq:Fj}
\end{align}
With the incorporation of this reduced form, the regularized version of \eqref{eq:PDE-constr-LS} now reads
\begin{equation}
		\min_{\mathbf{f}} \quad  \sum_{j \in \mathcal{J}} \mathcal{F}_{j} \! \left( \mathbf{f}\right) + \mathcal{R} \! \left( \mathbf{f} \right).
	\label{eq:PDE-constr-LS-reg}
\end{equation}
Since both the cost function and constraints are nonlinear, we resort to iterative methods to find a feasible solution to the regularized least-squares optimization problem shown above.

\subsection{Transmission vs Reflection}
\label{sec:TransvsRef}
A critical distinction between the transmission and reflection modes in inverse scattering manifests itself in the spatial frequency content that can be captured by the measured wavefields. In the transmission regime, the received measurements generally capture large amount of the lower spatial frequencies of the target distribution compared to the reflection regime. This is due to the fact that a probing pulse in the transmission mode is modulated by the complete object before reaching the receivers. On the other hand, the measured wavefields in the reflection mode are modulated by the discontinuities in the object permittivity that lead to reflections of the wavefields back to the receivers.

\begin{figure}[t]
\centering
\mbox{
\begin{subfigure}[b]{0.25\textwidth}
                \includegraphics[width=0.95\textwidth]{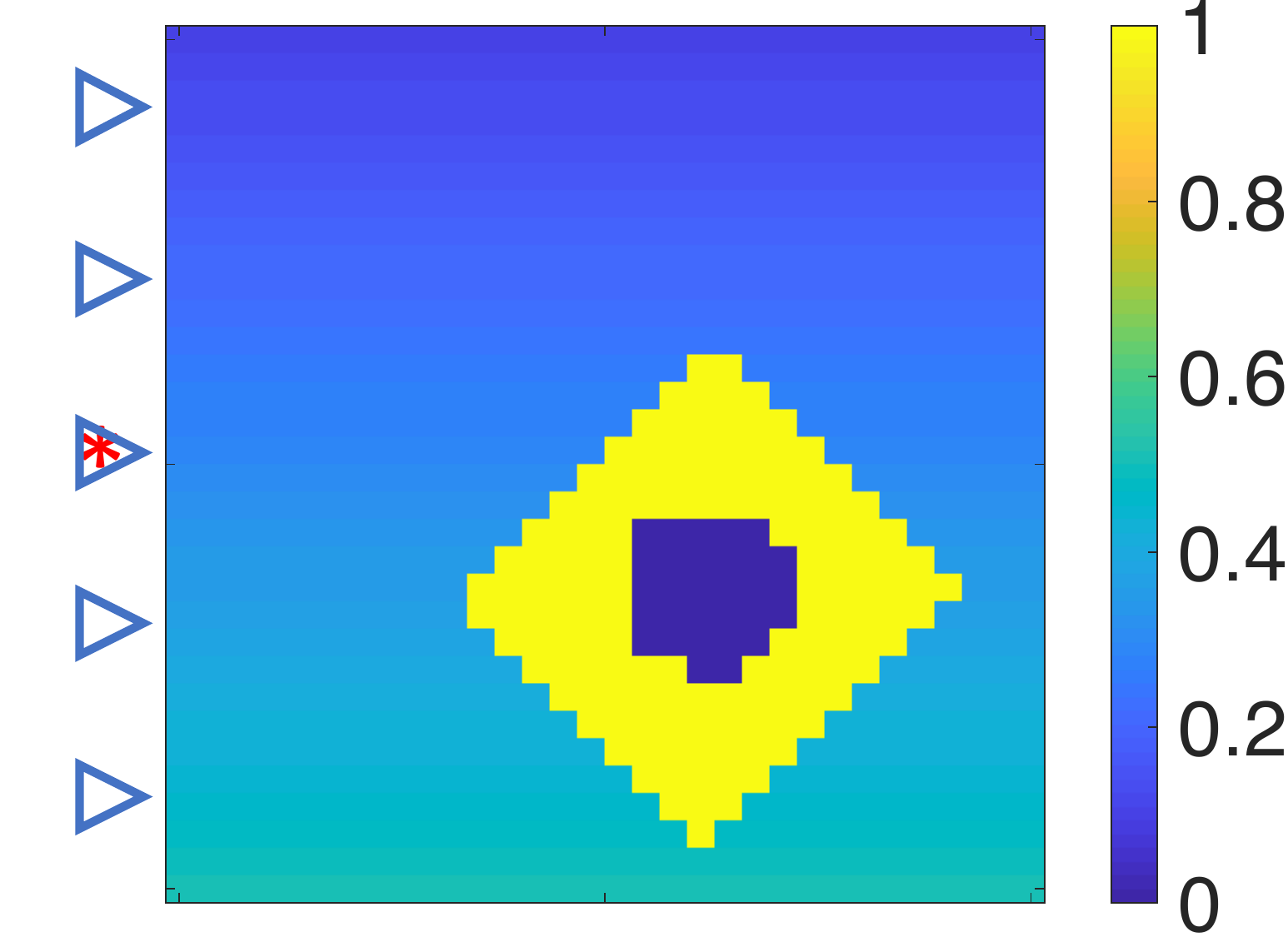}
        \end{subfigure}%
        \begin{subfigure}[b]{0.25\textwidth}
                \includegraphics[width=0.95\textwidth]{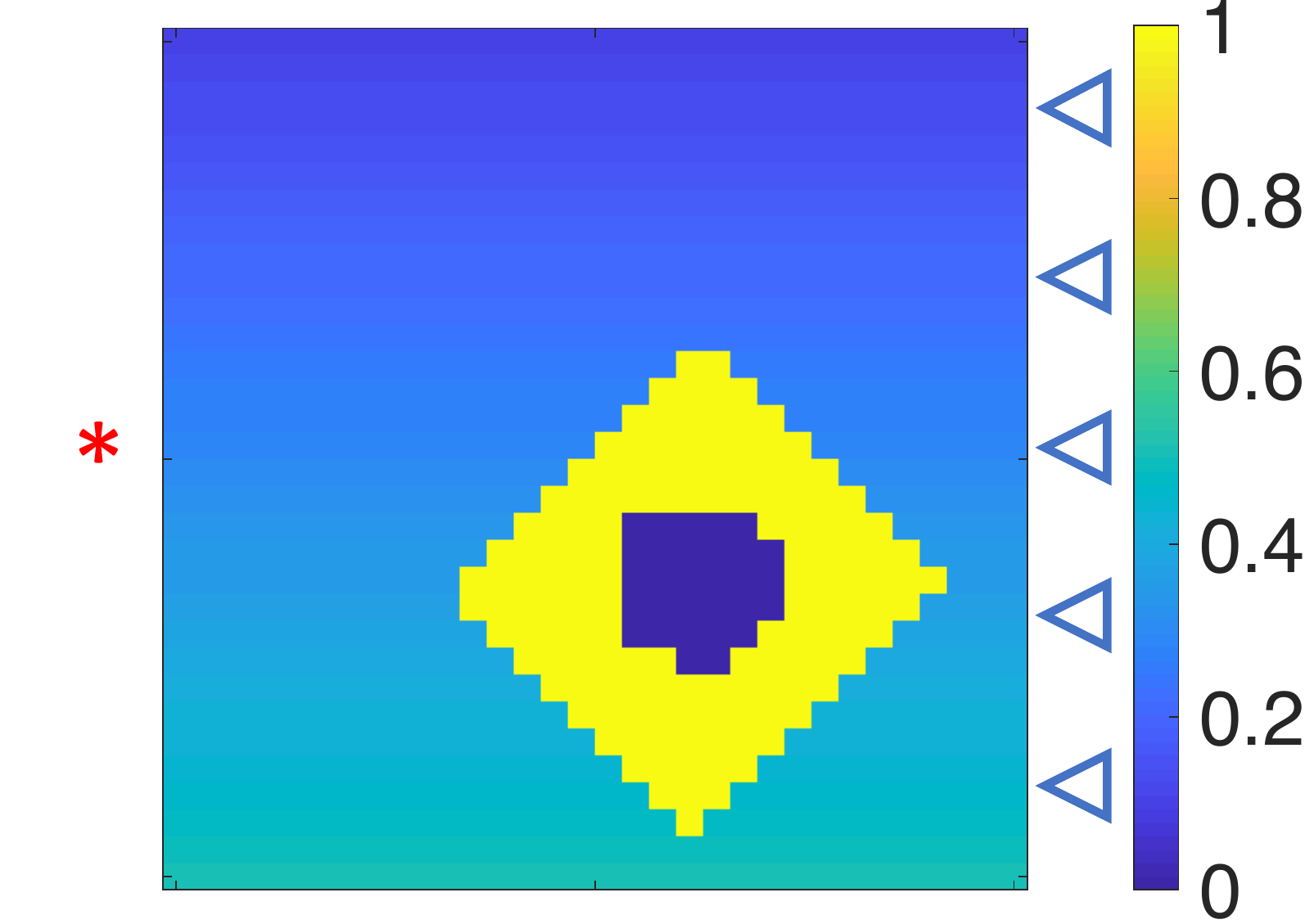}
        \end{subfigure}%
} 
\mbox{
\begin{subfigure}[b]{0.25\textwidth}
                \includegraphics[width=0.95\textwidth]{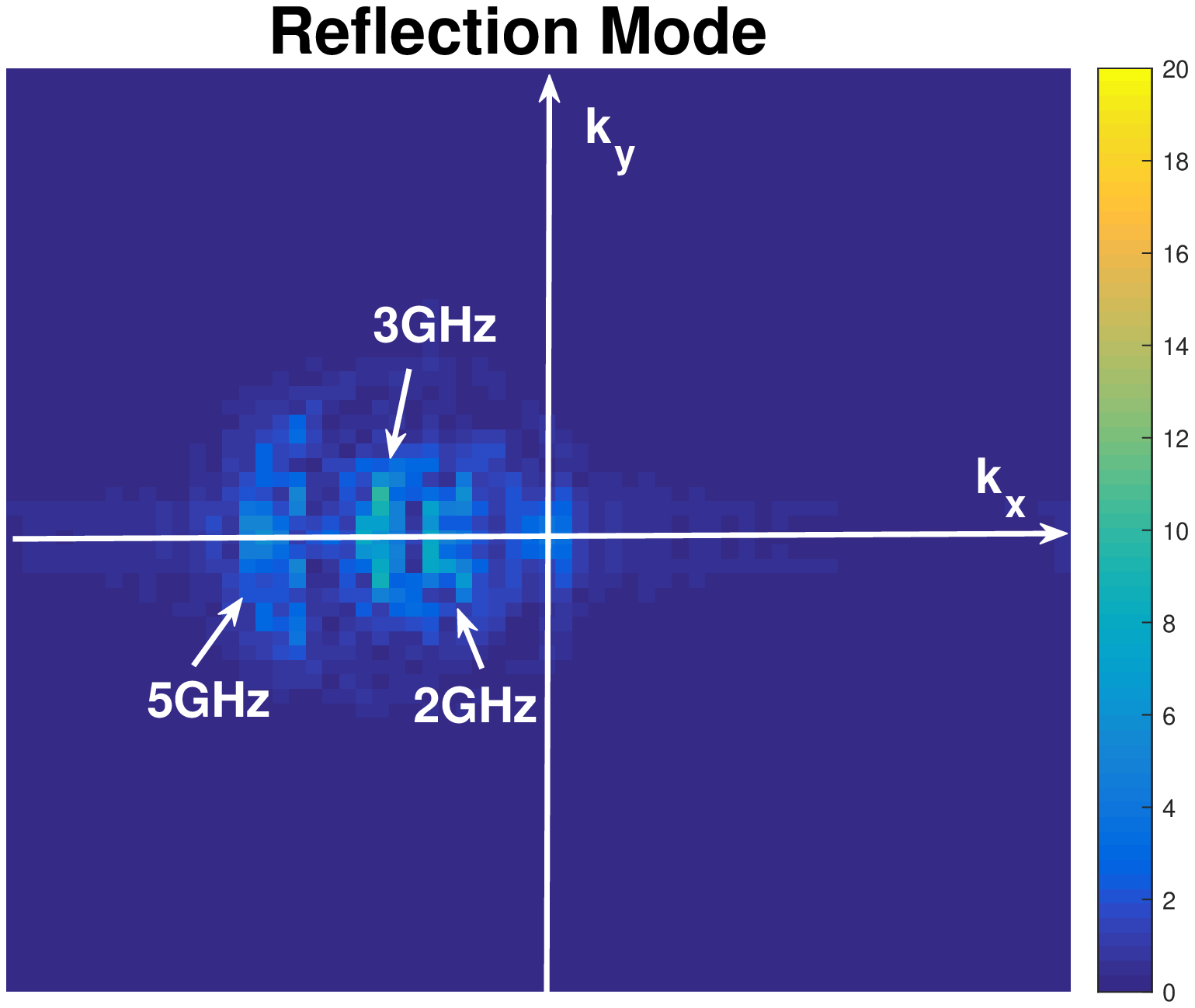}
        \end{subfigure}%
        \begin{subfigure}[b]{0.25\textwidth}
                \includegraphics[width=0.95\textwidth]{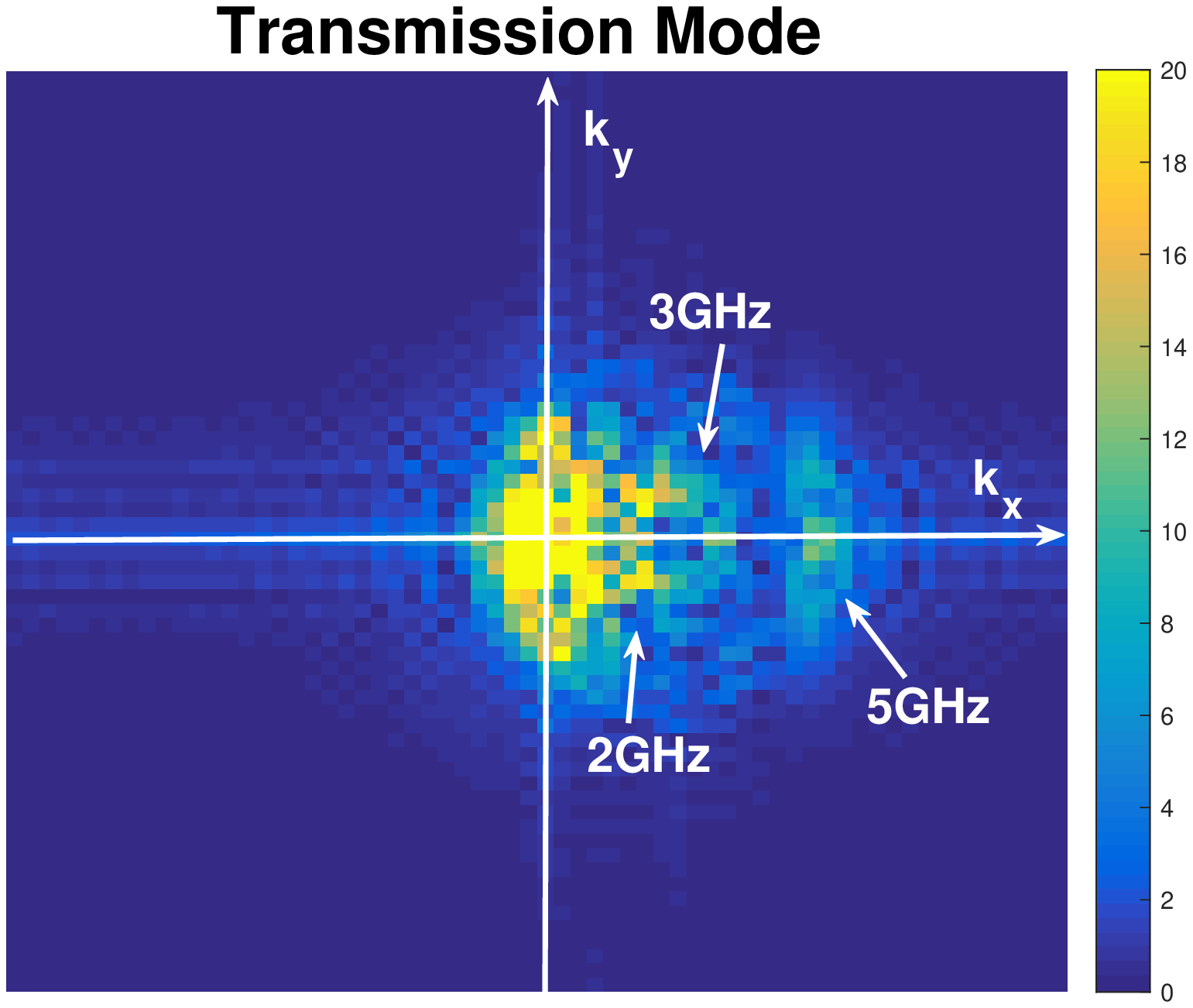}
        \end{subfigure}%
}
\caption{\small Comparison of the spatial frequency content of the received wavefields between the transmission mode and the reflection mode from a transmitted pulse containing 2GHz, 3GHz, and 5GHz frequency components.} \vspace{-0.4cm}
\label{fig:trans_vs_ref}
\end{figure}

In order to illustrate this phenomenon, we simulate two sets of measurements $\{\mathbf{y}_T, \mathbf{y}_R\}$ of the scattered wavefield from the same object, observed in the transmission and reflection modes through the measurement operators $\mathbf{H}_T$  and $\mathbf{H}_R$, respectively. Figure~\ref{fig:trans_vs_ref} illustrates the imaging setup where the object is illuminated from its left side by a transmitter, denoted by the red asterisk, with a flat spectrum pulse containing 2, 3, and 5GHz frequency components. Five receivers, denoted by blue triangles, are used to measure the scattered wavefield in both the reflection and tansmission regimes. We want to identify the amount of spatial frequency content that is encoded in each of $\mathbf{y}_T$ and  $\mathbf{y}_R$ without being affected by the nonconvexity of the inverse problem~\eqref{eq:PDE-constr-LS}. Therefore, we provide the true scattered wavefields $\mathbf{U}_{j}^{\star} = \left( \mathbf{I} - \mathbf{G}_j \diag\left( \mathbf{f}^{\star}\right) \right)^{-1} \mathbf{V}_{j}$ for each frequency, which reduces~\eqref{eq:PDE-constr-LS} to a convex linear inverse problem in $\mathbf{f}$. Consequently, we solve the convex form of~\eqref{eq:PDE-constr-LS} to compute $\mathbf{f}$ in each of the transmission and reflection modes and plot in Figure~\ref{fig:trans_vs_ref} the spatial frequency content (2D Fourier coefficients) of the reconstructed contrast $\mathbf{f}$ in each of the transmission and reflection modes. Notice how the recovered contrast in the reflection mode exhibits very little energy around the low spatial frequency subbands in the Fourier plane. This is in stark contrast to the transmission mode where a significant portion spectral energy of the recovered contrast corresponds to the low spatial frequencies. The illustration above helps motivate the argument that the received measurements of the scattered wavefields in the reflection tomography mode encode very little spatial frequency information about the target object. Since the goal of tomographic imaging is to estimate the spatial distribution the scattering potential of an object, the weak acquisition of spatial frequency information renders the problem severely ill-posed when compared to transmission tomography.

\section{Regularized Multiscale Approach}\label{sec:proposed}
In this section, we present an incremental frequency inversion method that does not require a smooth initial model of the target image for successful recovery. We also discuss the regularization and the optimization strategy to solve the resulting problem.

\vspace{-0.3cm}
\subsection{Sequential Workflow}
\label{sec:seqWorkflow}
\begin{figure*}[t]
\centering
\begin{tabular}{cccc}
(a) & (b) & (c) & (d) \\
\includegraphics[width=0.12\textwidth]{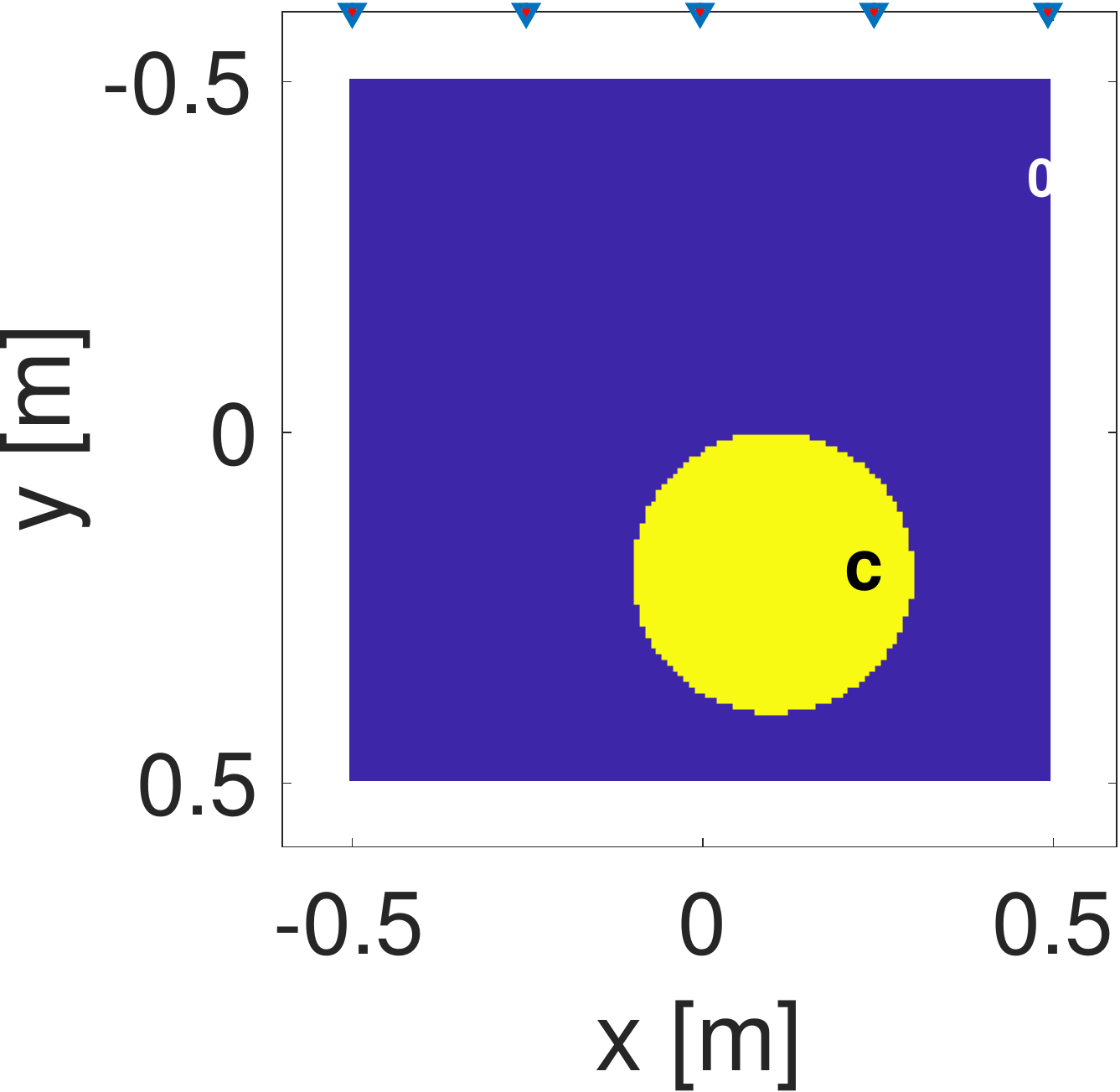} & \includegraphics[width=0.26\textwidth]{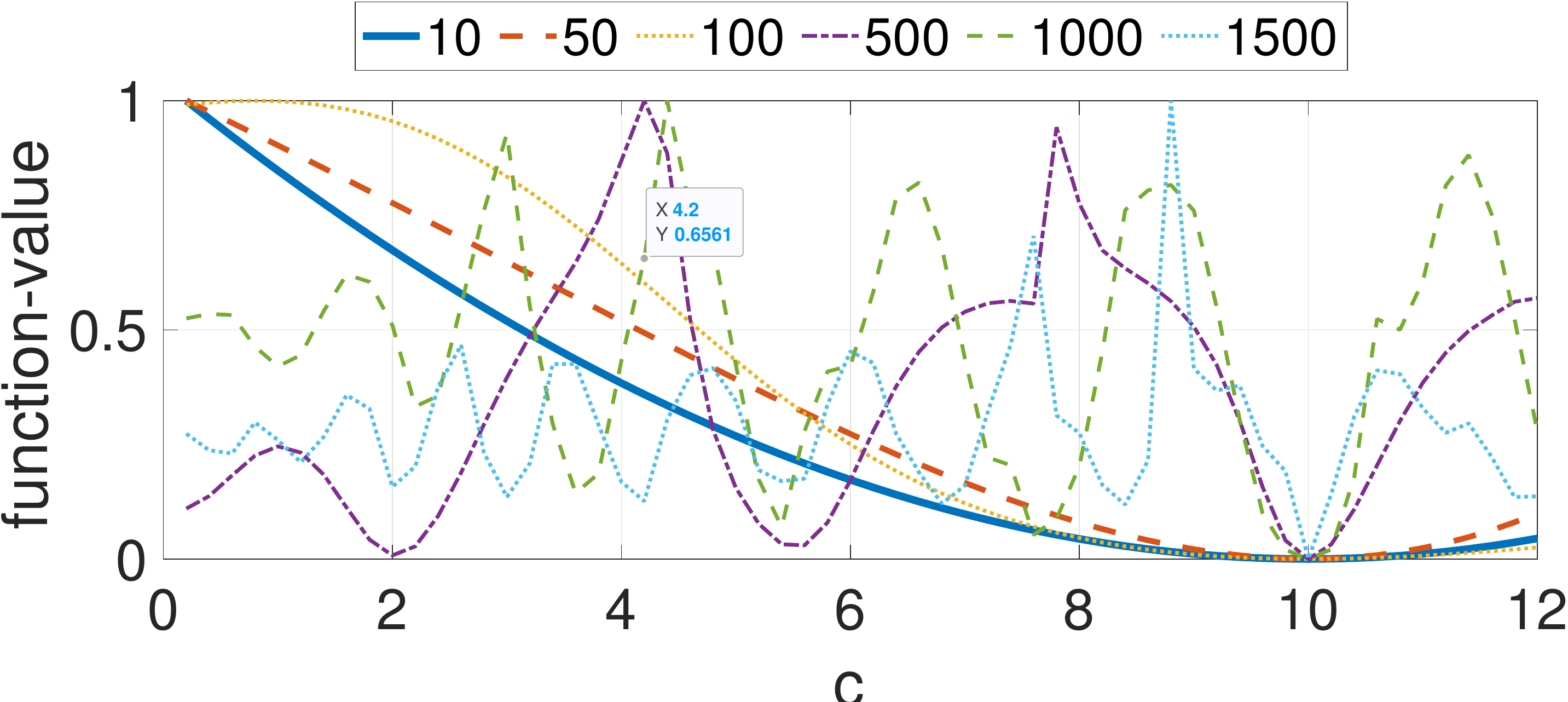} & \includegraphics[width=0.26\textwidth]{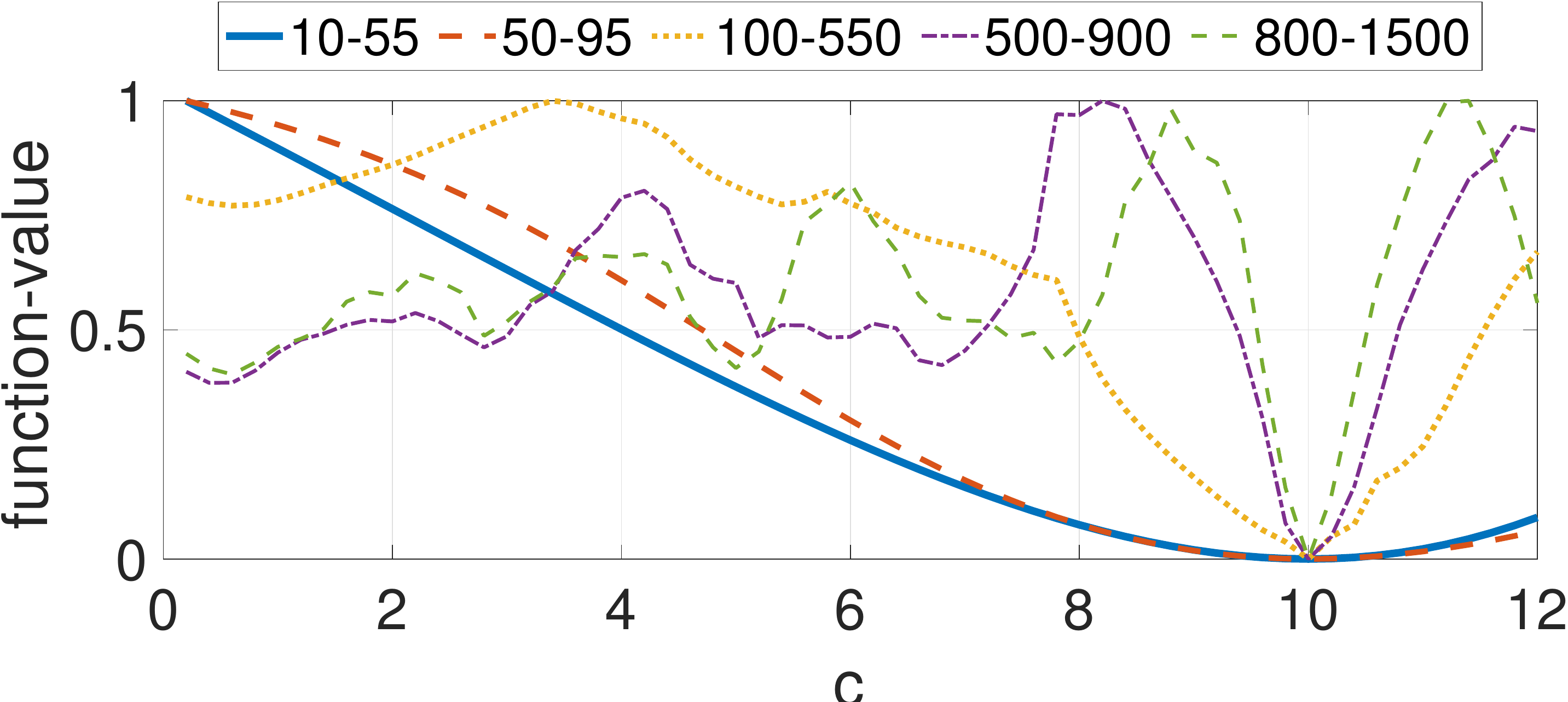} & \includegraphics[width=0.26\textwidth]{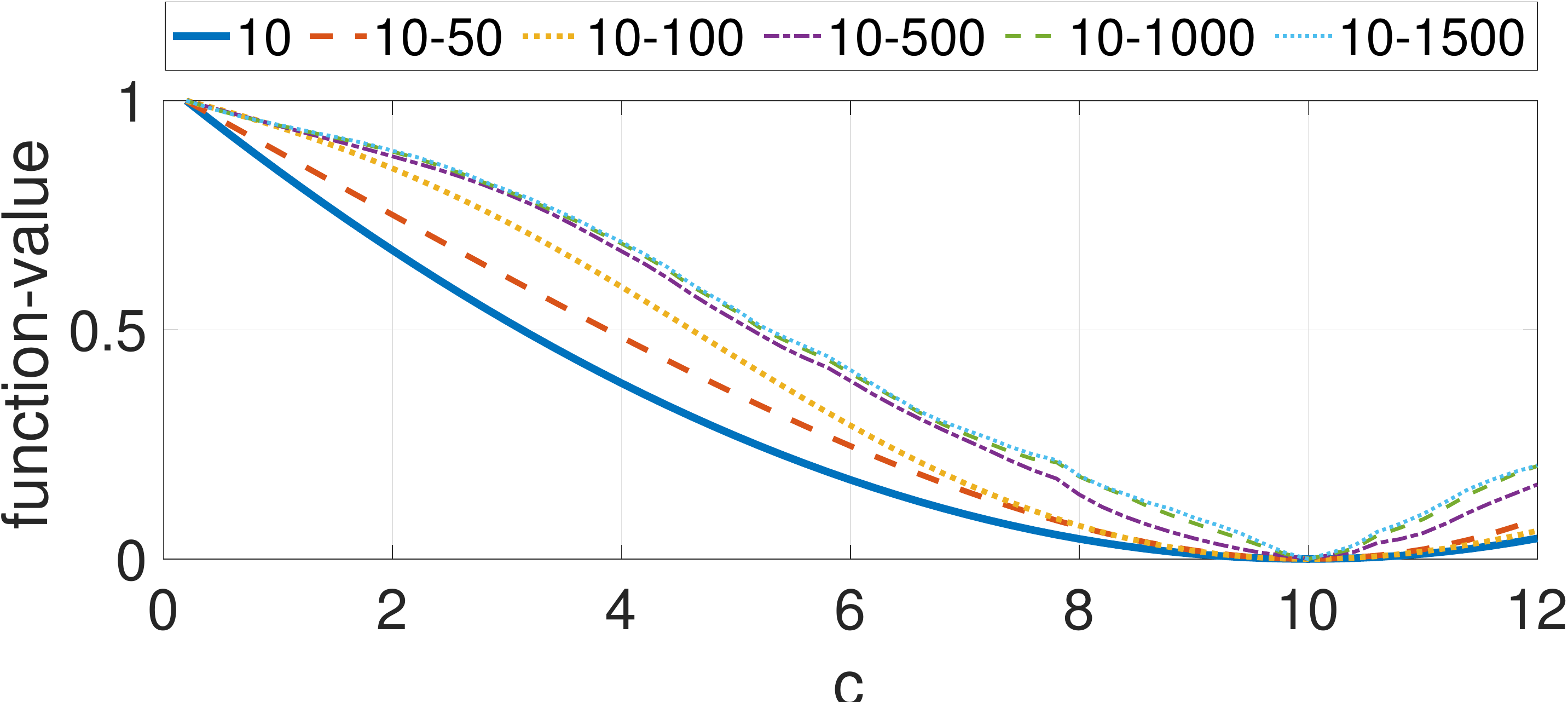}
\end{tabular}
\caption{\small (a) Illustration of a cylindrical object with true reflectivity equal to $c^\star = 10$ measured by five co-located transmitters and receivers. Topology of the cost function per frequency (b), per frequency batch of size 10 (c),  and incremental frequency batch (d) relative to the estimated reflectivity $c$. All the frequencies are in MHz.} \vspace{-6mm}
\label{fig:topology}
\end{figure*}

\vspace{-0.1cm}
The least-squares cost function in~\eqref{eq:PDE-constr-LS-reg} provides a natural separation across frequencies. Moreover, the topology of the non-convex cost function varies drastically between frequencies and can be leveraged to find a good local minimum. We illustrate this behavior using a simple cylindrical model for the target with a constant reflectivity $c$ as shown in Figure~\ref{fig:topology}(a). The true target has a reflectivity $c = 10$ and is illuminated with five transmitters located at a y-position of -0.6m. The transmitters and receivers are collocated and are equidistantly placed between $x=-0.5$m and $x=0.5$m. We plot in Figure~\ref{fig:topology}(b) the value of the data-fidelity cost function $\mathcal{F}_{j}( \mathbf{f})$ for various $j$ values.  Notice that for higher-frequency wavefields, the cost function exhibits many local minima that are farther away from the global minimizer than for the low-frequency wavefields.

A popular approach in the exploration geophysics community is to solve a sequence of inverse problems starting with a low-frequency batch, and then sliding linearly towards the high frequencies keeping the batch-size fixed. In Figure~\ref{fig:topology}(c), we plot such cost function $\left( \sum_{j \in \mathcal{J}_{\text{b}}} \mathcal{F}_j(\mathbf{f}) \text{ with } \mathcal{J}_{\text{b}} = \{ j_0, j_0+1, \dots, j_0+n_{b}-1 \}\right)$ for various frequency batches. We observe that the higher frequency batch has many local minima. The sliding approach works only when we get close to the global minimizer during the low-frequency batch inversions. A more robust approach would be to keep the low-frequencies as regularizer when inverting with high-frequency data. We plot the cost function $\left( \sum_{j=1}^{j_{\text{max}}} \mathcal{F}_j(\mathbf{f}) \right)$ in Figure~\ref{fig:topology}(d). Notice that the cost functions are almost convex even when dealing with high-frequncy data. However, the functions are well-behaved primarily due to the very simplistic setting of this example, where everything about the target is known except for the permittivity $c$. The function behavior will be significantly more erratic when the structure of the target and its surrounding medium are unknown.

The observations above led us to use an incremental frequency inversion framework where the model of the object's permittivity is sequentially updated as higher frequencies are included in the inversion. Given a measured wavefield containing $n_f$ frequency components indexed in increasing order from 1 to $n_f$, our framework iteratively estimates the model from low to high-frequency while keeping the low-frequency cost function as a regularizer for high-frequency inversions.
\begin{equation}
	\begin{split}
	\text{for } k = 1, &\dots, n_f : \\
		\mathbf{f}^{(k)} & \triangleq \, \underset{\mathbf{f}}{\argmin} \Bigg\lbrace \sum\nolimits_{j \in \mathcal{J}_k} \mathcal{F}_j(\mathbf{f})  + \, \mathcal{R}(\mathbf{f}) \Bigg\rbrace.
	\end{split}
	\label{eq:seqFramework}
\end{equation}
Therefore, instead of solving a single nonconvex minimization problem in~\eqref{eq:PDE-constr-LS-reg}, we solve $n_f$ subproblems sequentially according to~\eqref{eq:seqFramework}, where the sequence of solutions moves us closer to the global minimizer of~\eqref{eq:PDE-constr-LS-reg}.

\vspace{-0.3cm}
\subsection{Regularization}
\vspace{-0.1cm}
In this section we provide details on the total variation norm and the non-negativity constraints we use to regularize the problem, as well as their implementation through a proximal operator.
\subsubsection{Total-variation}
The Total-Variation (TV) norm of a compactly supported function $u : \Omega \mapsto \mathbb{R}$ is formally defined as
\begin{align*}
	TV(u) &\triangleq \sup \left\lbrace \int_{\Omega} u(\mathbf{x}) \; \mbox{div} \phi \; \mathrm{d}\mathbf{x} : \| \phi \|_{\infty} \leq 1 \right\rbrace,
\end{align*}
where $ \phi \in \mathcal{C}_c^{1} \left( \Omega, \mathbb{R}^d \right)$ denotes the set of continuously differentiable functions of compact support contained in $\Omega$. This norm measures the total change in the function over a finite domain~\cite{rudin1992nonlinear}. If $u$ is differentiable, then we can express the total-variation using an integral
\[
		TV(u) = \int_{\Omega} \| \nabla u(\mathbf{x}) \|_1 \; \mathrm{d}\mathbf{x},
\]
where $\| \cdot \|_1$ denotes the $L_1$ norm or Manhattan norm. As a result, regularization with a TV norm promotes piecewise constant approximation of the true model~\cite{unser2017splines}. In a discrete two dimensional setting, the TV-norm is represented  as
\begin{align}
	TV(\mathbf{f}) &= \| \mathbf{D} \mathbf{f} \|_1 \qquad \mbox{where} \quad \mathbf{D} = \begin{bmatrix}
	\mathbf{I}_x \otimes \mathbf{D}_y \\
	\mathbf{D}_x \otimes \mathbf{I}_y
	\end{bmatrix}.
	\label{eq:discretizedTV}
\end{align}
The $\mathbf{D}_x$ and $\mathbf{D}_y$ are the finite difference operators in x and y directions, and $\mathbf{I}_x, \mathbf{I}_y$ are the identity operators. We adopt the TV regularization in its constrained form \cite{esser2018total}, such that,
\begin{equation}
\begin{array}{ll}
	\mathcal{R}_{TV}(\mathbf{f}) & \triangleq \delta_{TV \leq \tau} \left( \mathbf{f} \right) = \delta_{\| \cdot \|_1 \leq \tau} \left( \mathbf{D} \mathbf{f} \right).
\end{array}
\label{eq:TVreg}
\end{equation}
where $\delta_\mathcal{C}(\cdot)$ is an indicator function to the set $\mathcal{C}$, and $\tau$ is a constraint parameter. The second line in~\eqref{eq:TVreg} expresses the discretized version of the TV regularization function using the constrained $\ell_1$-ball.
We note here that the proximal for an indicator function to set $\mathcal{C}$ corresponds to the projection of a vector onto the set $\mathcal{C}$. Efficient algorithms exist for the projection onto the $\ell_1$-norm ball (see, for example, \cite{condat2016fast}).

\subsubsection{Non-negative Constraints}
Since the contrast function is defined as the relative permittivity of an object (with respect to vacuum), it will always be non-negative. Hence, we impose this prior information using a regularization
\[
	\mathcal{R}_{\text{NN}} (\mathbf{f} ) = \delta_{\geq 0} (\mathbf{f}),
\]
where $\delta_{\geq 0}$ denotes the indicator to a non-negative orthant. 
The proximal operator for this function corresponds the projection of a vector onto a non-negative orthant. In specific, the projection operator is
\[
	\mathcal{P}_{\geq 0} (y) = \begin{cases}
	y & \quad \text{if} \;  y \geq 0 \\
	0 & \quad \text{otherwise}
	\end{cases}.
\]

\subsubsection{Implementation}
In order to impose the non-negative + TV constraints, we define the proximal operator:
\begin{equation}
	\prox_{\gamma \mathcal{R}}\!\! \left( \mathbf{w} \right) \triangleq  \argmin_{\mathbf{f}} \left\lbrace \tfrac{1}{2\gamma} \| \mathbf{f} - \mathbf{w} \|^2 + \mathcal{R}_{\text{TV}}\!(\mathbf{f})   + \mathcal{R}_{\text{NN}}\!(\mathbf{f}) \right\rbrace,
	\label{eq:projNNTV}
\end{equation}
with $\gamma >0$. The proximal operator becomes a projection onto the intersection of the sets: the TV-norm ball set and the non-negative orthant set. Although there is no simple analytical expression for this proximal operator, it can be evaluated efficiently using various splitting methods, \eg, the alternating direction method of multipliers (ADMM)~\cite{boyd2011distributed} and/or primal-dual method~\cite{chambolle2011first}. Here, we use the primal-dual method, which we derive for the sum of three convex functions in Appendix~C. 
Algorithm~\ref{alg:TVprojPD} describes the primal-dual method to solve \eqref{eq:projNNTV}.

\begin{algorithm}[!b]
 \setstretch{1.05}
 \caption{Proximal for Non-negative + Total-Variation}
 \begin{algorithmic}[1]
 \renewcommand{\algorithmicrequire}{\textbf{Input:}}
 \renewcommand{\algorithmicensure}{\textbf{Output:}}
 \Require $\mathbf{w} \in \mathbb{R}^{n}, \mathbf{D} \in \mathbb{R}^{m \times n}, \gamma > 0, \tau > 0, t_{\text{max}}$
 \Ensure  $\mathbf{f}^{\star} \approx \mathbf{f}_{t_{\text{max}}}$
  \State $ \mathbf{f}_{0} = \mathbf{0}, \mathbf{u}_{0} = \mathbf{0}, \mathbf{v}_{0} = \mathbf{0}$
  \State choose $\alpha \in \left(0, 1/\sqrt{\|\mathbf{D}^T \mathbf{D} + \mathbf{I} \|} \right)$
  \While {$t < t_{\text{max}}$}
  \State $\hat{\mathbf{f}} = \mathbf{f}_{t} - \alpha \left( \mathbf{D}^T \mathbf{u}_{t} + \mathbf{v}_{t} \right) $
  \State $\mathbf{f}_{t+1} = ( \gamma \hat{\mathbf{f}} + \alpha \mathbf{w} ) / \left( \alpha + \gamma \right)$
  \State $\mathbf{u}_{t+1} = \mathbf{u}_{t} - \alpha \mathcal{P}_{\| \cdot \|_1 \leq \tau}\left( \mathbf{u}_{t}/\alpha + \mathbf{D} \left( 2\mathbf{f}_{t+1} - \mathbf{f}_{t} \right) \right) $
  \State $\mathbf{v}_{t+1} = \mathbf{v}_{t} - \alpha \mathcal{P}_{\geq 0}\left( \mathbf{v}_{t}/\alpha + \left( 2\mathbf{f}_{t+1} - \mathbf{f}_{t}\right)\right) $
  \State $t = t + 1$
  \EndWhile
 \end{algorithmic}
 \label{alg:TVprojPD}
\end{algorithm}

\subsection{Proximal Quasi-Newton Method}
To solve each subproblem in~\eqref{eq:seqFramework}, \ie,  
\begin{align}
	\mathbf{f}^\star = \argmin_{\mathbf{f}} \Big\lbrace \sum\nolimits_{j \in \mathcal{J}_k}\mathcal{F}_j \left( \mathbf{f} \right) + \mathcal{R} \left( \mathbf{f}\right) \Big\rbrace.
	\label{eq:seqSubProblem}
\end{align}
we propose a proximal Quasi-Newton (prox-QN) method. For simplicity of illustration, we enumerate the steps in Algorithm~\ref{alg:prox-QN}, but provide a complete derivation in Appendix~D. 

\begin{algorithm}[!t]
 \caption{Prox-QN method for solving \eqref{eq:seqSubProblem}}
 \begin{algorithmic}[1]
 \renewcommand{\algorithmicrequire}{\textbf{Input:}}
 \renewcommand{\algorithmicensure}{\textbf{Output:}}
 \Require $\mathbf{f}^{(0)}, \tau > 0, \gamma \in (0, 1)$
 \Ensure  $\mathbf{f}^\star$
  \For {$i = 0$ to $i_{\text{max}}$}
  \State compute the gradient $g_i = \sum_{j=1}^{k} \nabla \mathcal{F}_{j} \left( \mathbf{f}^{(i)} \right) $.
  \State compute the approximate Hessian $\mathbf{H}_i$
  \State compute $\mathbf{s}_i$ from equation~\eqref{eq:prox-QN:iterates}(a).
  \State define $\hat{\mathbf{f}}(\alpha) = \prox_{\gamma \mathcal{R}} \left(  \mathbf{f}^{(i)} + \alpha \mathbf{s}_i \right)$
  \State $\alpha_i = \mbox{linesearch}_{\alpha} \left( \hat{\mathbf{f}}(\alpha) \right)$.
  \State $\mathbf{f}^{(i+1)} := \prox_{\gamma \mathcal{R}} \left(  \mathbf{f}^{(i)} + \alpha_i \mathbf{s}_i \right)$ using Algorithm~\ref{alg:TVprojPD}.
  \State check optimality conditions
  \EndFor
 \end{algorithmic}
 \label{alg:prox-QN}
\end{algorithm}

 The algorithm consists of two loops. The inner loop, implicit in step 4 and described in Appendix~\ref{sec:prox-QN}, finds the search direction, while the outer loop computes the next iterate based on the computed search direction and step length.  At every sequence of the outer loop, we compute the gradient using an adjoint-state method, and form the approximate Hessian with the L-BFGS procedure. The procedure to compute the gradient is explained in Appendix~B. 
 Once we have the gradient and approximate Hessian at the current iterate, we compute the search direction using the primal-dual method (see \eqref{eq:prox-QN:primalDual}). Next, we search for the feasible step length using the backtracking linesearch. Finally, we compute the next iterate using Algorithm~\ref{alg:TVprojPD}. 

The computational complexity of Prox-QN method relies on step 2 of Algorithm~\ref{alg:prox-QN}. The gradient computation involves solving an adjoint of Lippmann-Schwinger equation. We use GMRES method which has complexity of $\mathcal{O}(nt^2)$, where $n$ is the size of image, and $t$ is the number of iterations. For $n_t$ transmitters and $n_f$ frequencies, the step 2 involves $\mathcal{O}(nt^2 n_t n_f)$ floating-point operations. Steps 3 to 8 have lower complexity order than that of step 2. Hence, Prox-QN has $\mathcal{O}(nt^2 n_t n_f i_{\text{max}})$ computational complexity. For a single frequency of 100 Hz ($n_f = 1$) with 5 transmitters on a $32 \times 32$ image, it took approximately 2 minutes to run 100 iterations of prox-QN on a 2.7 GHz Intel Core i5 processor with 8 GB RAM.
\section{Estimating the constraint parameter $\tau$}
Recall that for each subproblem~\eqref{eq:seqSubProblem} in our proposed framework, we are solving a TV-constrained nonlinear least-squares problem where the constraint parameter $\tau_k$ should bound the total variation of the solution. Naturally, the choice of constraint parameter $\tau_k$ would significantly affect the reconstruction performance.

In order to estimate $\tau_k$ for each new subproblem, we develop a parameter estimation routine inspired by the approach in~\cite{BergFriedlander:2011} for sparse optimization with linear least squares constraints. Suppose that we have an initial estimate of $\mathbf{f}^k$ obtained at the frequency corresponding to the $k^{\text{th}}$ subproblem for which the TV norm $\tau_k = TV(\mathbf{f}^k)$, specifically,
\begin{equation}
	\mathbf{f}^k = \argmin_{\mathbf{f}} \Big\lbrace \sum\nolimits_{j \in \mathcal{J}_k}\mathcal{F}_j \left( \mathbf{f} \right) \ \textrm{s.t.} \ || \mathbf{D}\mathbf{f}||_1 \leq \tau_k \Big\rbrace,
	\label{eq:seqSubProblem_k}
\end{equation}
where $\mathcal{F}_j$ is as defined in~\eqref{eq:Fj}, and the constraints $\mathbf{C}_j(\mathbf{f}^k, \mathbf{U}_j) = \mathbf{0}$ are satisfied for all $j \in \mathcal{J}_k$. At subproblem $k+1$, the cost $\mathcal{F}_{k+1} \left( \mathbf{f} \right)$ is added to the objective function, resulting in the potentially unsatisfied constraint
\begin{align}
	\mathbf{V}_{k+1} = \mathbf{A}_{k} \mathbf{U}_{k+1},
\label{eq:constraint_k}
\end{align}
where $\mathbf{A}_{k} \triangleq \mathbf{I} - \mathbf{G} \diag( \mathbf{f}^k)$. To overcome the nonconvexity of the objective function due to~\eqref{eq:constraint_k}, we linearize the objective function around $\mathbf{f}^k$ by estimating $\mathbf{U}_{k+1}^\star = \mathbf{A}_{k}^{-1} \mathbf{V}_{k+1}$, thus reducing $\mathcal{F}_{k+1} \left( \mathbf{f} \right)$ to a convex least squares cost function in $\mathbf{f}$, i.e.,
\[
	\mathcal{F}_{k+1} \left( \mathbf{f} \right) \approx  \mathcal{D}_{k+1} \left( \mathbf{f},\mathbf{U}_{k+1}^\star \right),
\]
where $\mathcal{D}_{k+1} \left( \mathbf{f},\mathbf{U}_{k+1}^\star \right)$ is the data mismatch cost function defined in~\eqref{eq:Fj}. Consequently, we may now define a value function $\Phi(\tau)$ for the $\left( k+1 \right)^{\text{th}}$ subproblem as
\begin{equation}
	\begin{array}{ll}
	\Phi(\tau) \! \! & = \argmin\limits_{\mathbf{f}} \Big\lbrace \! \sum\nolimits_{j \in \mathcal{J}_{k+1} }\! \! \mathcal{D}_{j} \left( \mathbf{f},\mathbf{U}_{j}^\star \right) \ \textrm{s.t.} \ || \mathbf{D}\mathbf{f}||_1 \leq \tau \Big\rbrace \\
	& = \argmax\limits_{\lambda} \Bigg\lbrace   \sum\nolimits_{j \in \mathcal{J}_{k+1} } \! \! \mathbf{r_j}^H \mathbf{Y}_{j} / \|\mathbf{r}^{k+1}\| - \tau\lambda  \\
	& \quad \textrm{s.t.} \ \text{TV}_{\text{polar}} \left( \frac{\sum\nolimits_{j \in \mathcal{J}_{k+1} }  \diag \left( \mathbf{U}_{j}^\star \right) \mathbf{H}_{j}^H \mathbf{r_j} }{\|\mathbf{r}^{k+1}\|} \right)\leq \lambda \Bigg\rbrace
\end{array}
\label{eq:ValueFunction}
\end{equation}
where $\mathbf{r}_{j} = \mathbf{Y}_{j} - \mathbf{H}_{j} \diag \left( \mathbf{U}_{j}^\star \right) \mathbf{f}_k$ is the data residual at the $j^{\text{th}}$ frequency, and $\mathbf{r}_{k+1}$ is the vector formed by concatenating all the vectors $\mathbf{r}_j$, such that, $\|\mathbf{r}^{k+1}\|^2 = \sum_{j \in \mathcal{J}_{k+1} } ||\mathbf{r}_j||^2 $. The $\text{TV}_{\text{polar}}$ function is defined as $\text{TV}_{\text{polar}} (\mathbf{x}) = \| \mathbf{D}^{-T} \mathbf{x} \|_{\infty}$, with $\mathbf{D}^{-T}$ being the transposed pseudo-inverse of the finite difference operator $\mathbf{D}$ defined in~\eqref{eq:discretizedTV}. Note that~\eqref{eq:ValueFunction} shows the primal and dual problems for computing the value function $\Phi(\tau)$.

The dual problem in~\eqref{eq:ValueFunction} conveniently shows that the maximum is achieved when $\lambda$ is at its minimum $\lambda^{\star} = \text{TV}_{\text{polar}} \left( \sum_{j \in \mathcal{J}_{k+1} }  \diag \left( \mathbf{U}_{j}^\star \right) \mathbf{H}_{j}^H \mathbf{r_j} / \|\mathbf{r}\| \right)$. Moreover, the gradient of $\Phi(\tau)$ with respect to $\tau$ is easily computed as $\nabla_{\tau}\Phi(\tau) = \lambda^{\star}$. Therefore, we can compute the update for $\tau$ using a Newton root finding step, such that,
\begin{equation}
	\tau_{k+1} = \tau_{k} + \frac{\| \mathbf{r}^{k+1} \| \left( \| \mathbf{r}^{k+1} \|  - \sigma_{k+1}\right)}{\text{TV}_{\text{polar}} \left( \sum\nolimits_{j \in \mathcal{J}_{k+1} }  \diag \left( \mathbf{U}_{j}^\star \right) \mathbf{H}_{j}^H \mathbf{r_j} \right)},
	\label{eq:tauEst}
\end{equation}
where $\sigma_{k+1}$ is the upper bound on the $\ell_2$ norm of the noise up to the $k+1$ frequency bin. Finally, we note that at the zeroth iteration, the parameter $\tau$ can be set to zero, resulting in a homogeneous solution for $\mathbf{f}_0$.

\section{Numerical Experiment}\label{sec:results}
In this section, we describe the experimental setup for the reflection tomography. We evaluate our method on two numerical phantoms and compare it with two other approaches. We also experiment with a partially non-inverse-crime dataset in Section~\ref{sec:nonInvCrime}.

\begin{figure*}[!htbp]
	\centering
	\begin{tabular}{cccc}
		\includegraphics[height=0.14\textheight]{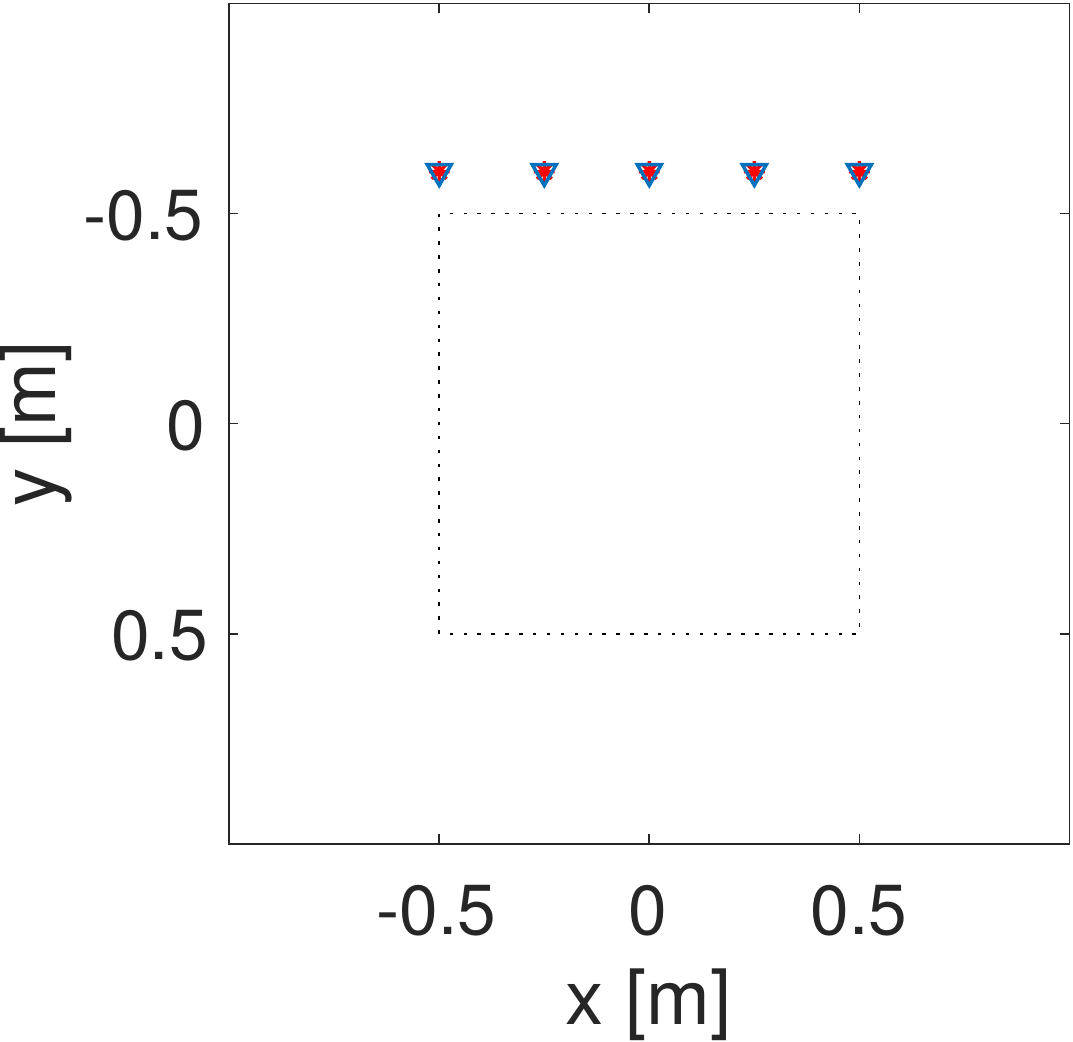} &  \includegraphics[height=0.14\textheight]{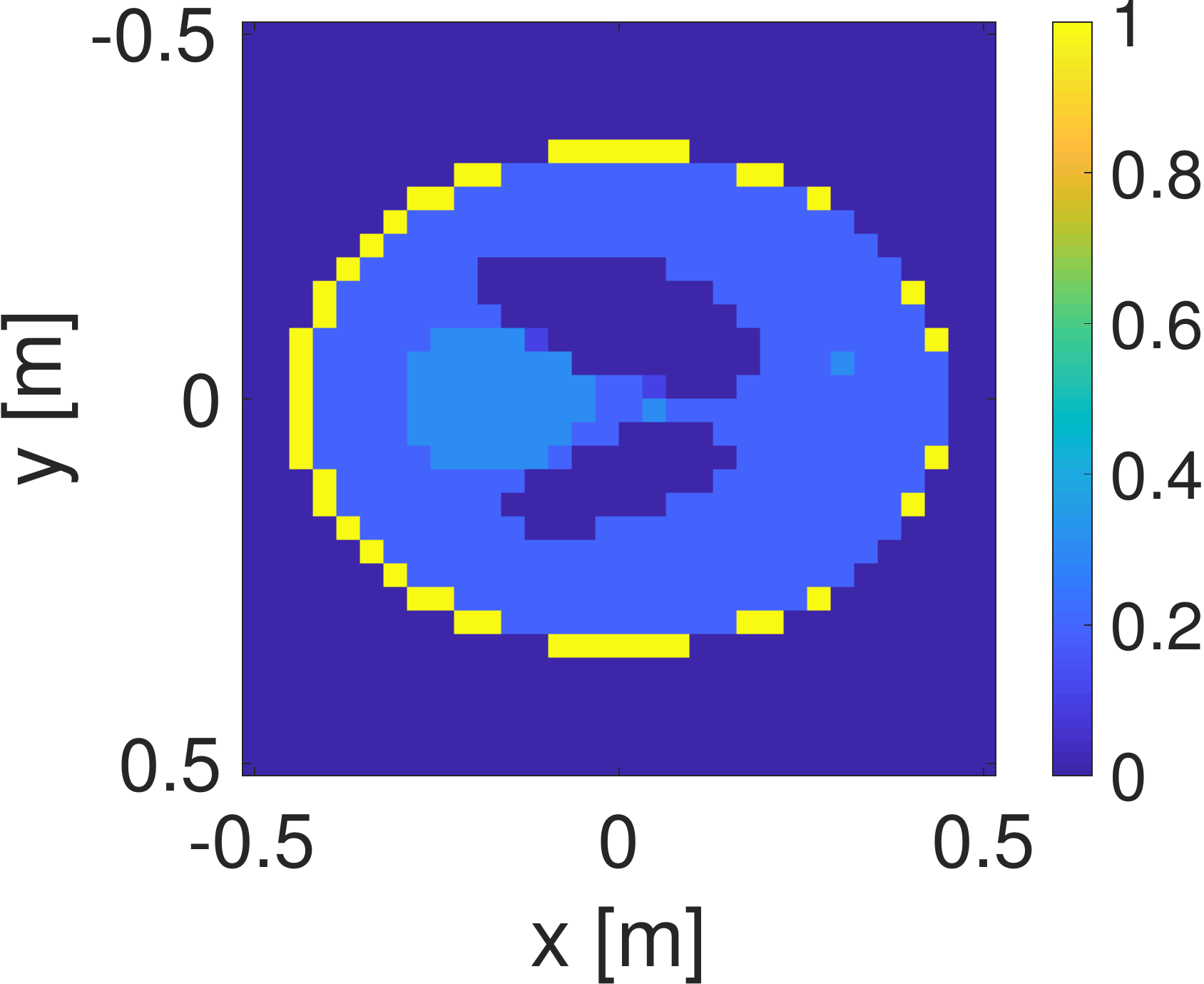} & \includegraphics[height=0.14\textheight]{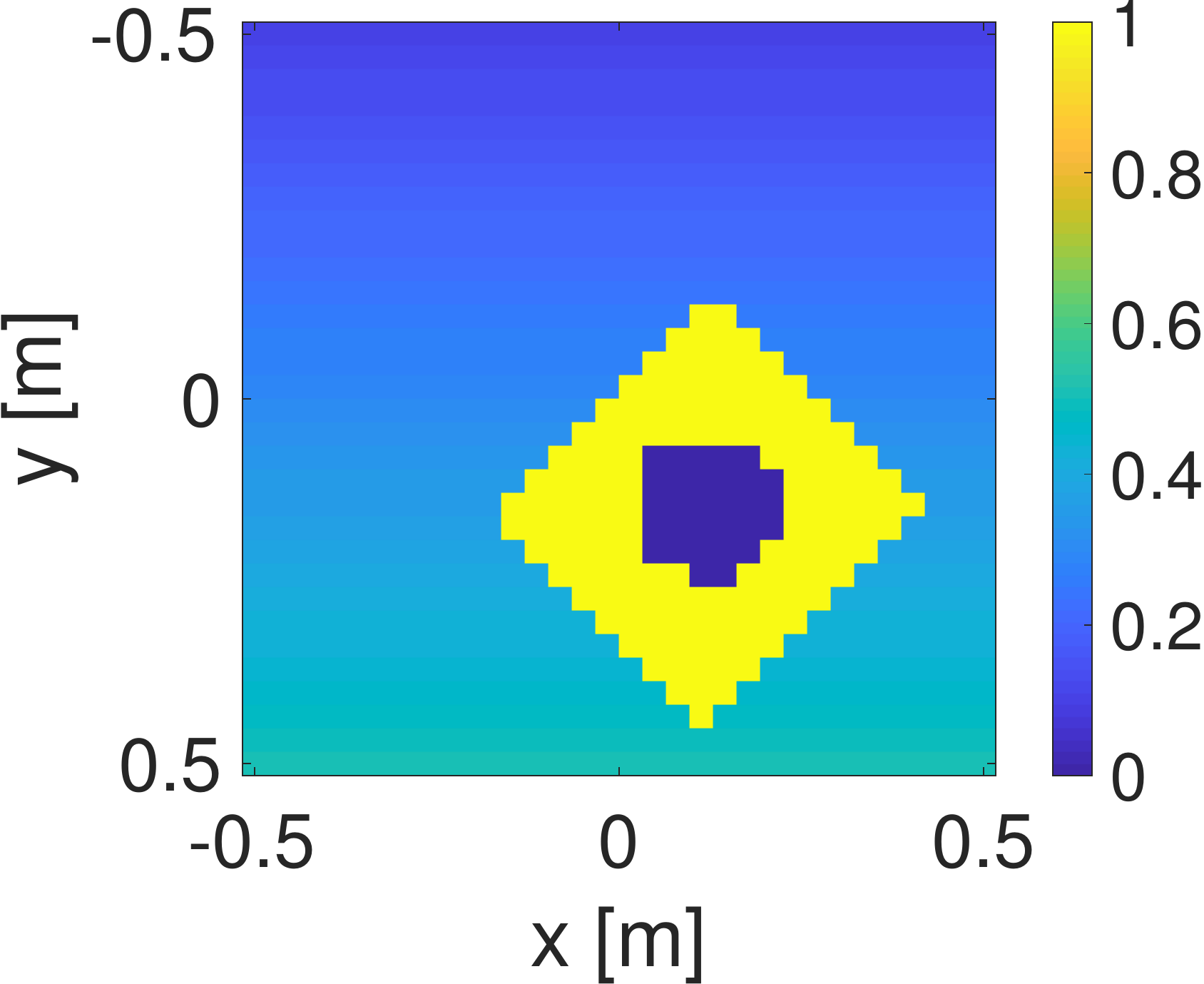} & \includegraphics[height=0.14\textheight]{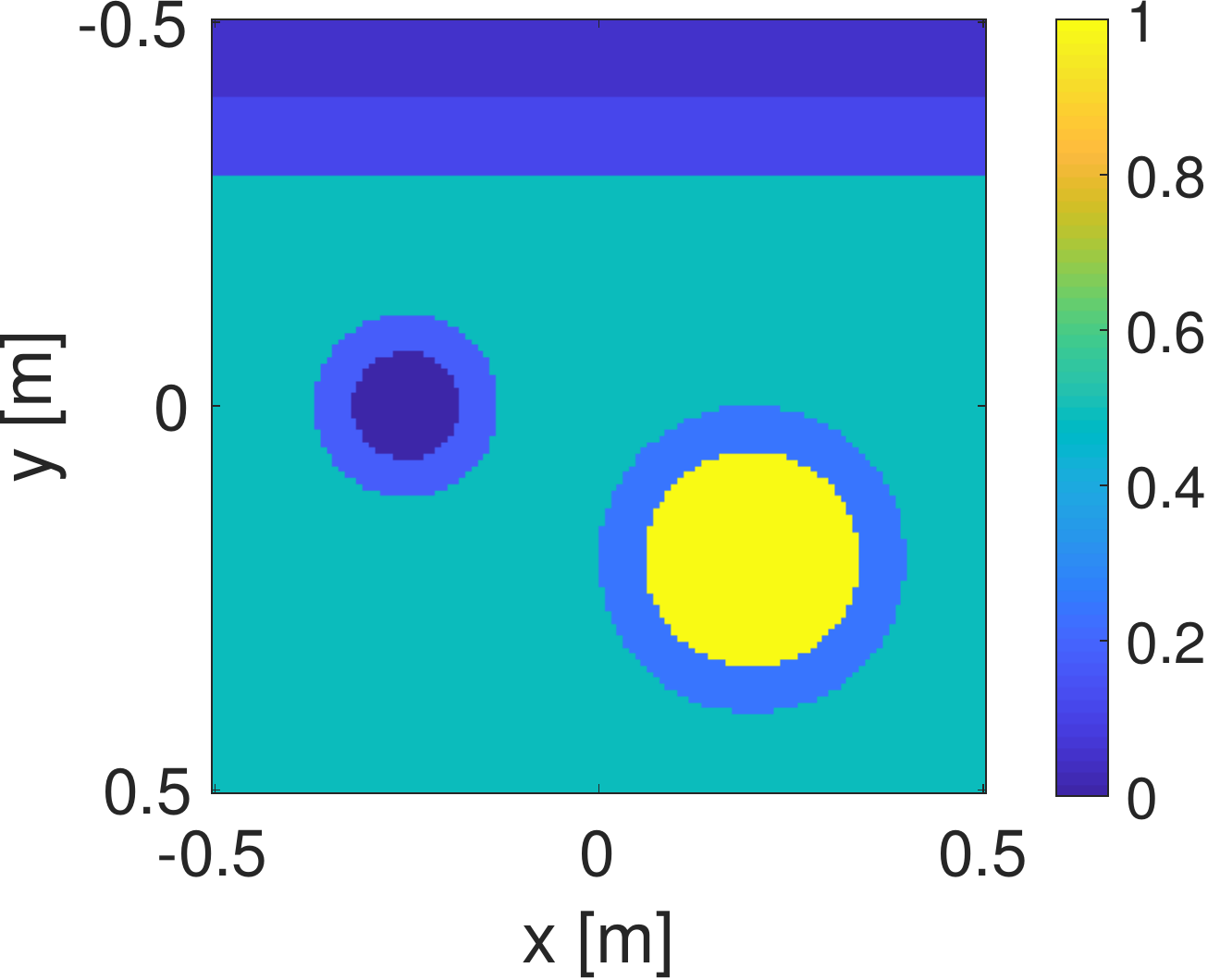} \\
			(a) Setup & (b) Phantom 1 & (c) Phantom 2 & (d) Phantom 3
	\end{tabular}
	\caption{(a) Tomography setup for all the numerical experiments. The dotted region denotes the object domain $\Omega$. The transmitters and receivers are collocated at $y = -0.6$~m. (b), (c), (d) are the three numerical phantoms used for the experimentation.}
	\label{fig:setupPh}
\end{figure*}

\subsection{Experimental details}
We consider an experimental setup illustrated in Figure~\ref{fig:setupPh}(a). The domain is 1 m $\times$ 1 m and extends in x-direction from $x = -0.5$~m to $0.5$~m and in y-direction from $y=-0.5$~m to $0.5$~m. There are total of five transmitters and receivers located on a line $y = -0.6$~m. Each transmitter illuminates a flat spectrum pulse occupying the frequency band $[10, 2000]$ MHz. All 5 receivers are activated for each transmitter. We consider three frequency bands: \textit{i)} a low frequency band consisting of $\{ 10 + 5 j \}$~MHz with $j = 0, \dots, 17$, \textit{ii)} a medium frequency band consisting of $\{ 100 + 50 j \}$~MHz with $j = 0, \dots, 17$, and \textit{iii)} a high frequency band consisting of $\{ 1000 + 100 j \}$~MHz with $j = 0, \dots, 10$. Hence, in total, we consider 47 frequencies between 10~MHz and 2000~MHz.

We work with 3 phantoms shown in Figure~\ref{fig:setupPh}(b)-(d). All phantoms have a length of 1~m in both x and y directions. Phantom1 is a Shepp-Logan phantom which resembles the brain. It is a well-known phantom in the image processing and tomography community. Here, we discretize it on $32 \times 32$ grid. It has total of 4 contrast values, namely $\{0, 0.2, 0.3, 1\}$. Phantom 2 resembles an underground scene. It has layer structure in the background whose contrast ranges from 0.1 to 0.5. A square-type hole (of contrast of 0) is embedded in a rhombus-type structure with a contrast of 1. This phantom also has a resolution of $32 \times 32$. We use these low resolution phantoms to compare our method with other exisiting methods and to check the robustness against the noise.

Phantom 3 is a high-resolution phantom depicting another underground scene. It has a resolution of $128 \times 128$. It contains 3 horizontal layers of contrast $\{0.05, 0.1235,0.5\}$. The phantom consists of two circular pipes of outer diameter 0.4~m and 0.24~m with a thickness of 0.6~m and 0.5~m respectively. A large pipe has an inner region filled with a high contrast material of permittivity 1 and a small pipe has a vaccum inside. We use this phantom to perform a partially non-inverse-crime test as described in Section~\ref{sec:nonInvCrime}.

\subsection{Comparison with other methods}
We restrict ourself to the two classical methods. For fair comparison we modify these methods to add the prescribed regularization. We do not compare with linearized methods like Born approximation and Rytov approximation as these methods have shown to fail for high-contrast imaging \cite{ma2018accelerated}.
\begin{description}
	\item[CISOR]: The CISOR algorithm aims to solve~\eqref{eq:PDE-constr-LS} by taking all frequencies at once~\cite{ma2018accelerated}. As opposed to TV-norm penalization, we use the proposed regularization, \ie, we regularize it with non-negative and total-variation constraints with known $\tau$ value. The TV constraint parameter is set to the total-variation of the true model. The problem is solved using a prox-QN method with a maximum of 5000 iterations or until convergence (norm of the gradient below $10^{-6}$).
	\item[RL:] Recursive linearization (RL) method was introduced in \cite{chen1997inverse}, and has been a standard while working with multi-frequency data. The method enjoys the computational benefit of solving a single constraint (\ie, solving a single linear system of equations) at a time, but might suffer in the high-contrast regime. It solves the sequence of problems
	\begin{equation*}
		\begin{split}
			\mathbf{f}^{(j)} \! \triangleq \underset{\mathbf{f}}{\argmin} \bigg\lbrace \mathcal{D}_{j}  ( \mathbf{f},\mathbf{U}_{j}) \quad \mbox{s.t.} \quad  \mathcal{\mathbf{C}}_j  (\mathbf{f}, \mathbf{U}_j) = \mathbf{0} \bigg\rbrace,
		\end{split}
	\end{equation*}
	with an initial guess to each subproblem being the solution of the previous subproblem. We modify the cost function to include the regularization. Similarly to the CISOR, we consider non-negative and TV regularization with known $\tau$ value. Each subproblem is solved using a prox-QN method with a maximum of 500 iterations.
	\item[SF-$\tau$:] This method corresponds to the proposed sequential framework with known $\tau$ value. It solves the problem described in \eqref{eq:seqFramework}. We use a prox-QN method to solve each subproblem with a maximum of 500 iterations or until converge.	\item[SF-$\sigma$:] This method corresponds to the proposed sequential framework with estimation of $\tau$ at each iteration. It solves the problem described in \eqref{eq:seqFramework} with the $\tau$ estimation from \eqref{eq:tauEst}. Here, we assume that the noise-level $\sigma$ is known. We use a prox-QN method to solve each subproblem with a maximum of 500 iterations or until converge.
\end{description}
For all the methods the initial model corresponds to a contrast of 0 everywhere.

\begin{figure}[!b]
	\centering
	\setlength\tabcolsep{1.5pt}
	\begin{tabular}{c | c cccc}
	& & {\small \bf CISOR} &  {\small \bf RL} &  {\small \bf SF-$\tau$} & {\small \bf SF-$\sigma$} \\ \midrule
	\rotatebox[origin=l]{90}{{\small $ f_\text{max}=1$ }} & & \includegraphics[width=0.8in]{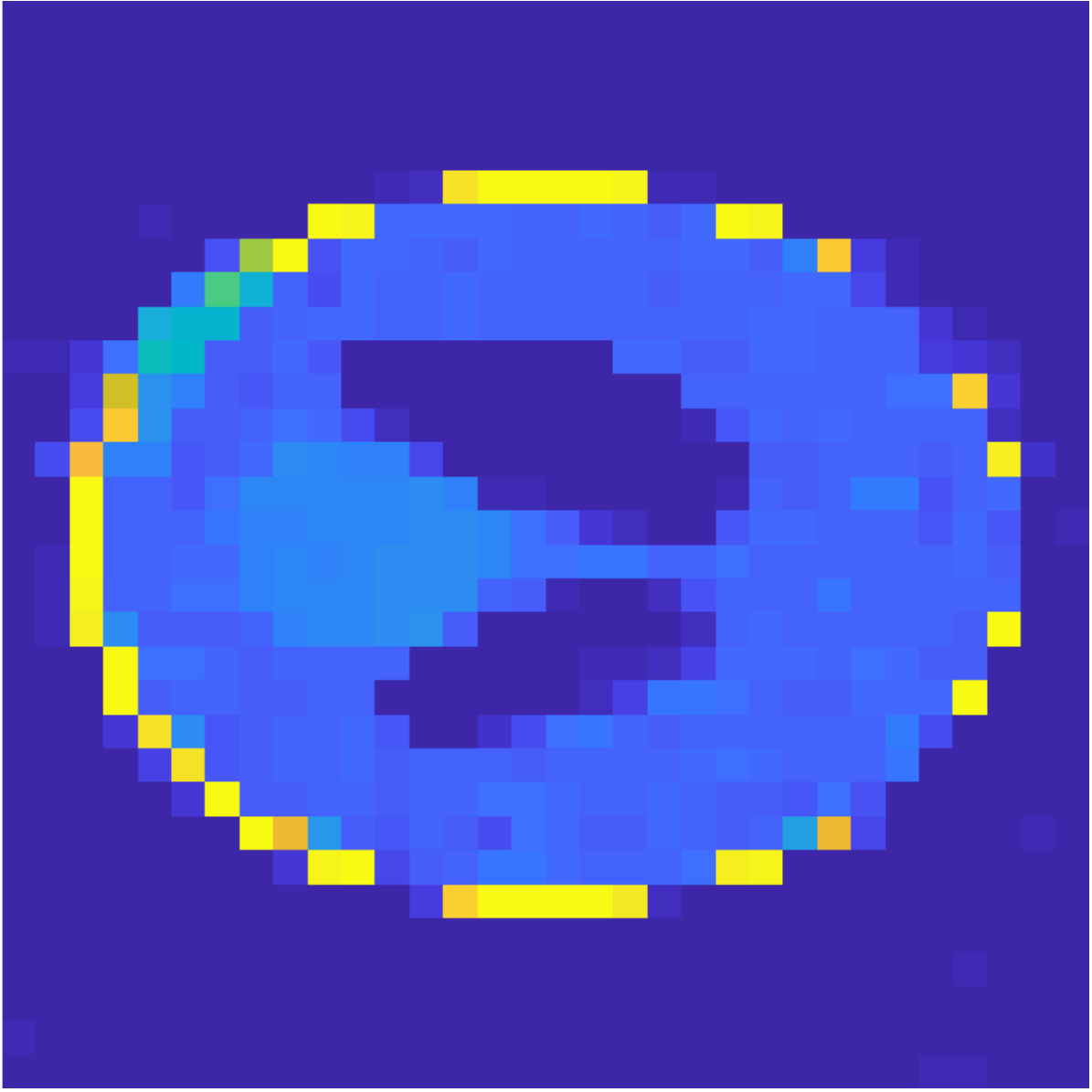} & \includegraphics[width=0.8in]{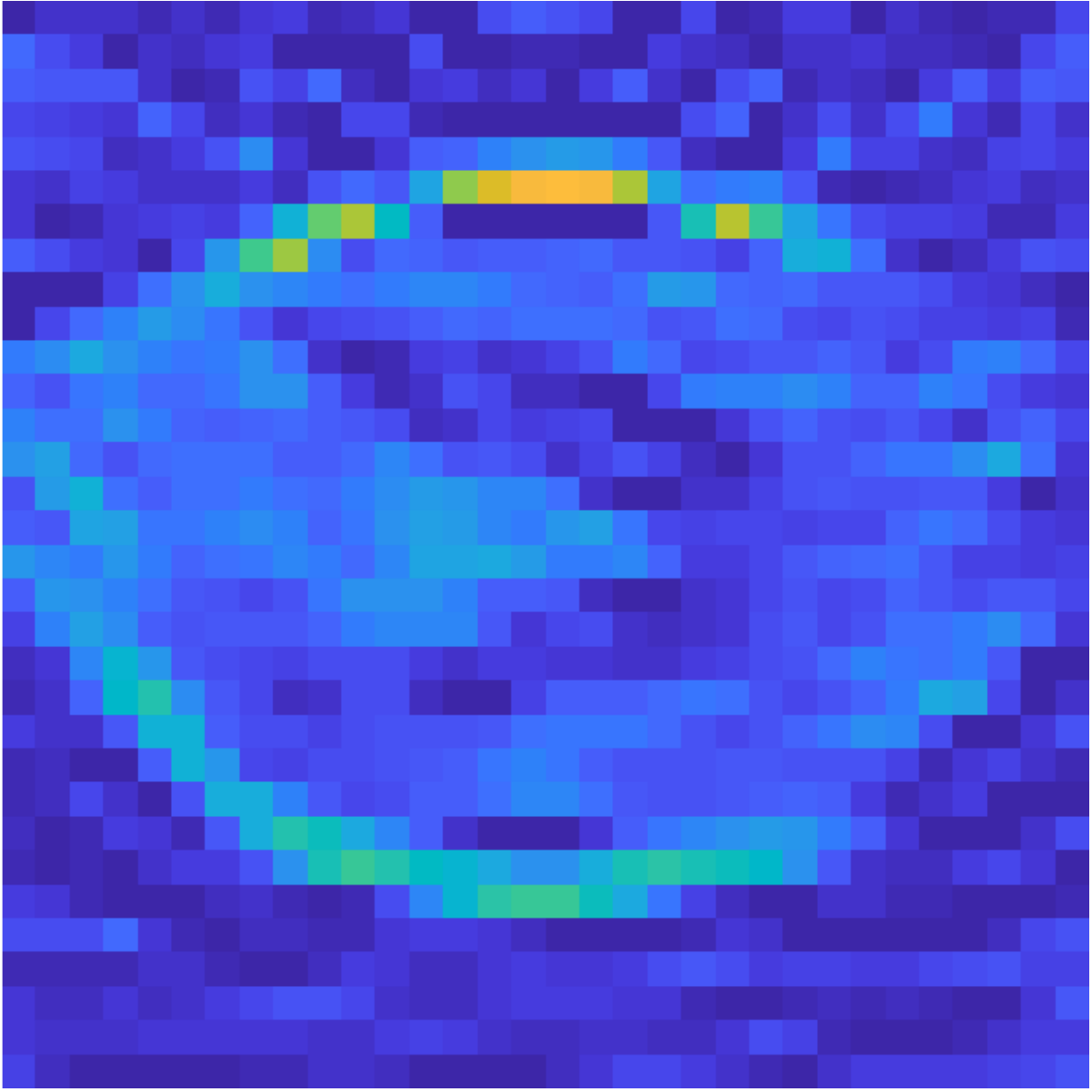} & \includegraphics[width=0.8in]{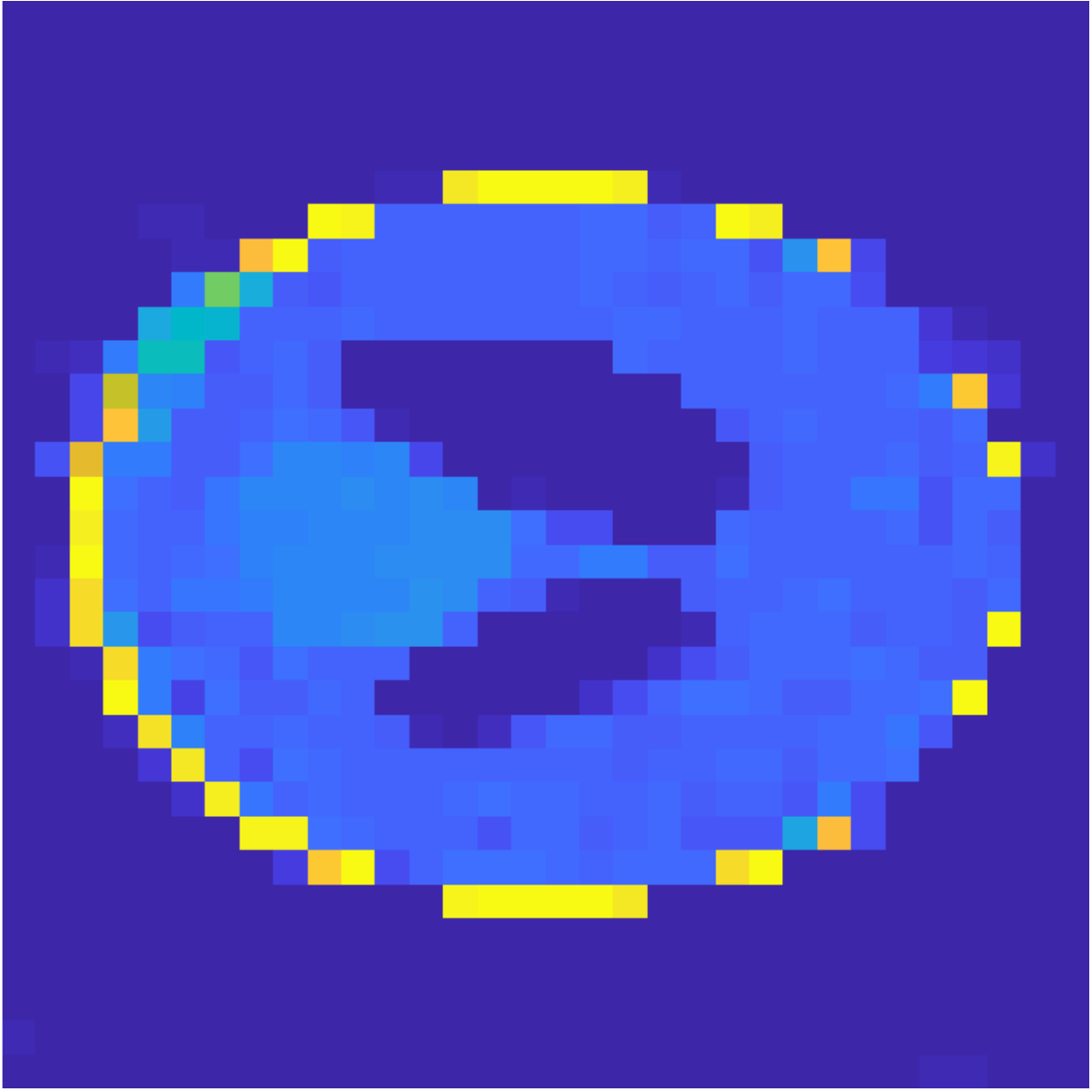} & \includegraphics[width=0.8in]{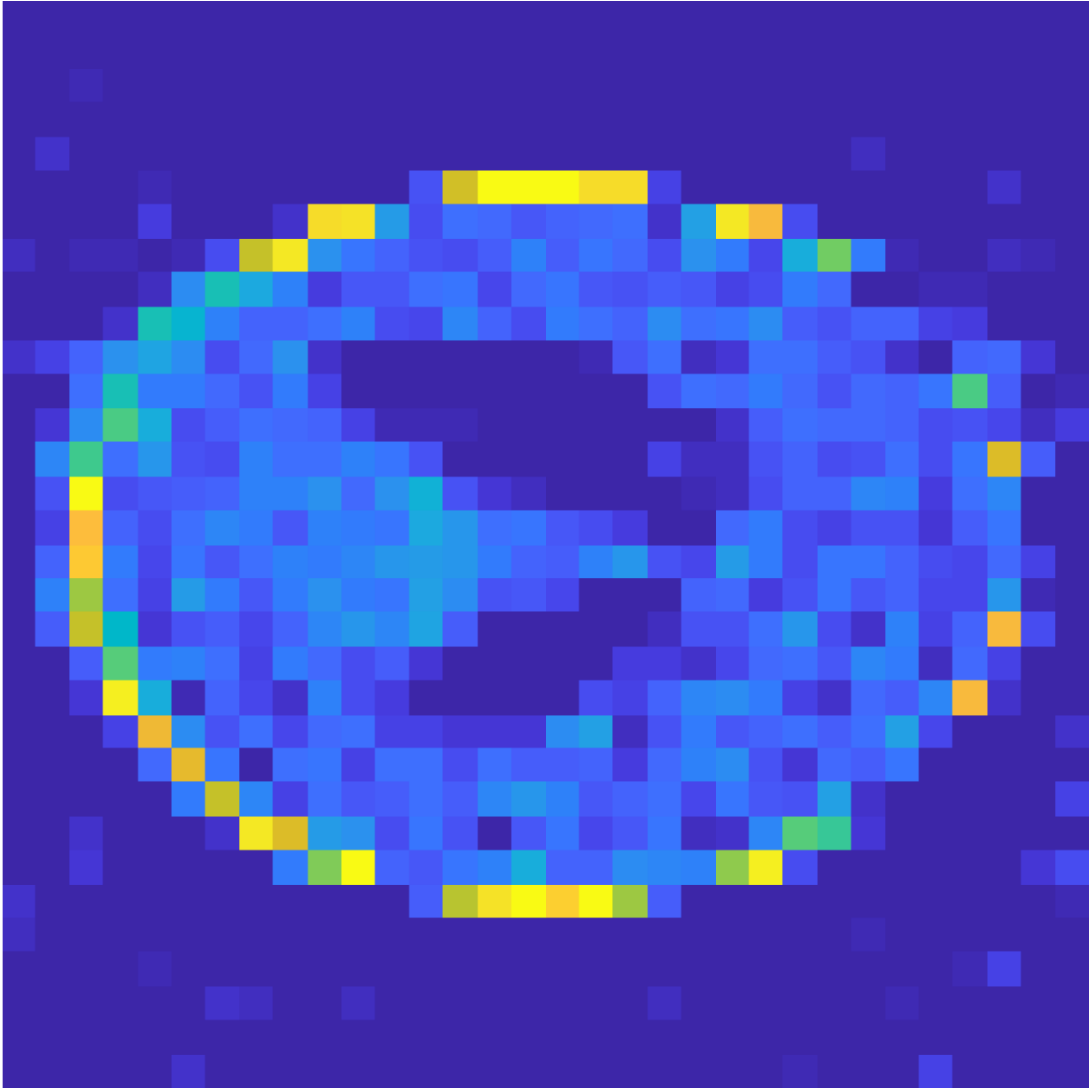} \\
	\rotatebox[origin=l]{90}{{\small $ f_\text{max}=10$ }} & & \includegraphics[width=0.8in]{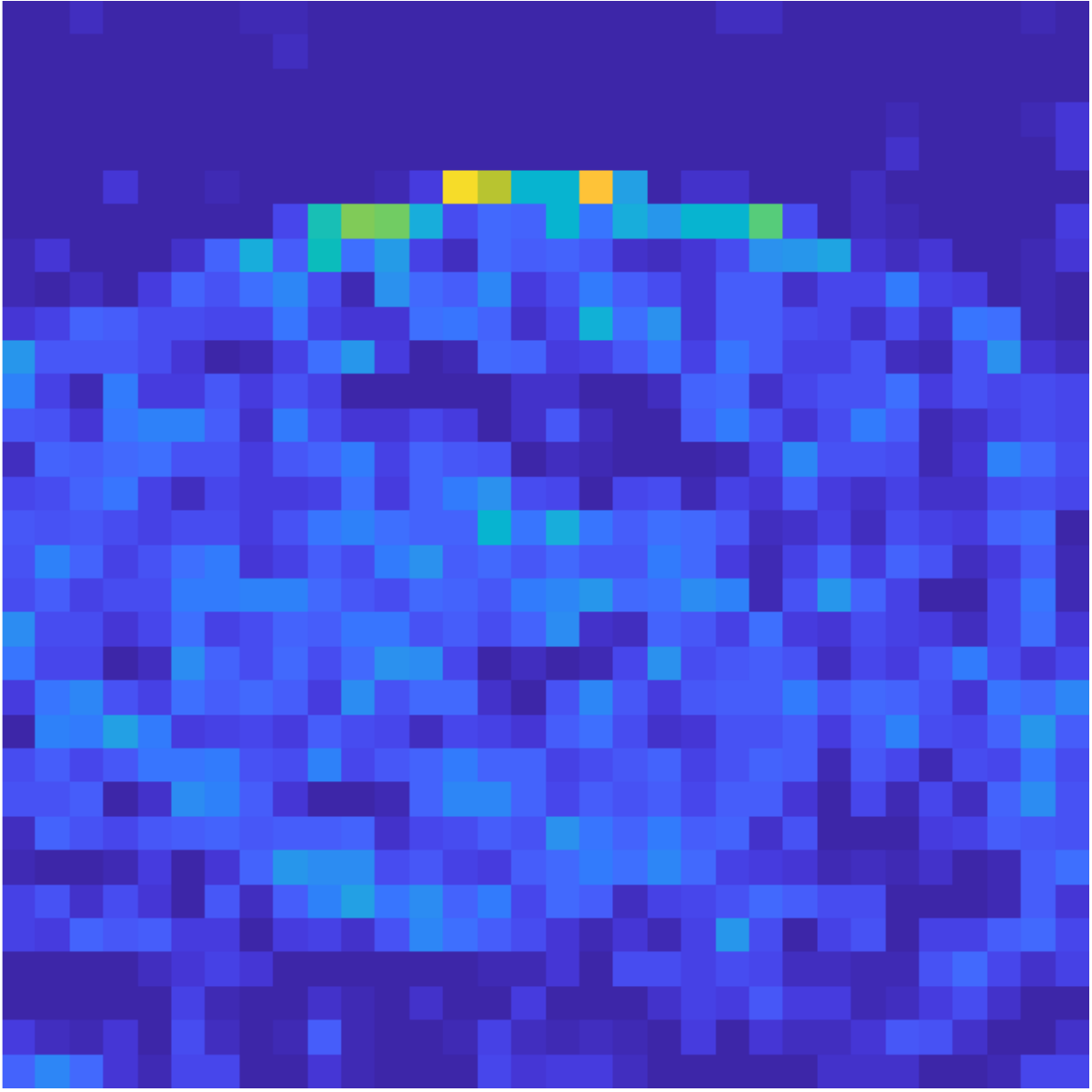} & \includegraphics[width=0.8in]{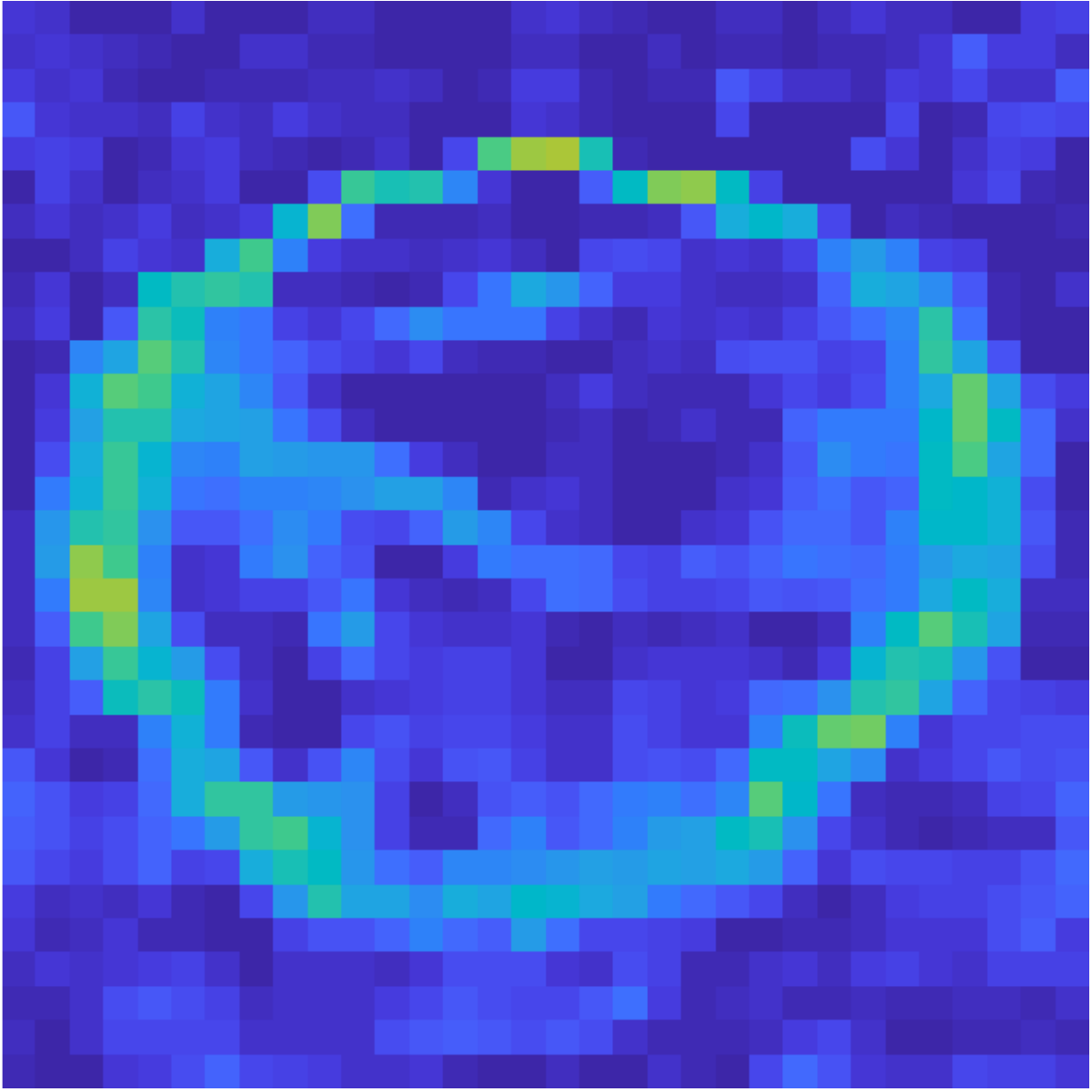} & \includegraphics[width=0.8in]{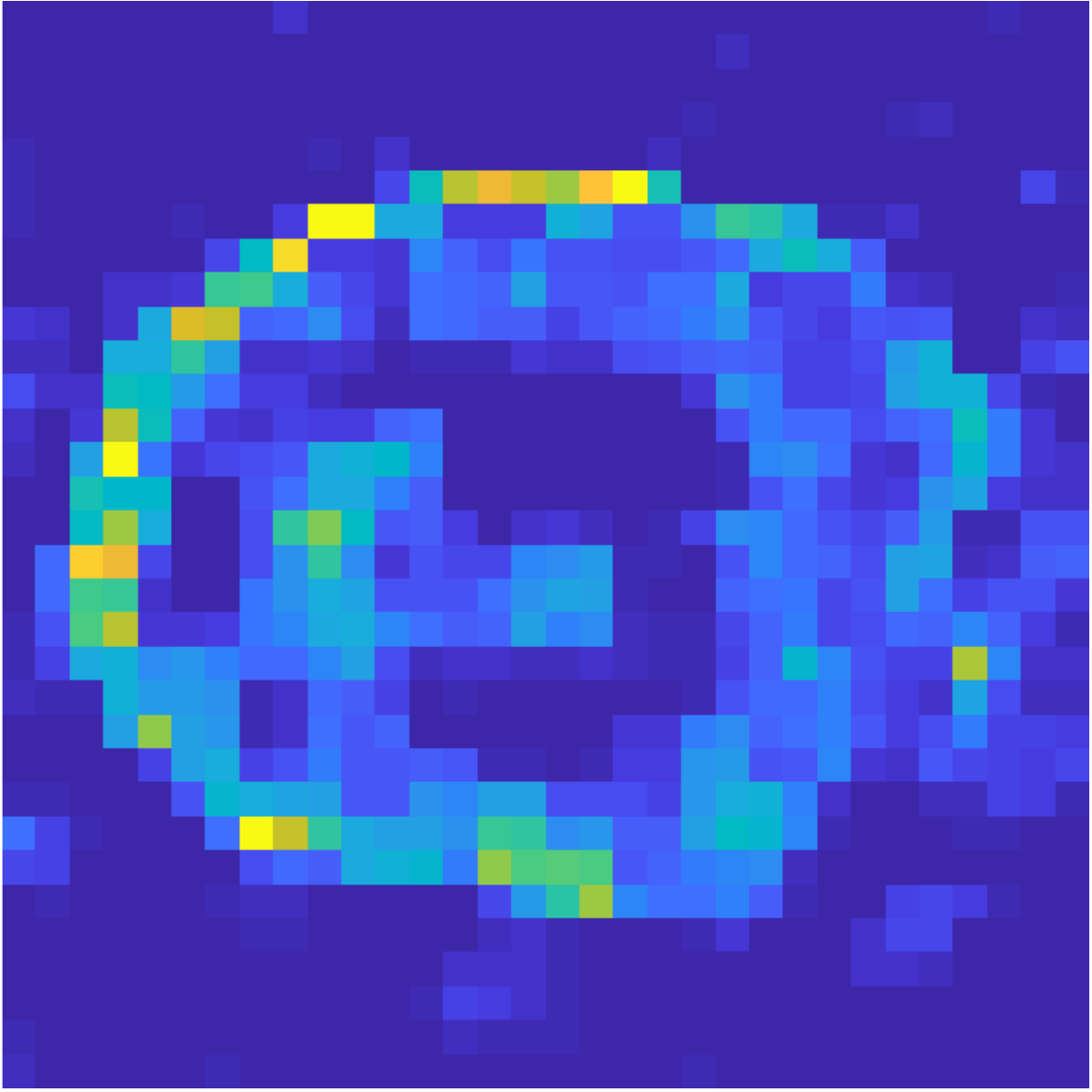} & \includegraphics[width=0.8in]{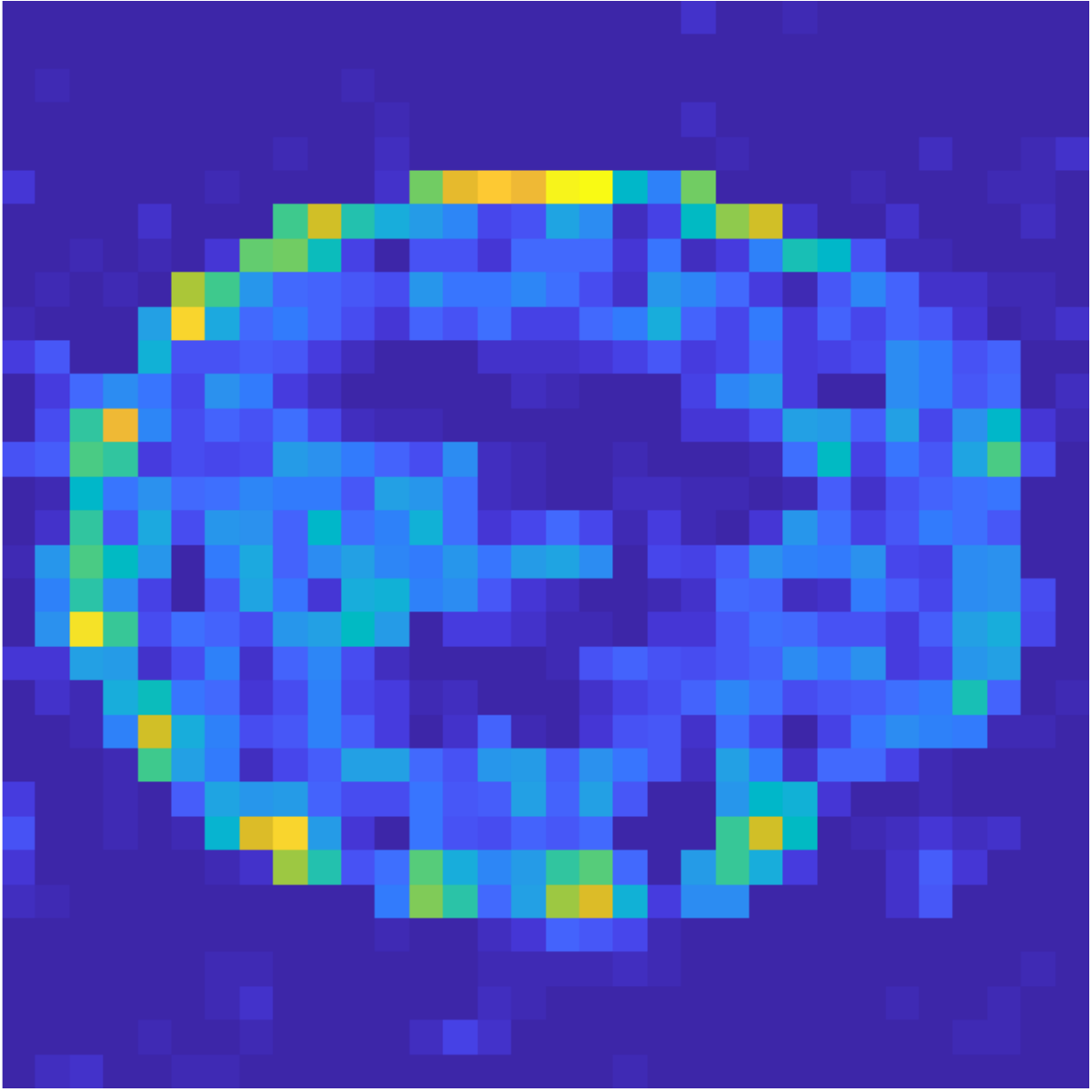} \\
	\rotatebox[origin=l]{90}{{\small $ f_\text{max}=100$ }} & & \includegraphics[width=0.8in]{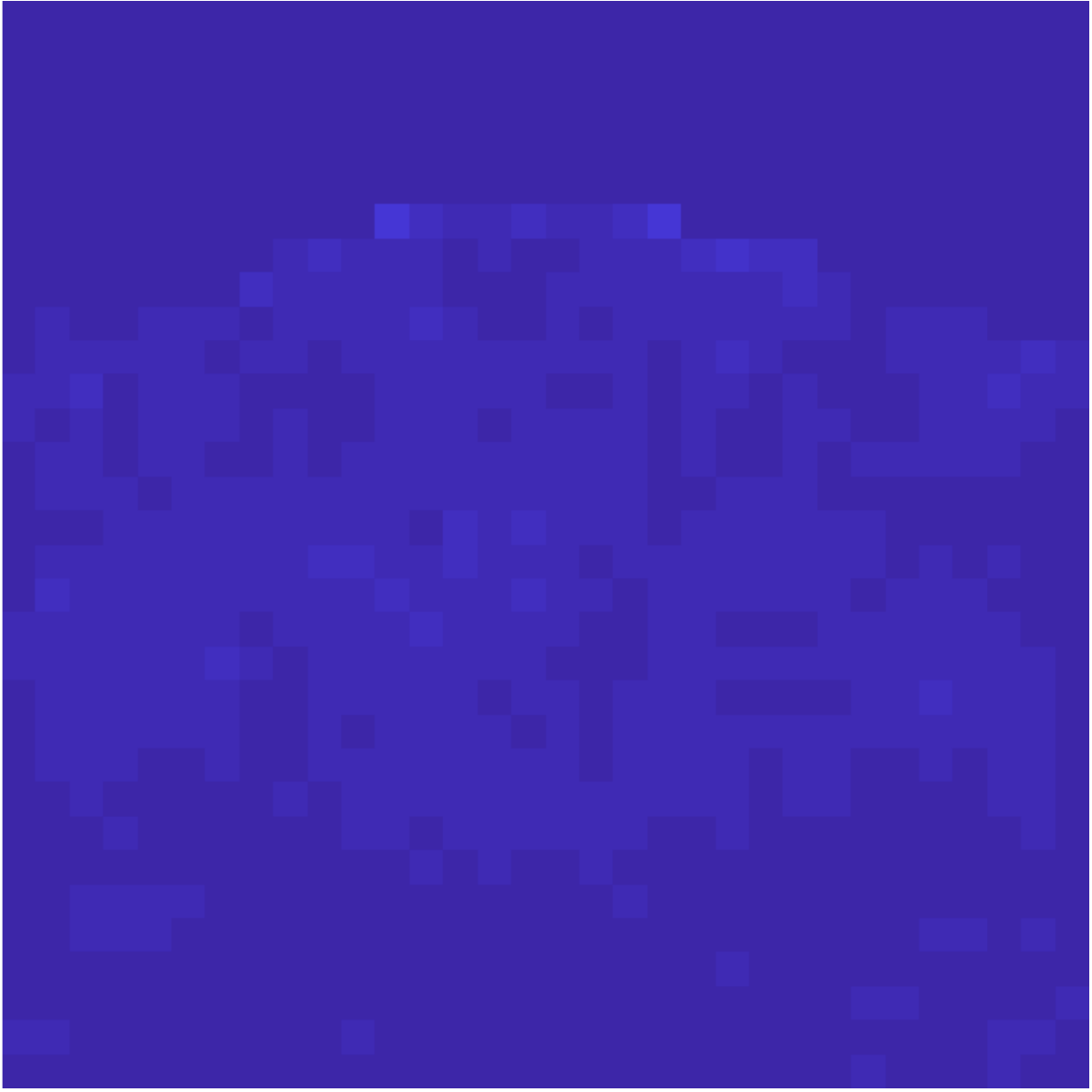} & \includegraphics[width=0.8in]{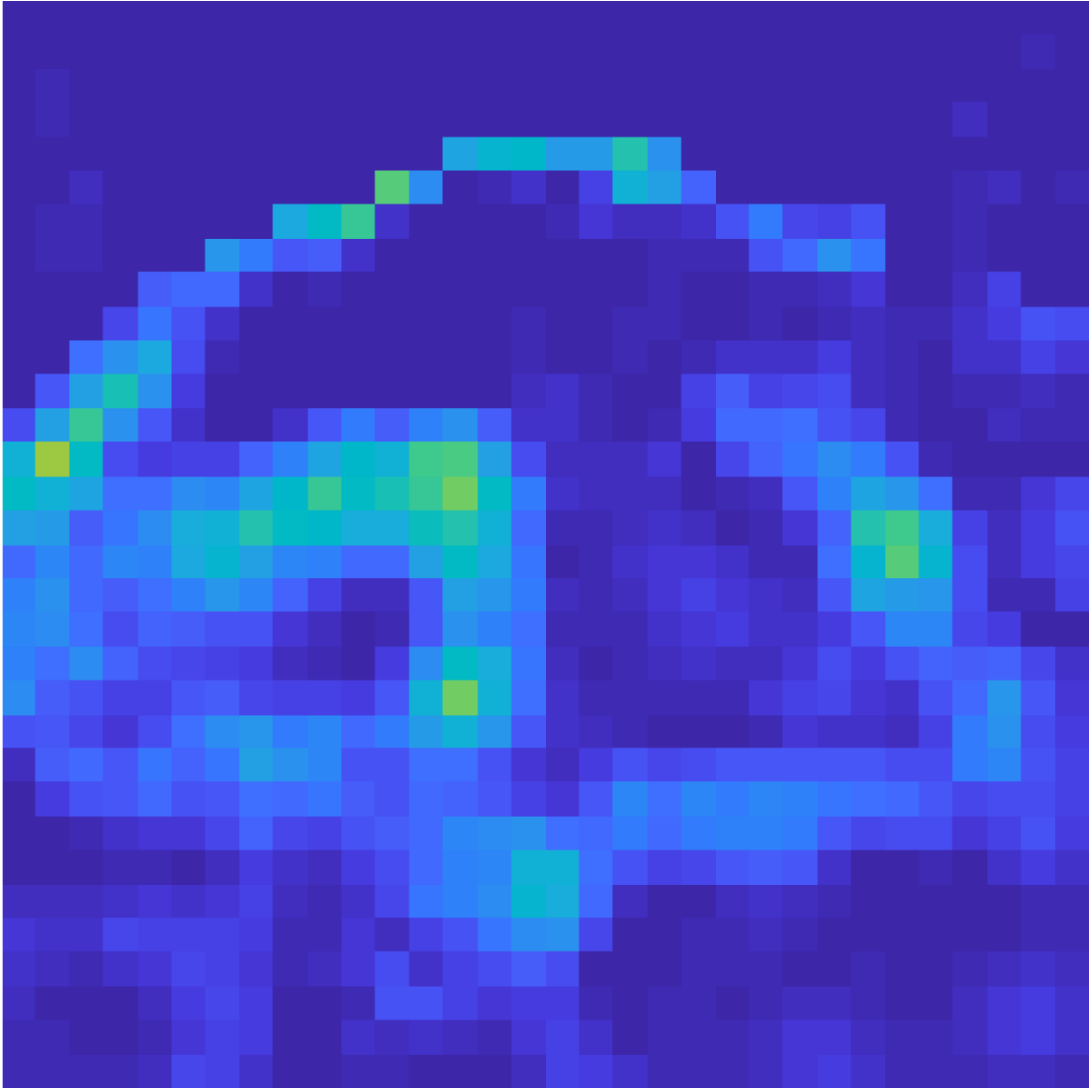} & \includegraphics[width=0.8in]{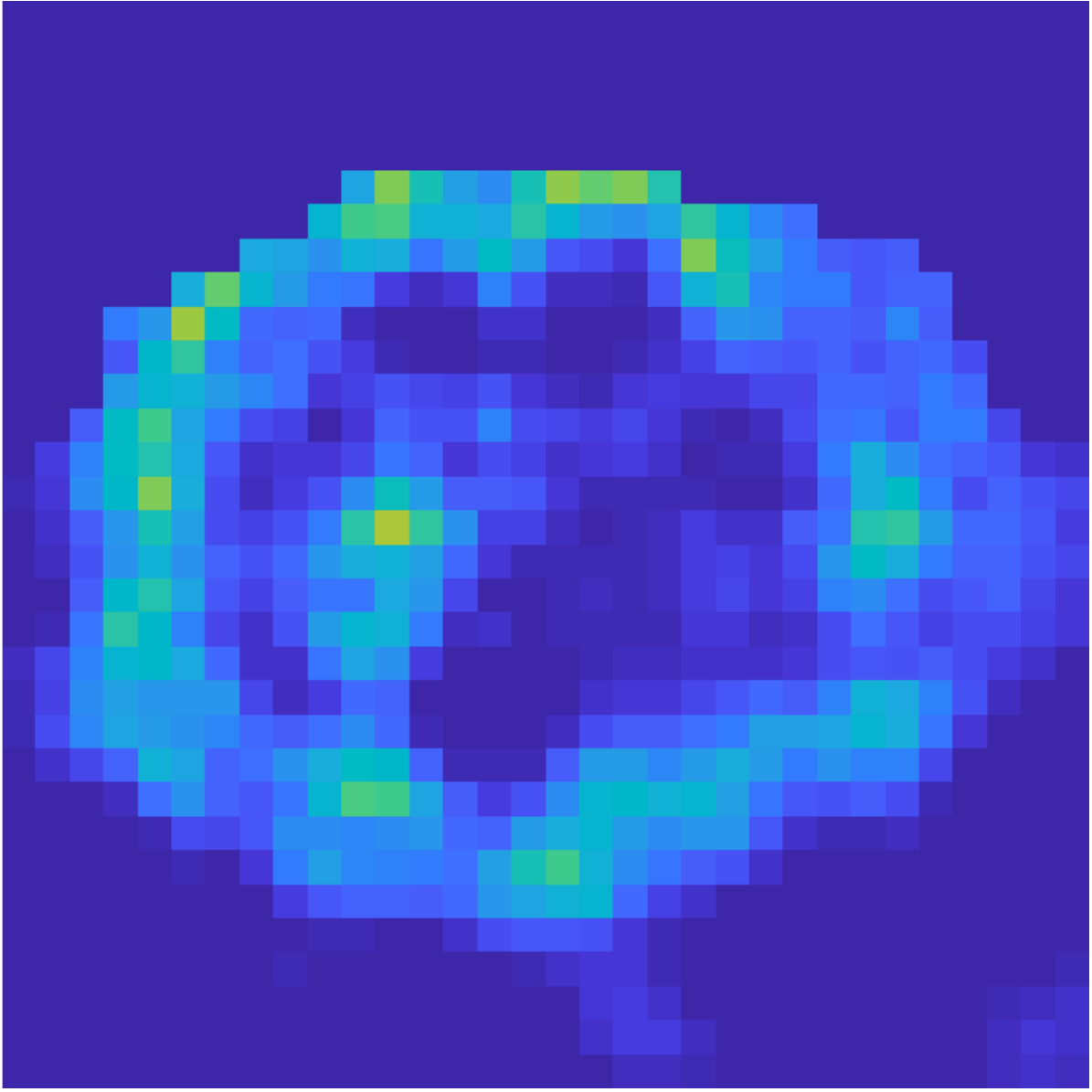} & \includegraphics[width=0.8in]{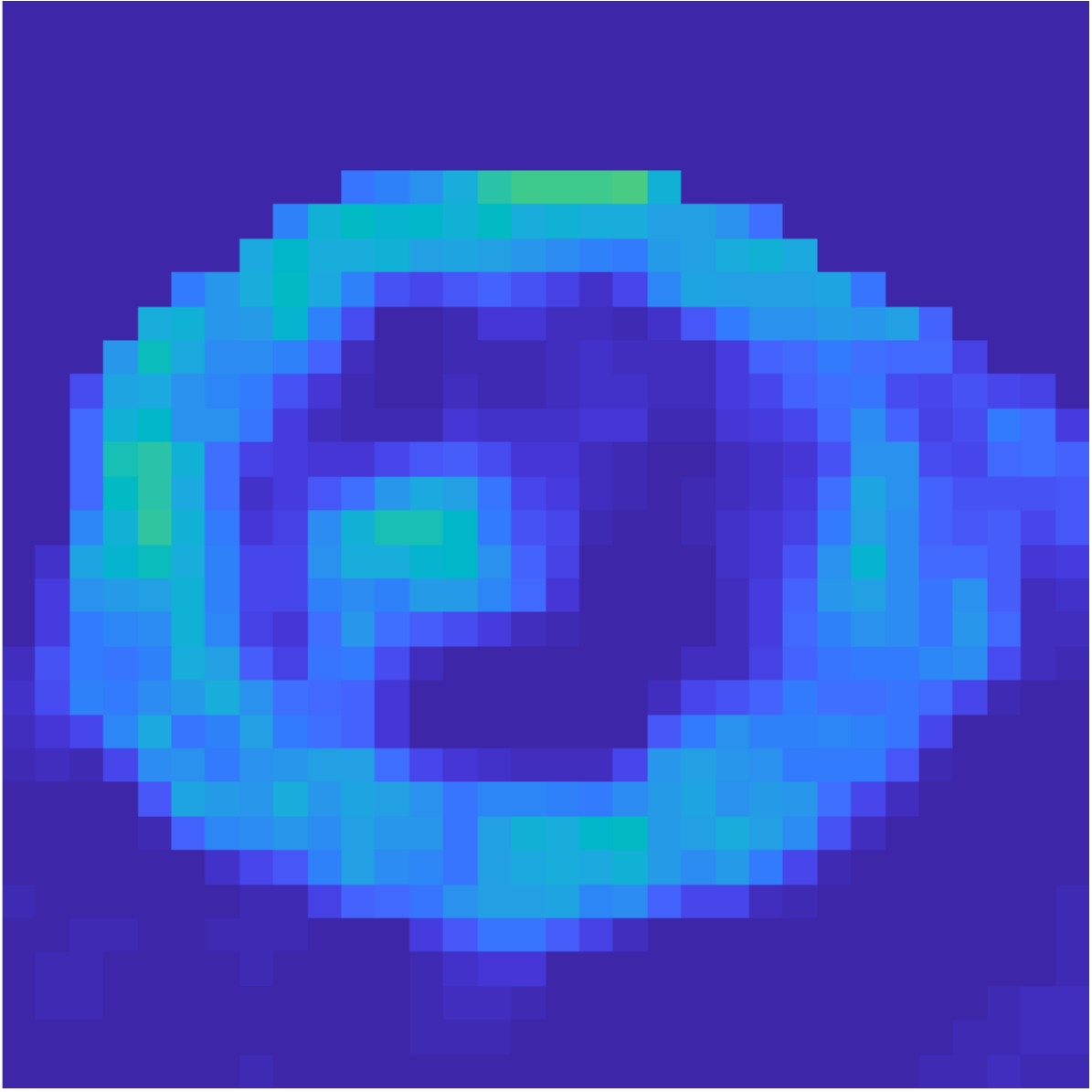}
	\end{tabular}
	\caption{\small Comparison of methods on Phantom 1.}
	\label{fig:Phantom1:comp}
\end{figure}

\begin{figure}[!b]
	\centering
	\setlength\tabcolsep{1.5pt}
	\begin{tabular}{c| c cccc}
	& & {\small \bf CISOR} &  {\small \bf  RL} &  {\small \bf SF-$\tau$} & {\small \bf SF-$\sigma$} \\ \midrule
	\rotatebox[origin=l]{90}{{\small \quad $ f_\text{max}=1$ }} & & \includegraphics[width=0.8in]{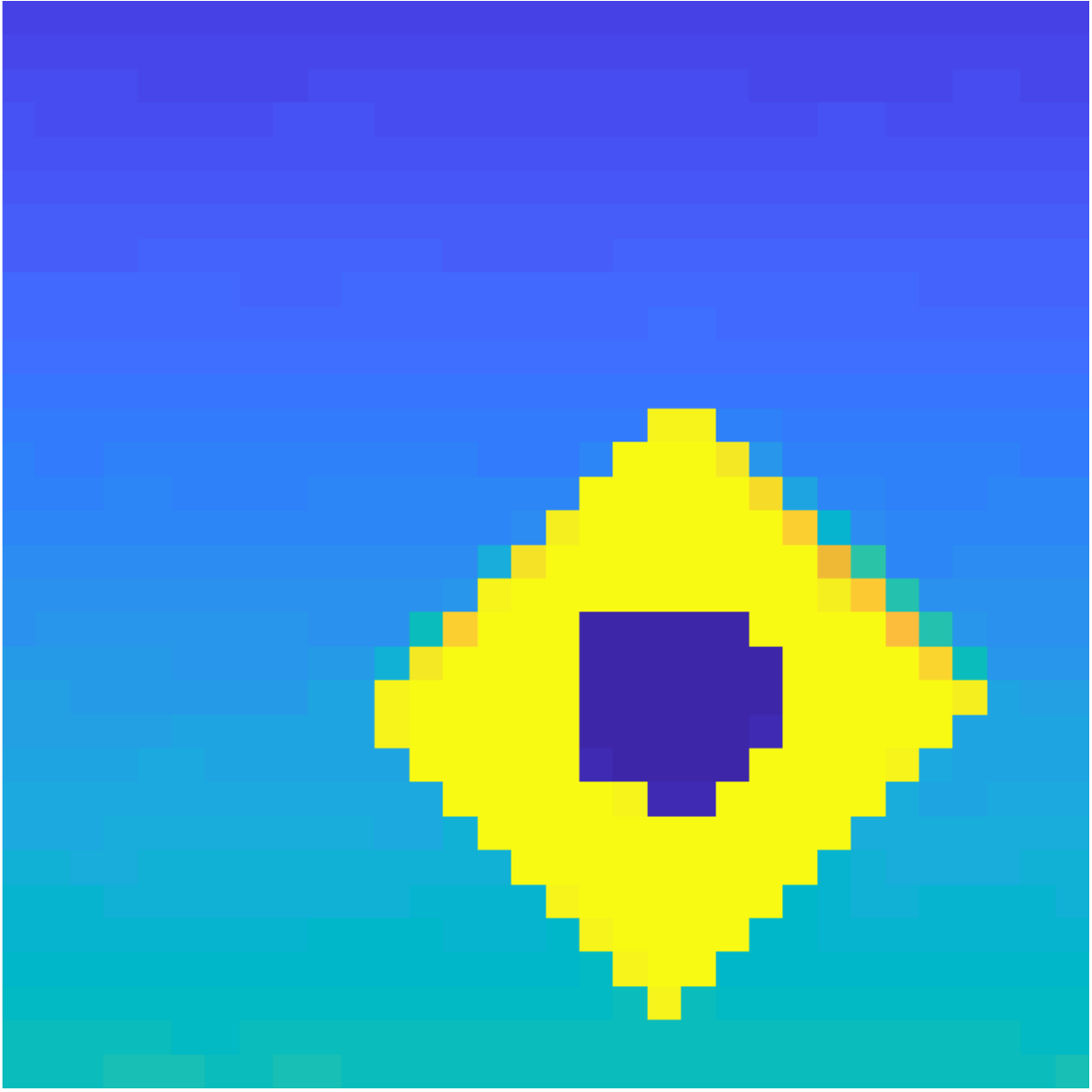} & \includegraphics[width=0.8in]{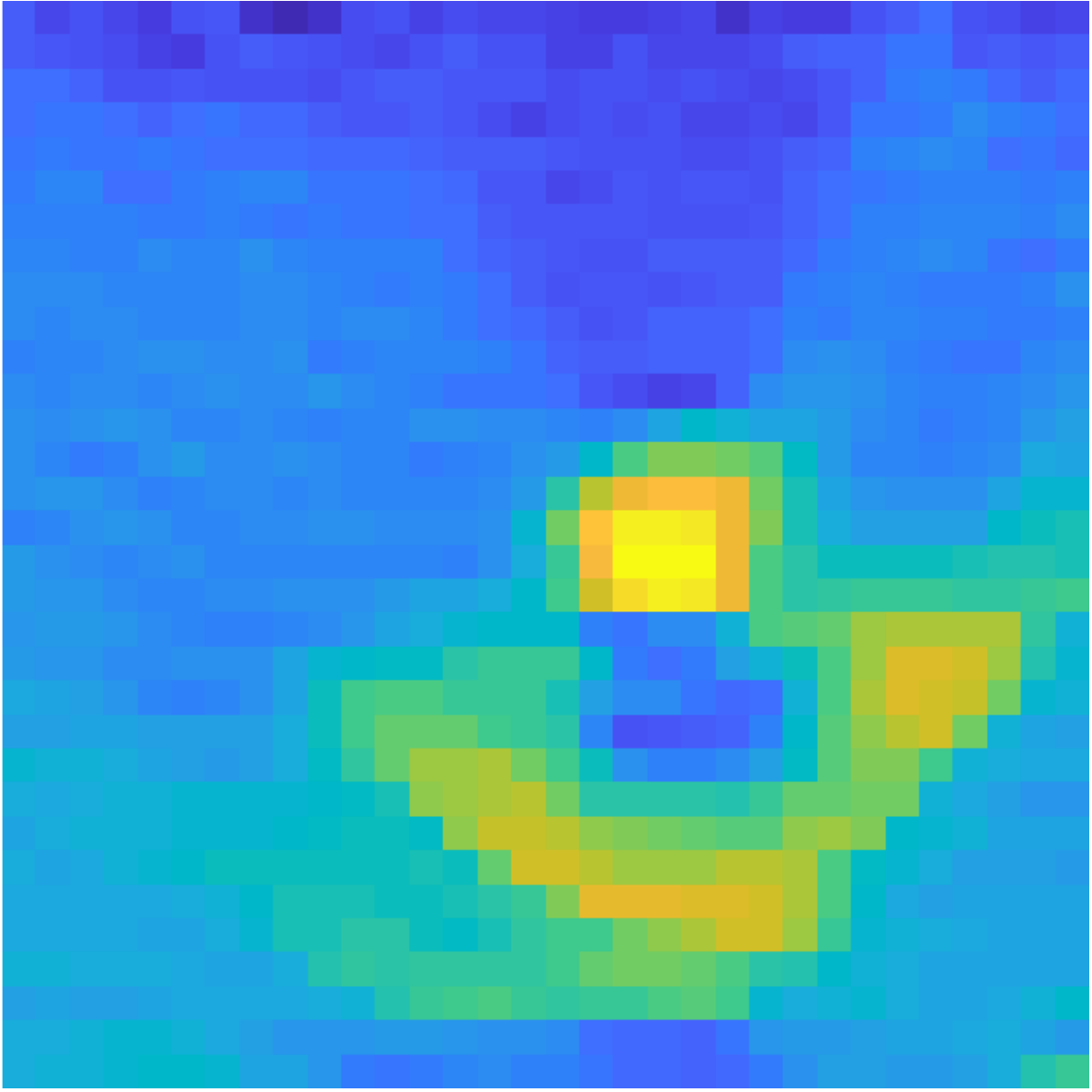} & \includegraphics[width=0.8in]{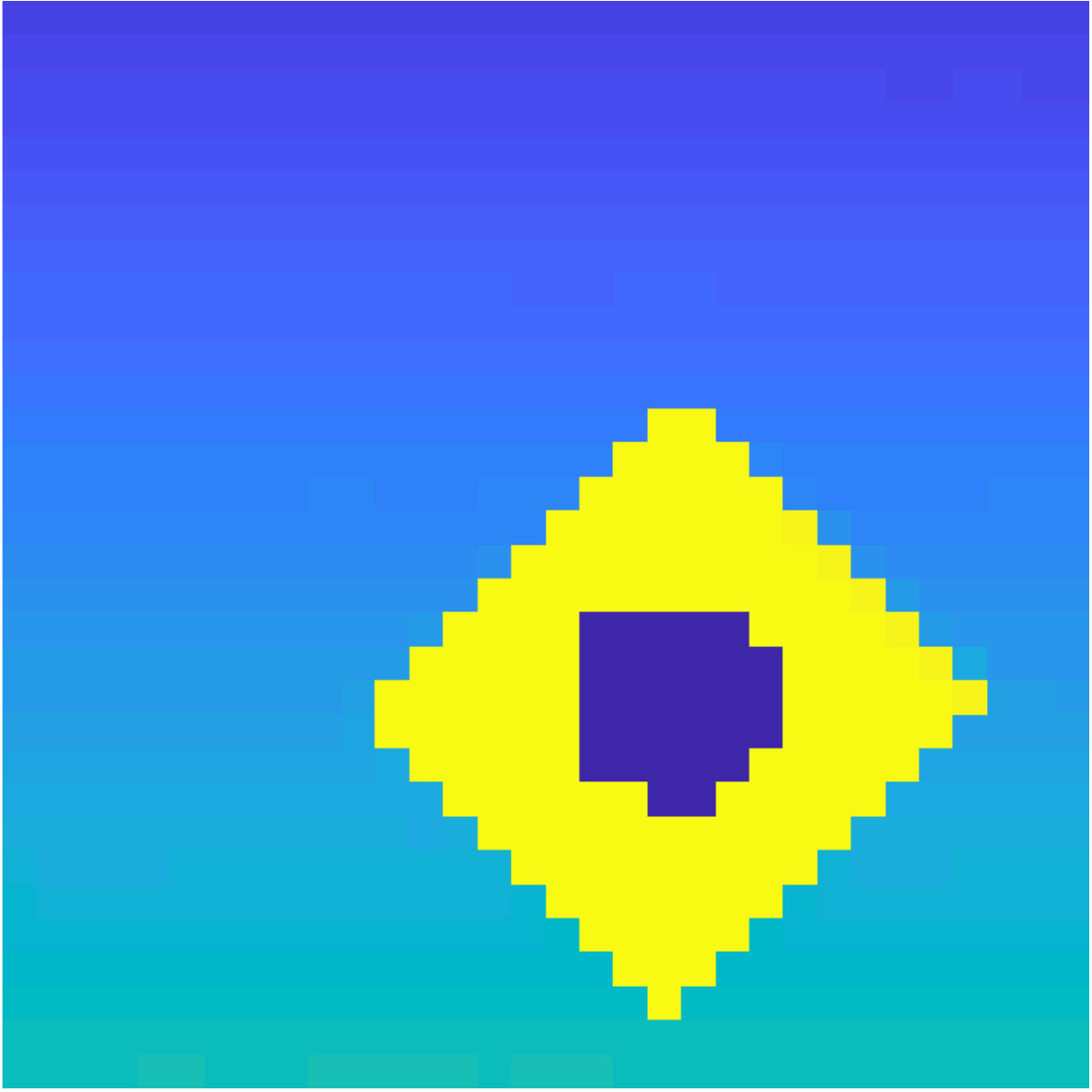} & \includegraphics[width=0.8in]{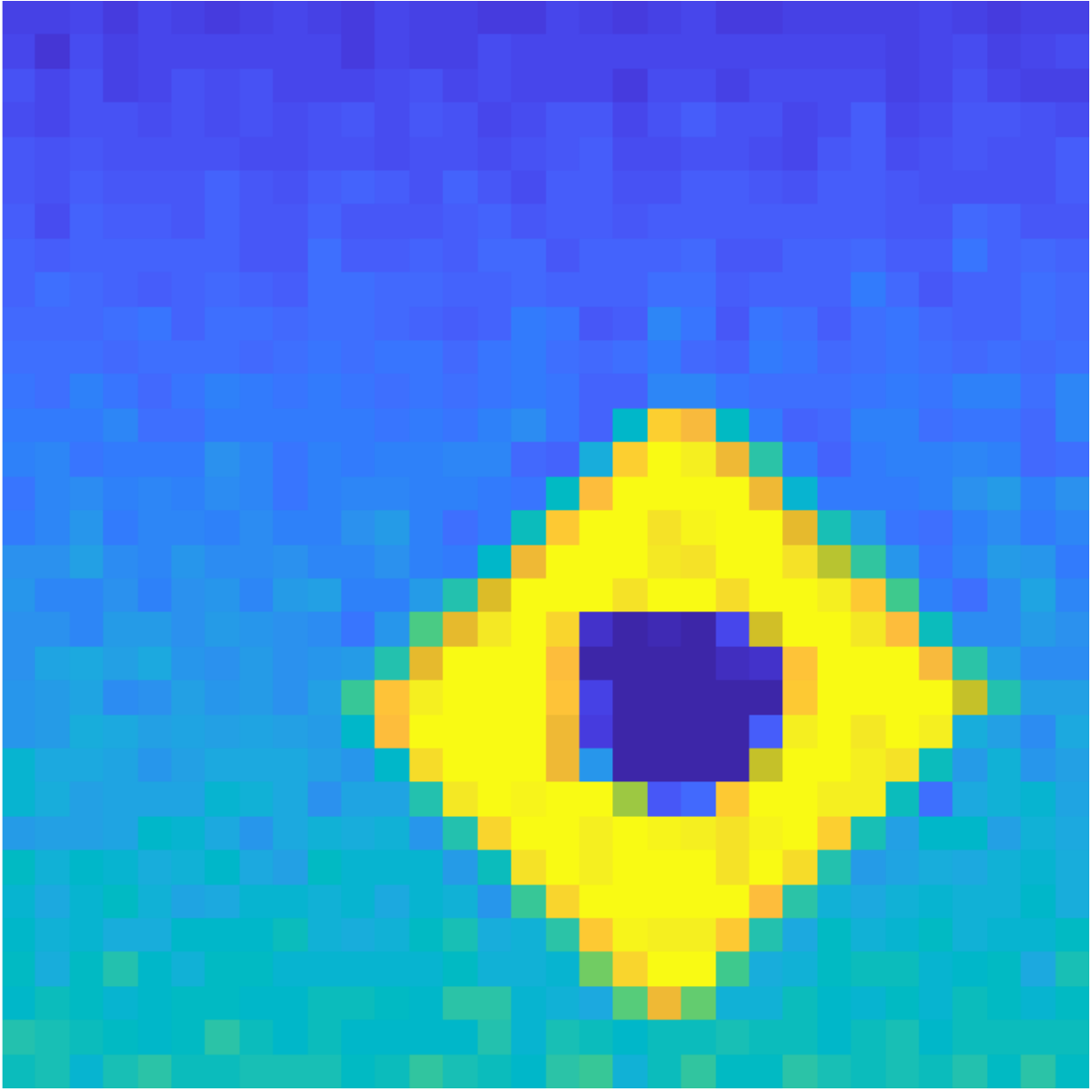} \\
	\rotatebox[origin=l]{90}{ {\small \quad $ f_\text{max}=10$} } & &\includegraphics[width=0.8in]{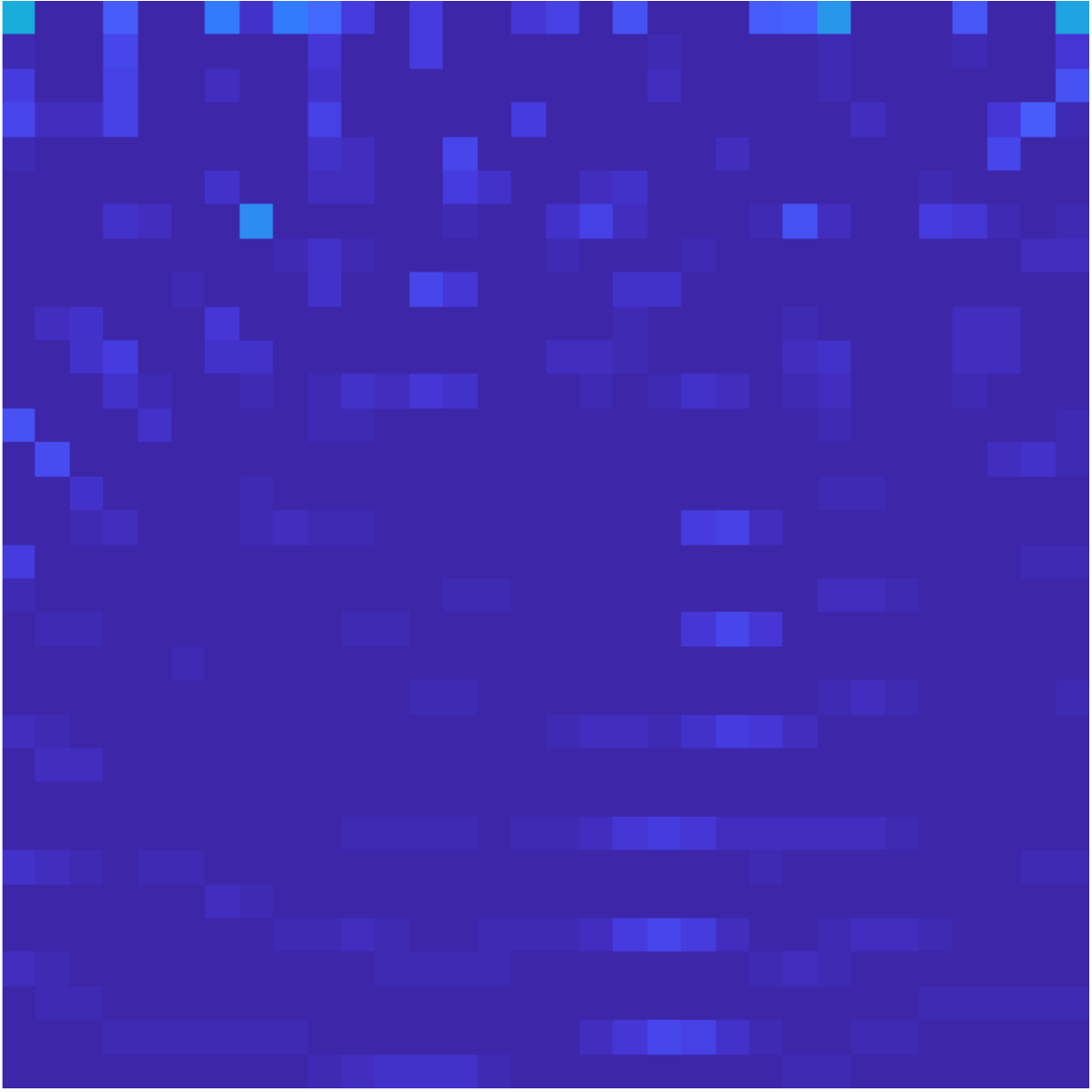} & \includegraphics[width=0.8in]{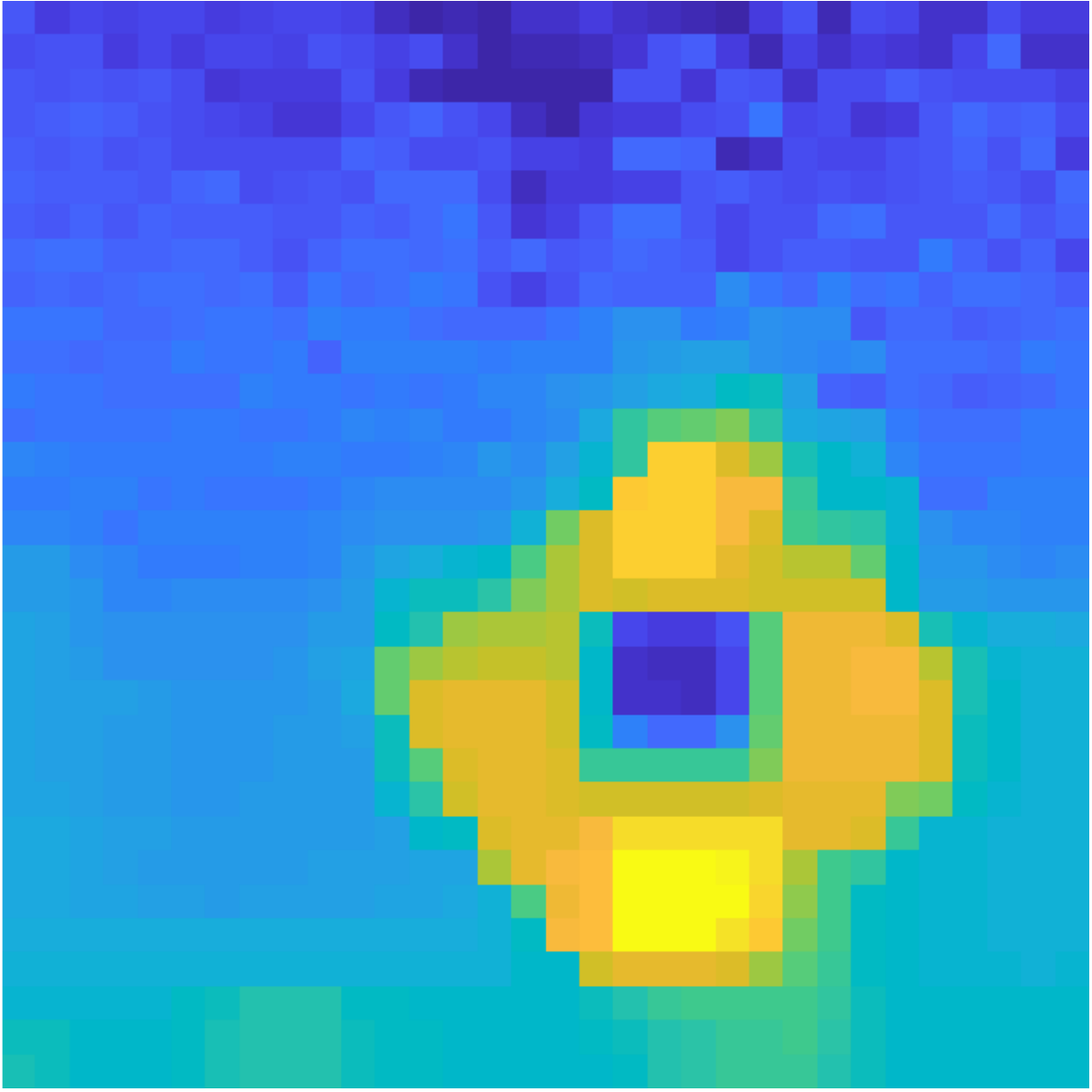} & \includegraphics[width=0.8in]{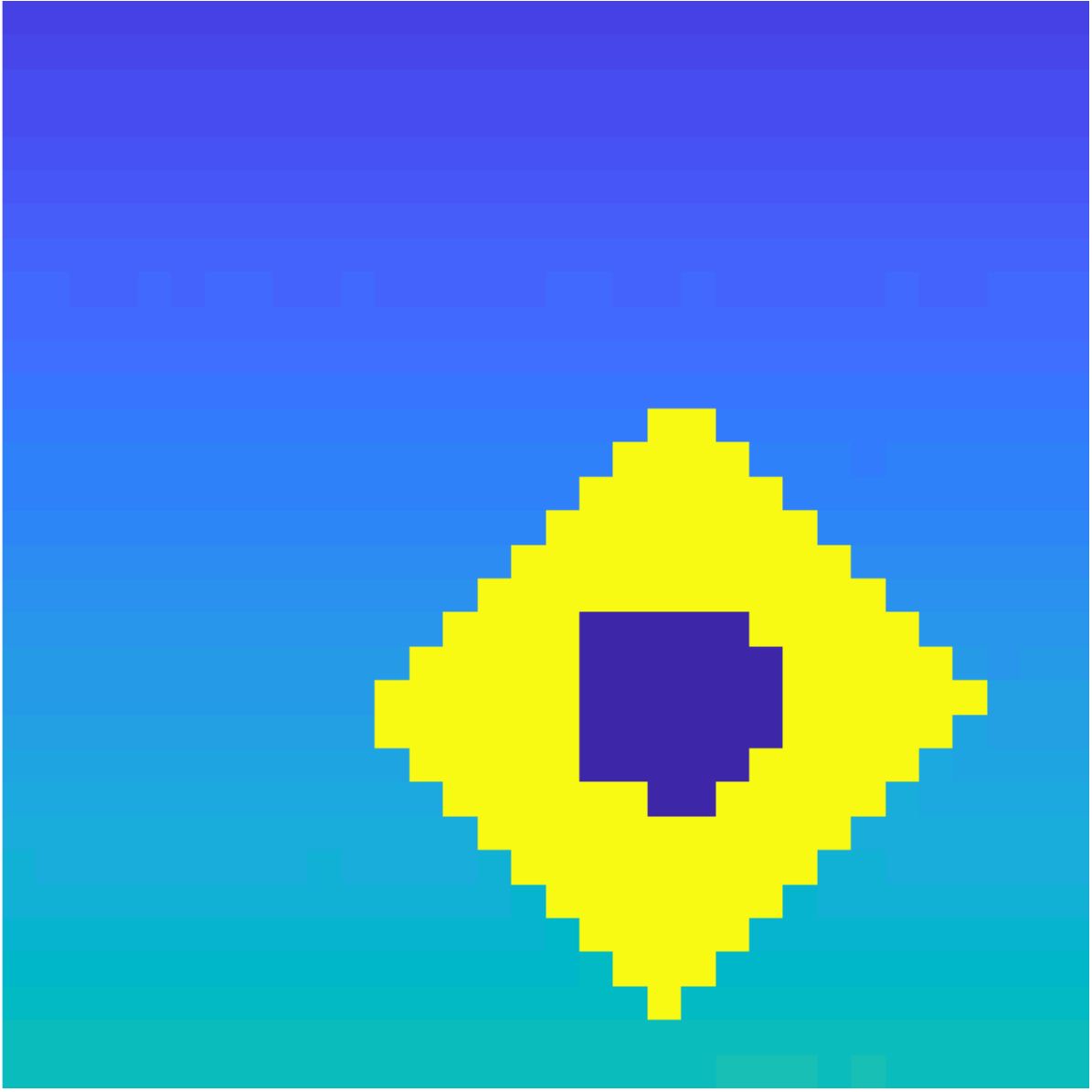} & \includegraphics[width=0.8in]{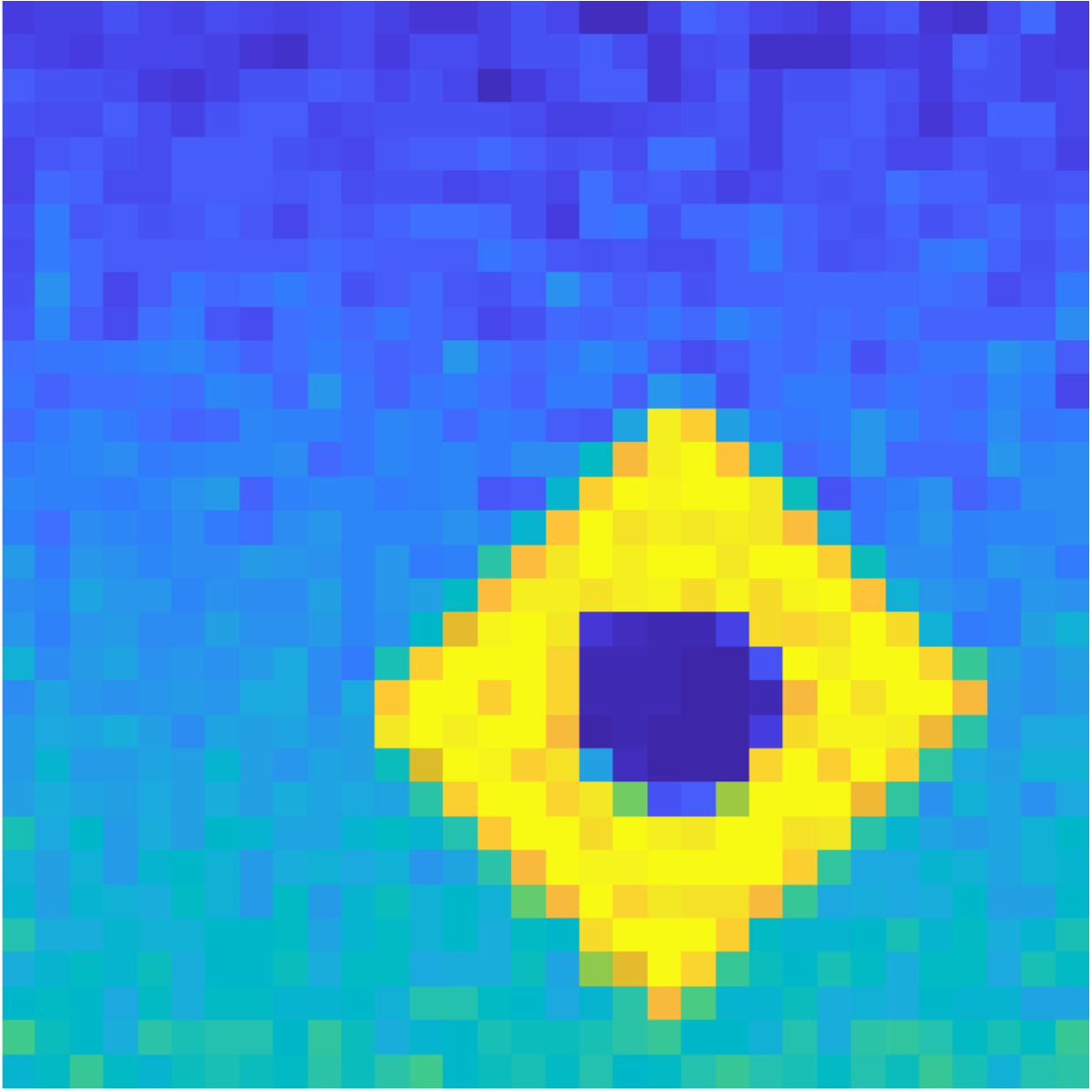} \\
	\rotatebox[origin=l]{90}{ {\small  $ f_\text{max}=100$} }  & & \includegraphics[width=0.8in]{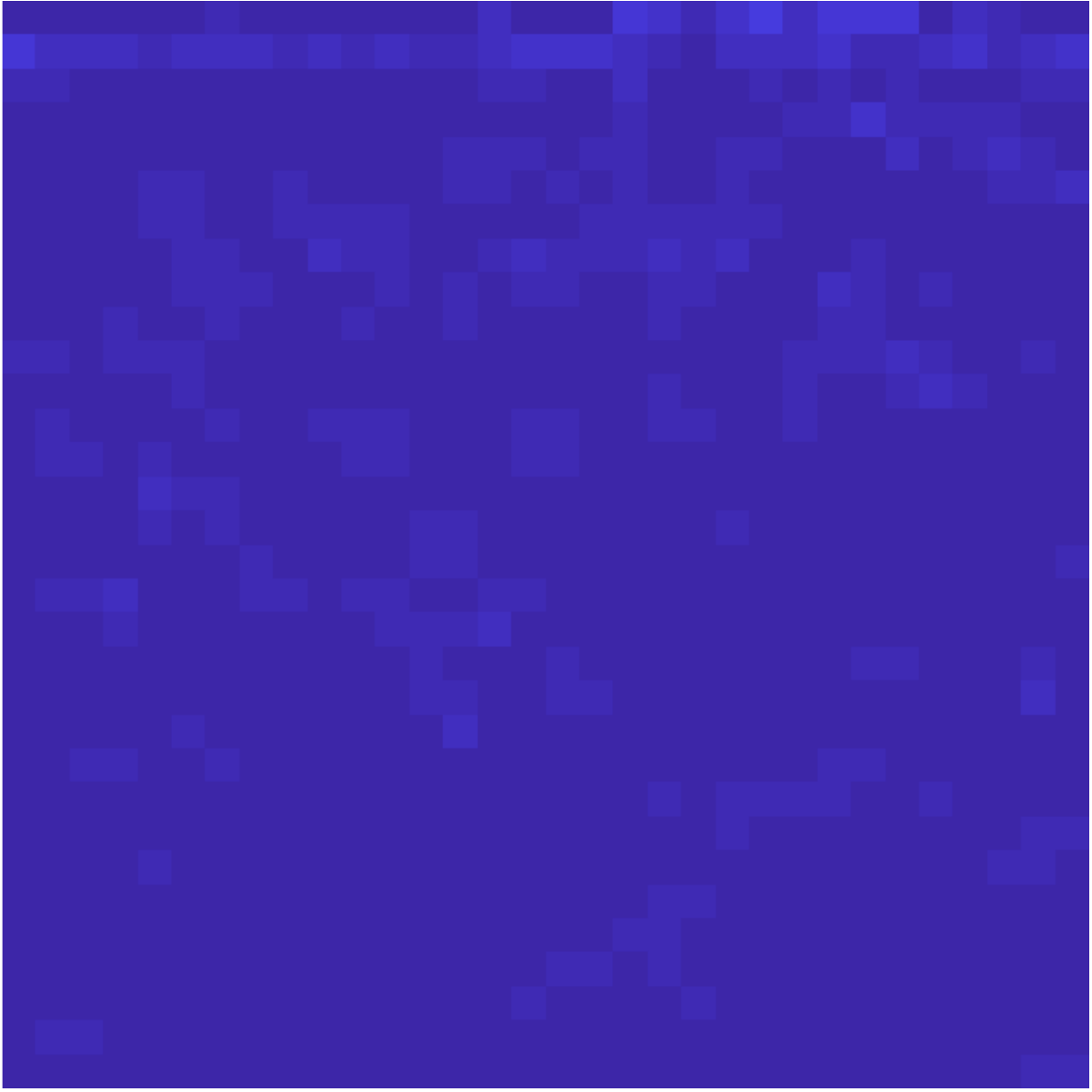} & \includegraphics[width=0.8in]{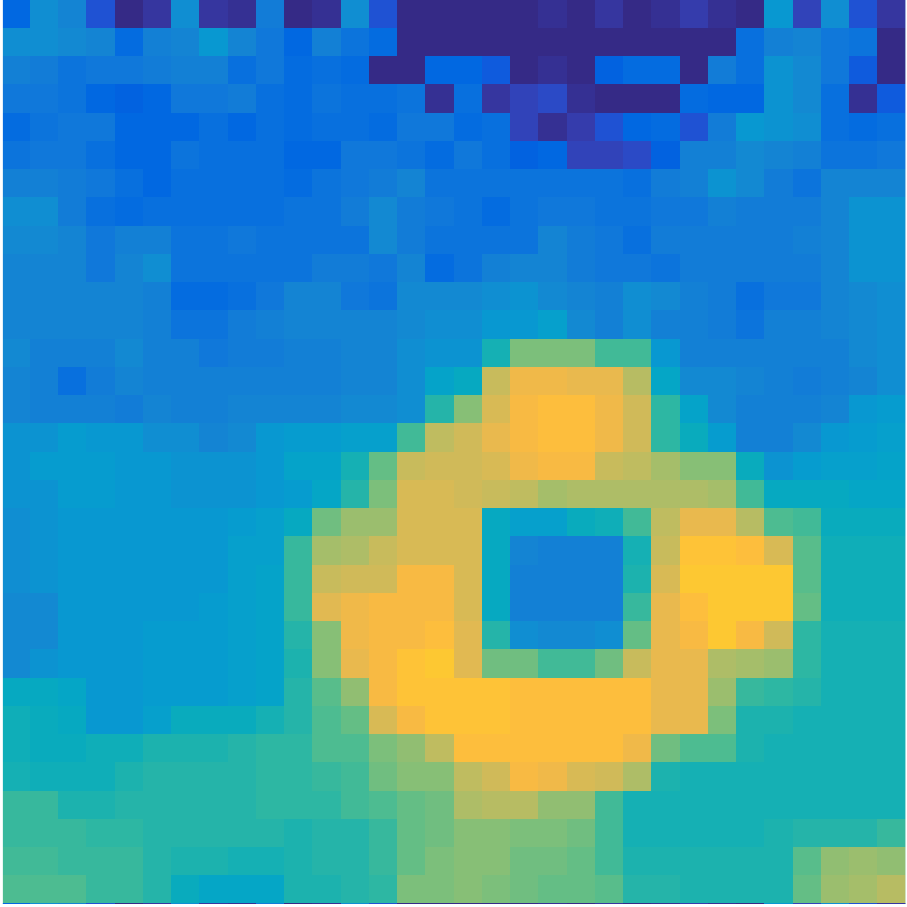} & \includegraphics[width=0.8in]{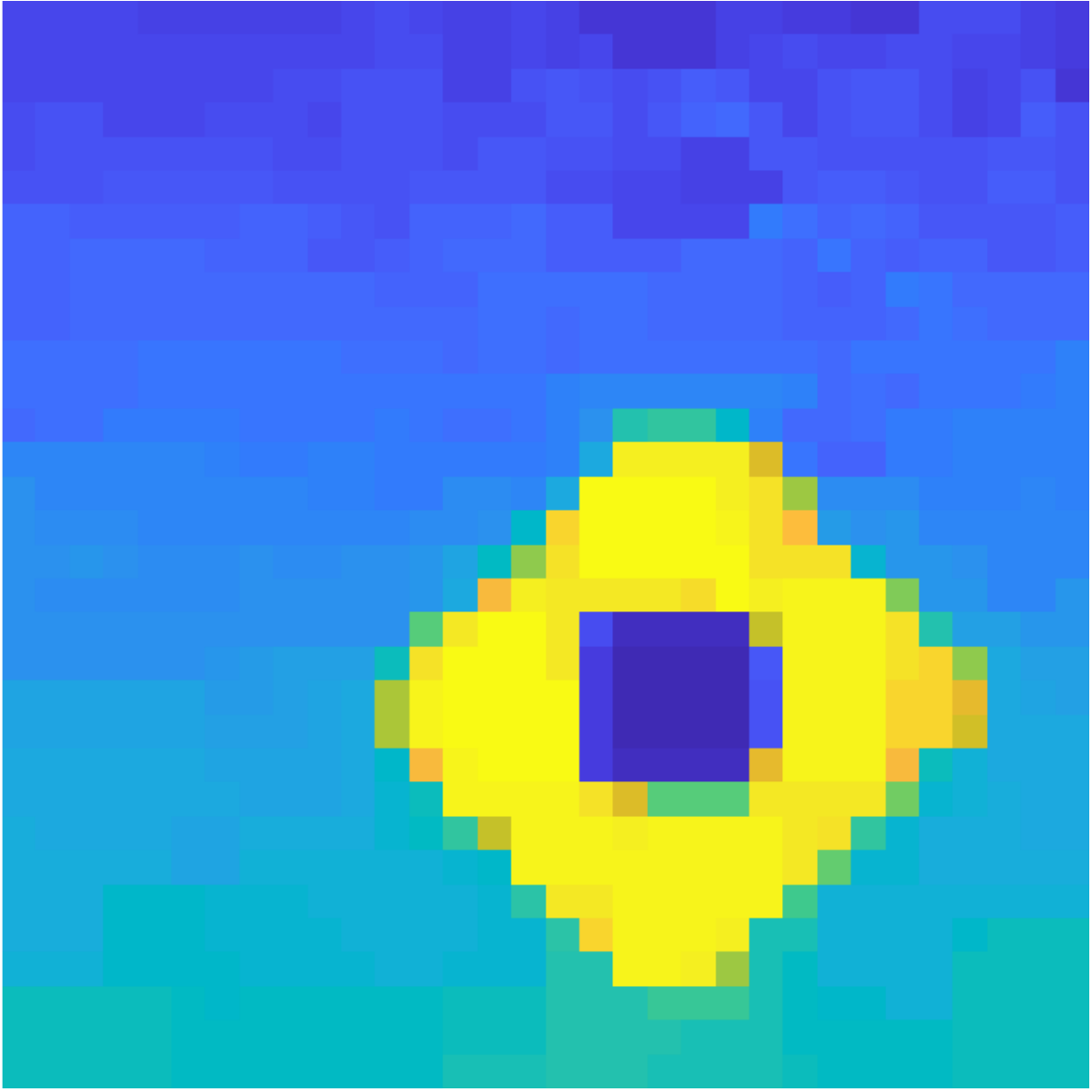} & \includegraphics[width=0.8in]{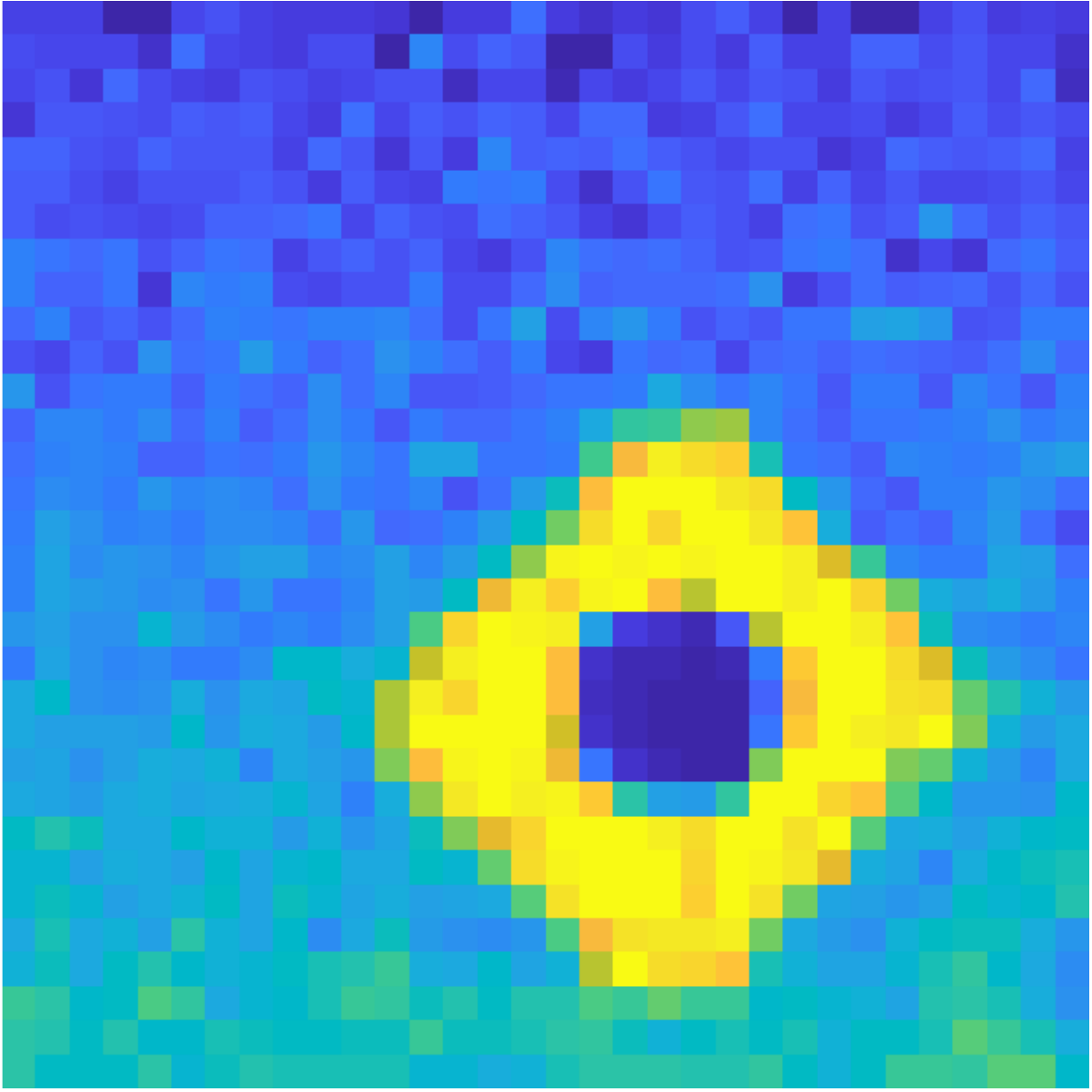}
	\end{tabular}
	\caption{\small Comparison of methods on Phantom 2.}
	\label{fig:Phantom2:comp}
\end{figure}

\begin{figure}[!b]
	\centering
	\setlength\tabcolsep{1.5pt}
	\begin{tabular}{c| c c | cc}
	 & {\small \bf SF-$\tau$} & {\small \bf SF-$\sigma$} &  {\small \bf SF-$\tau$} & {\small \bf SF-$\sigma$} \\ \midrule
	 \rotatebox[origin=l]{90}{{\small \quad $ f_\text{max}=1$ }} & \includegraphics[width=0.8in]{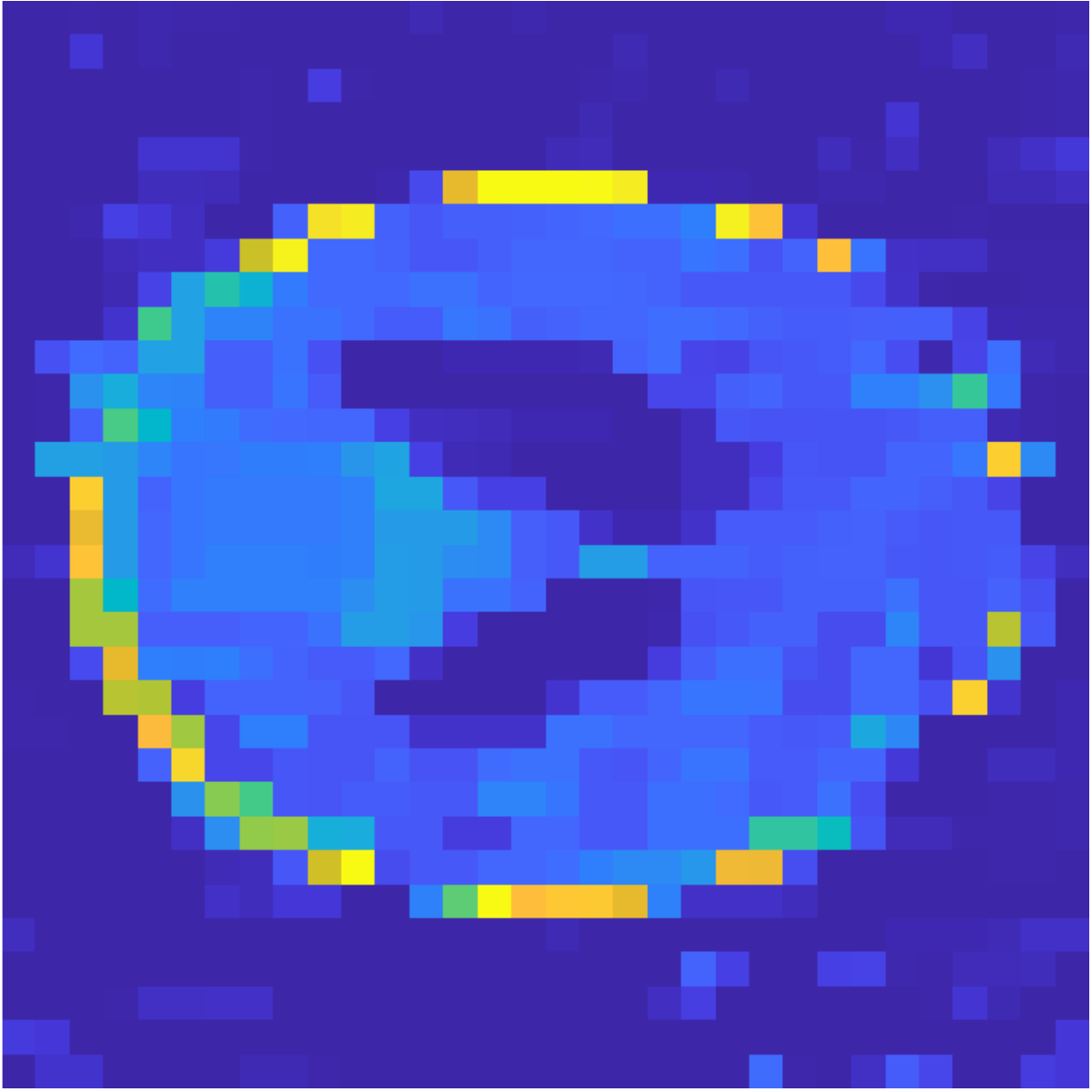} & \includegraphics[width=0.8in]{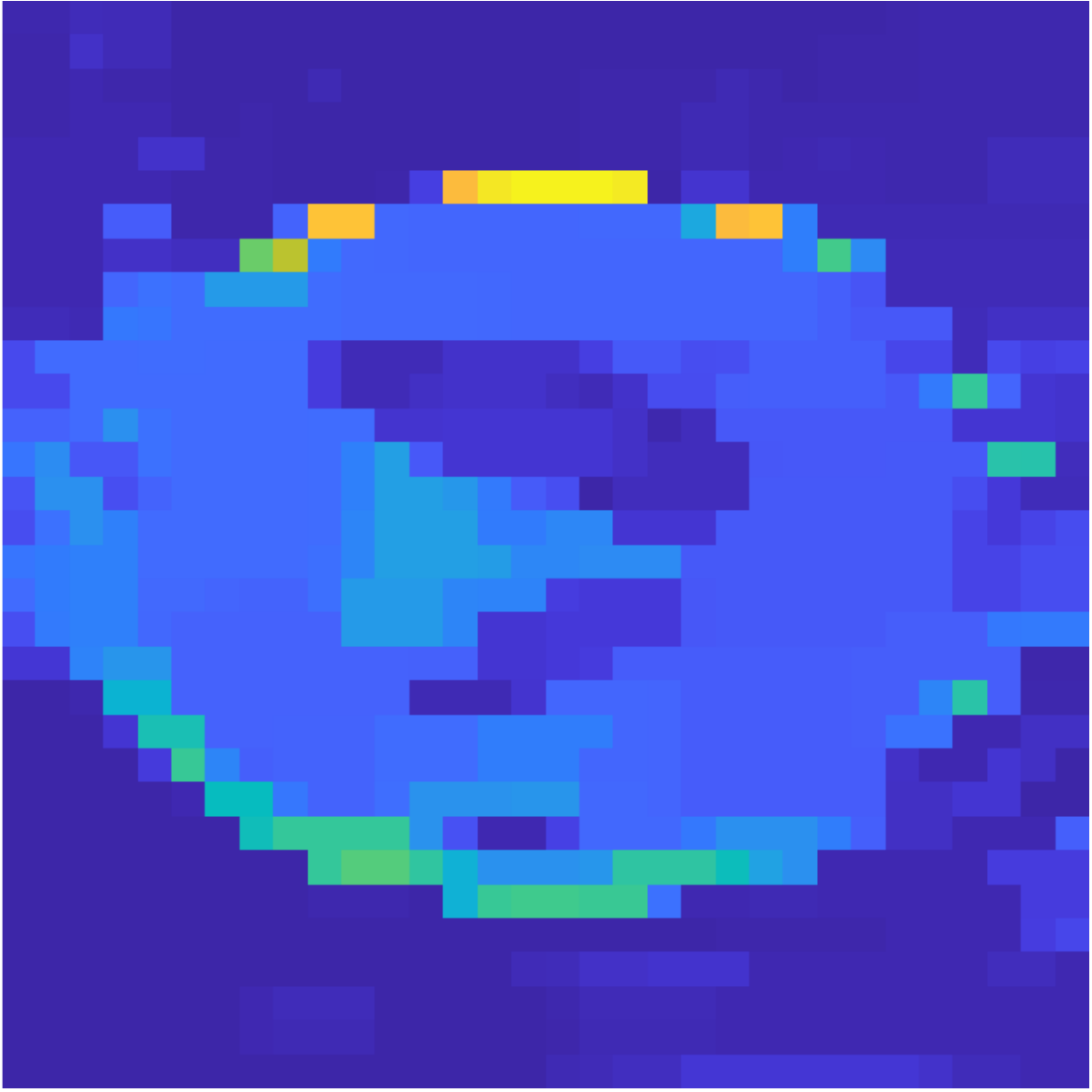} & \includegraphics[width=0.8in]{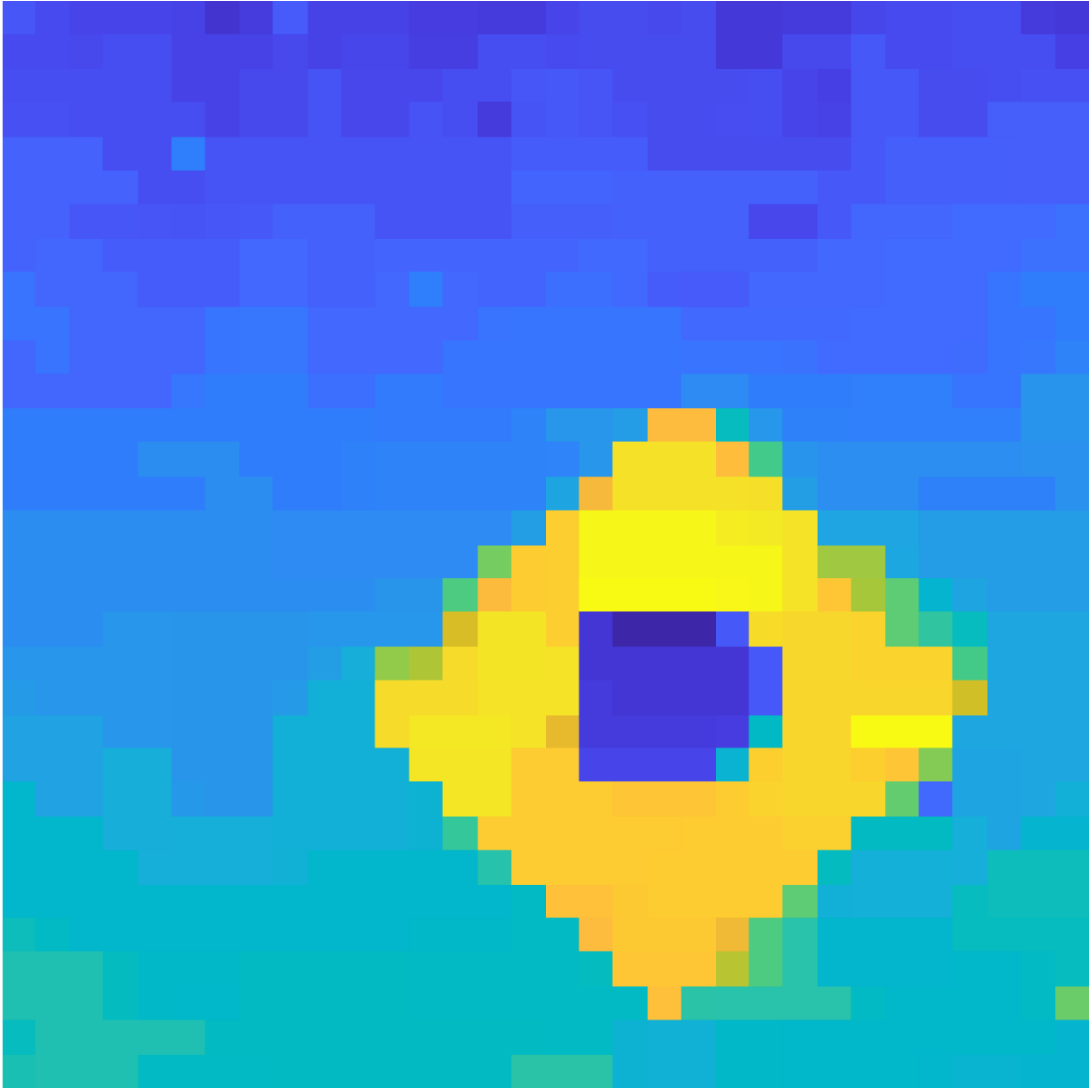} & \includegraphics[width=0.8in]{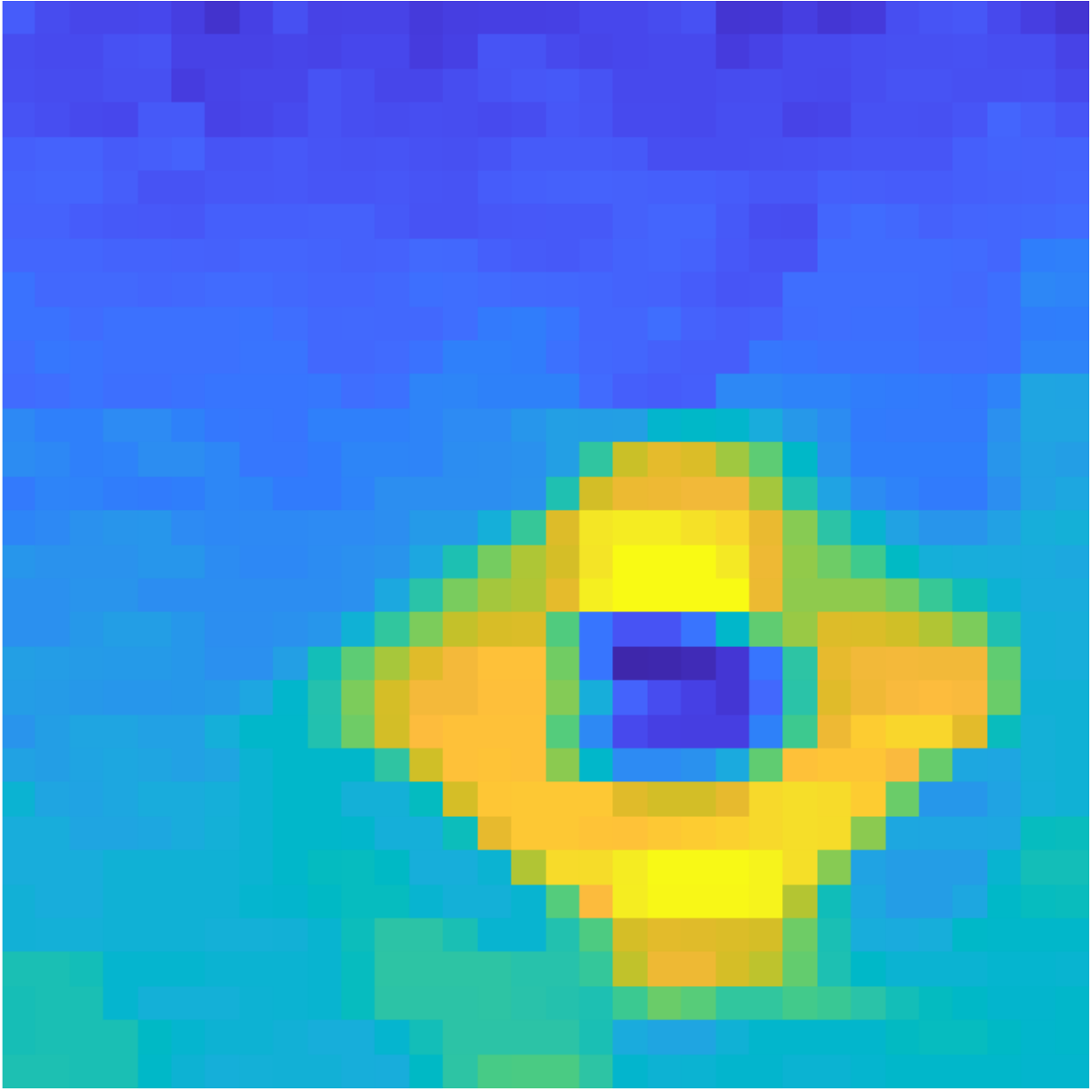} \\
	 \rotatebox[origin=l]{90}{{\small \quad $ f_\text{max}=10$ }} & \includegraphics[width=0.8in]{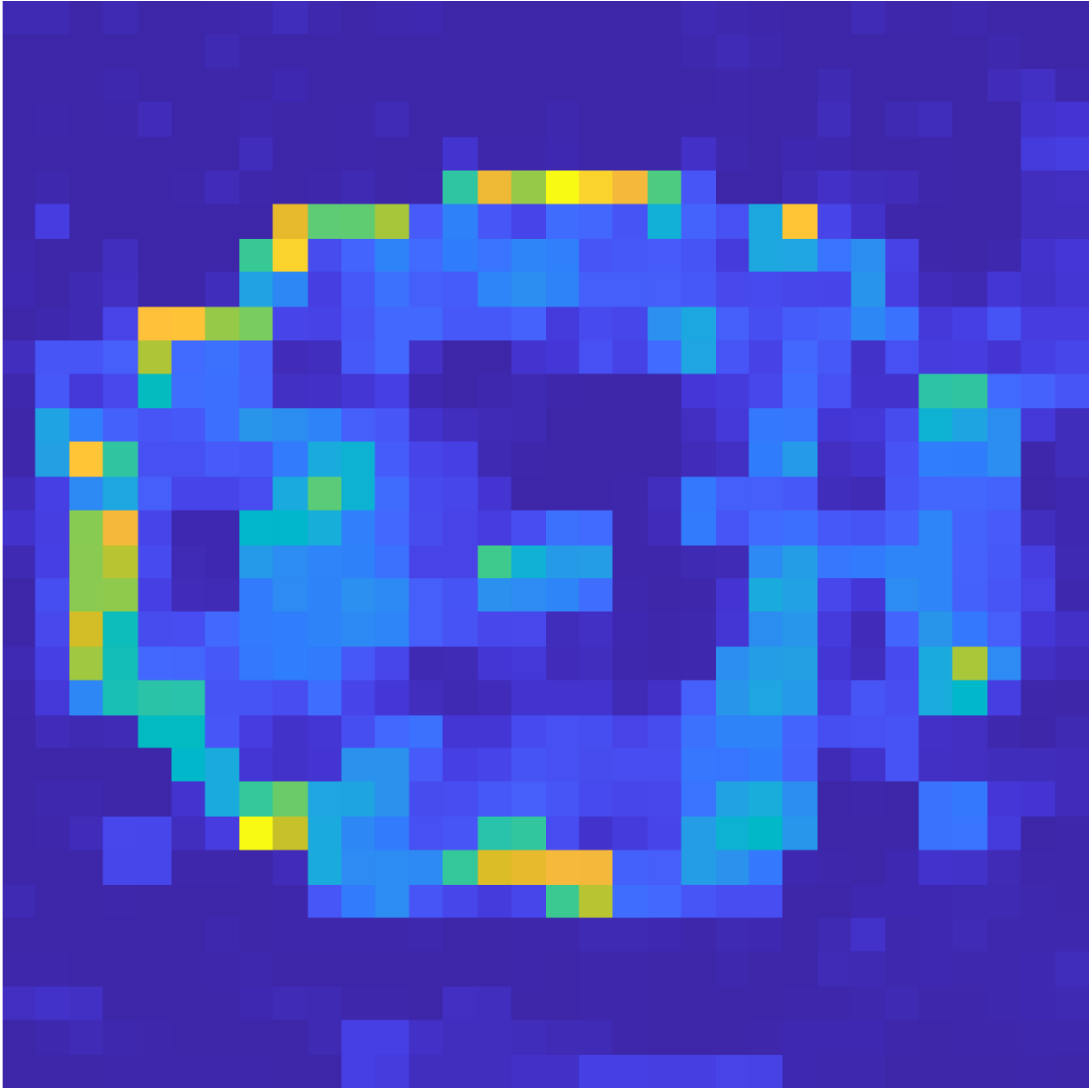} & \includegraphics[width=0.8in]{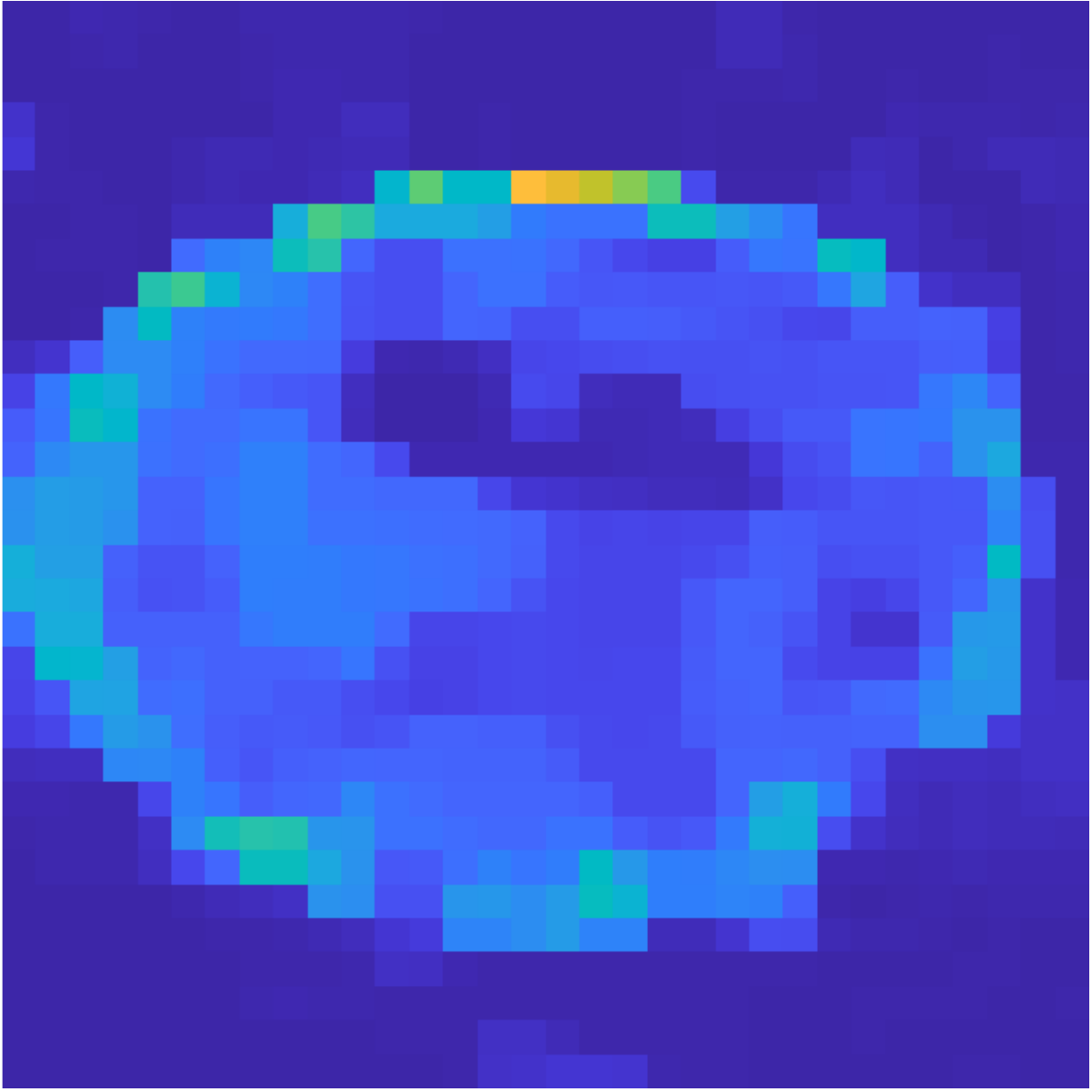} &  \includegraphics[width=0.8in]{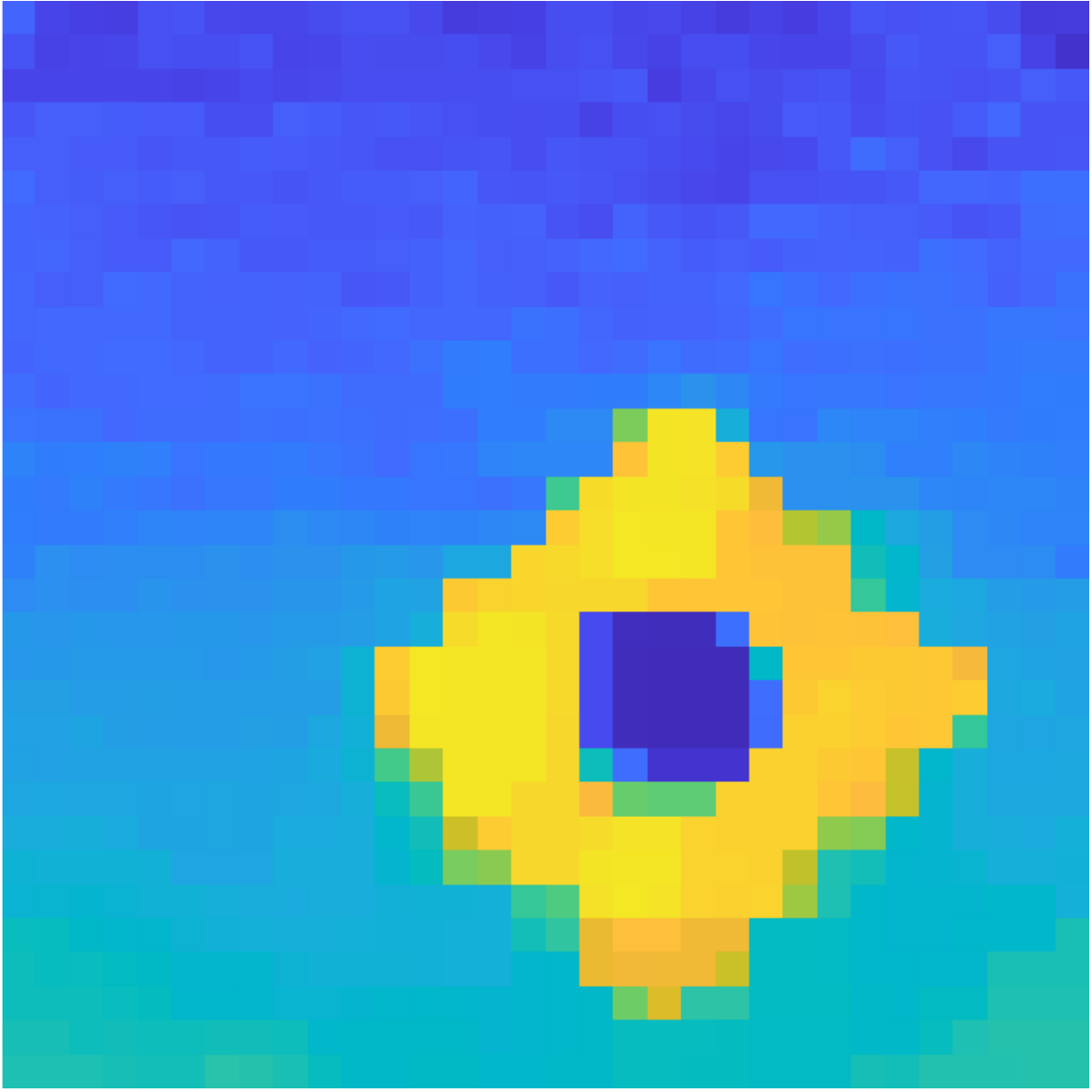} & \includegraphics[width=0.8in]{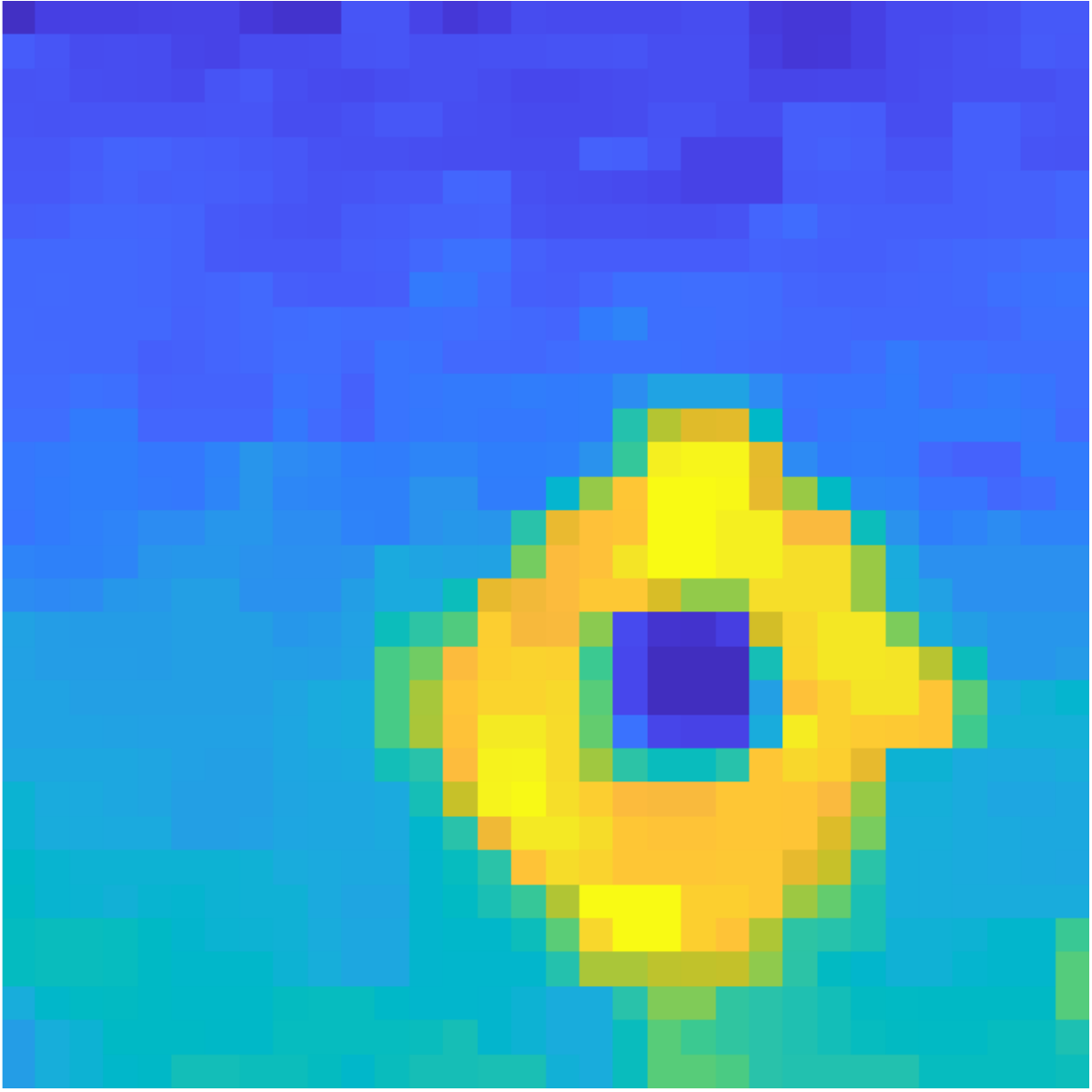} \\
	 \rotatebox[origin=l]{90}{{\small \quad $ f_\text{max}=100$ }} & \includegraphics[width=0.8in]{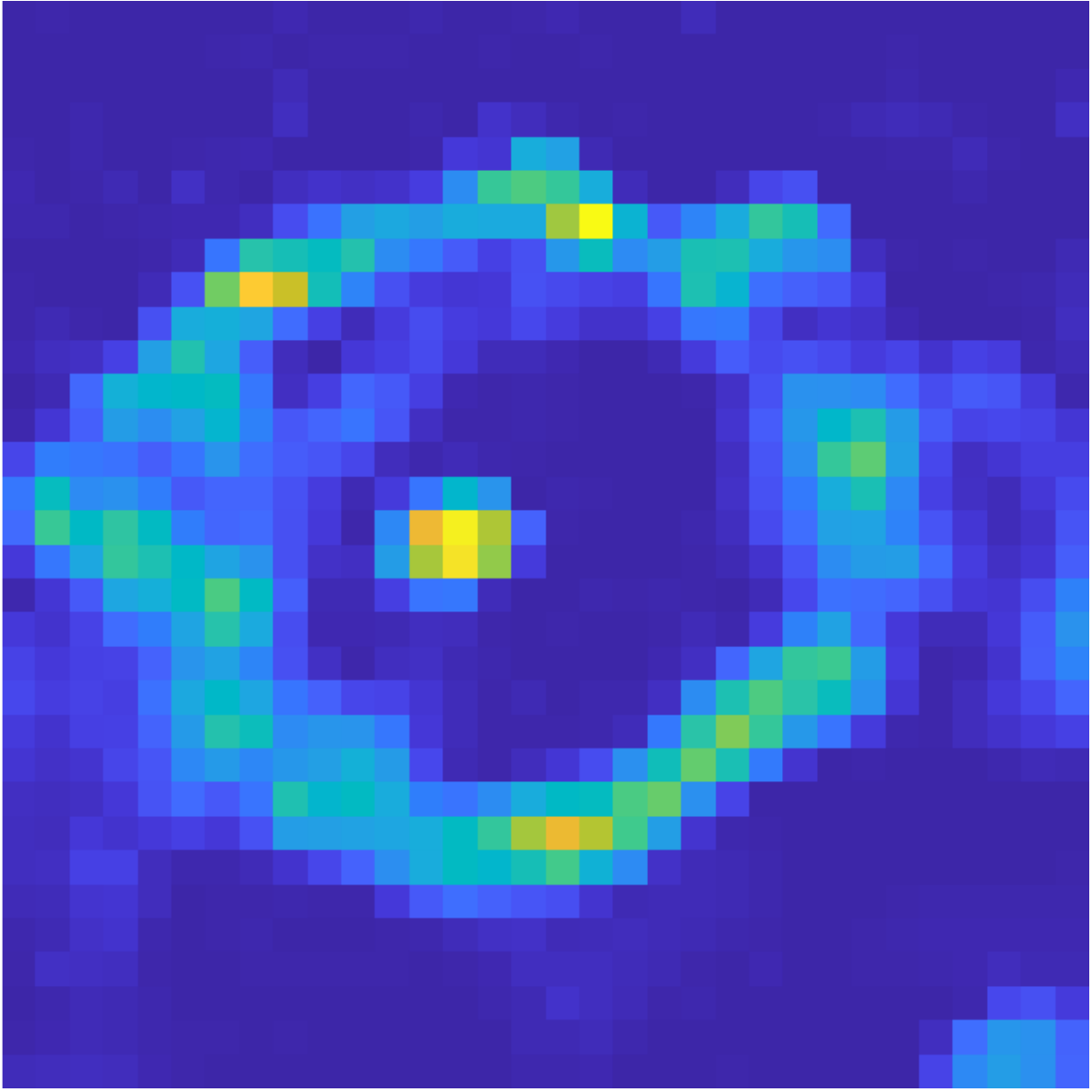} & \includegraphics[width=0.8in]{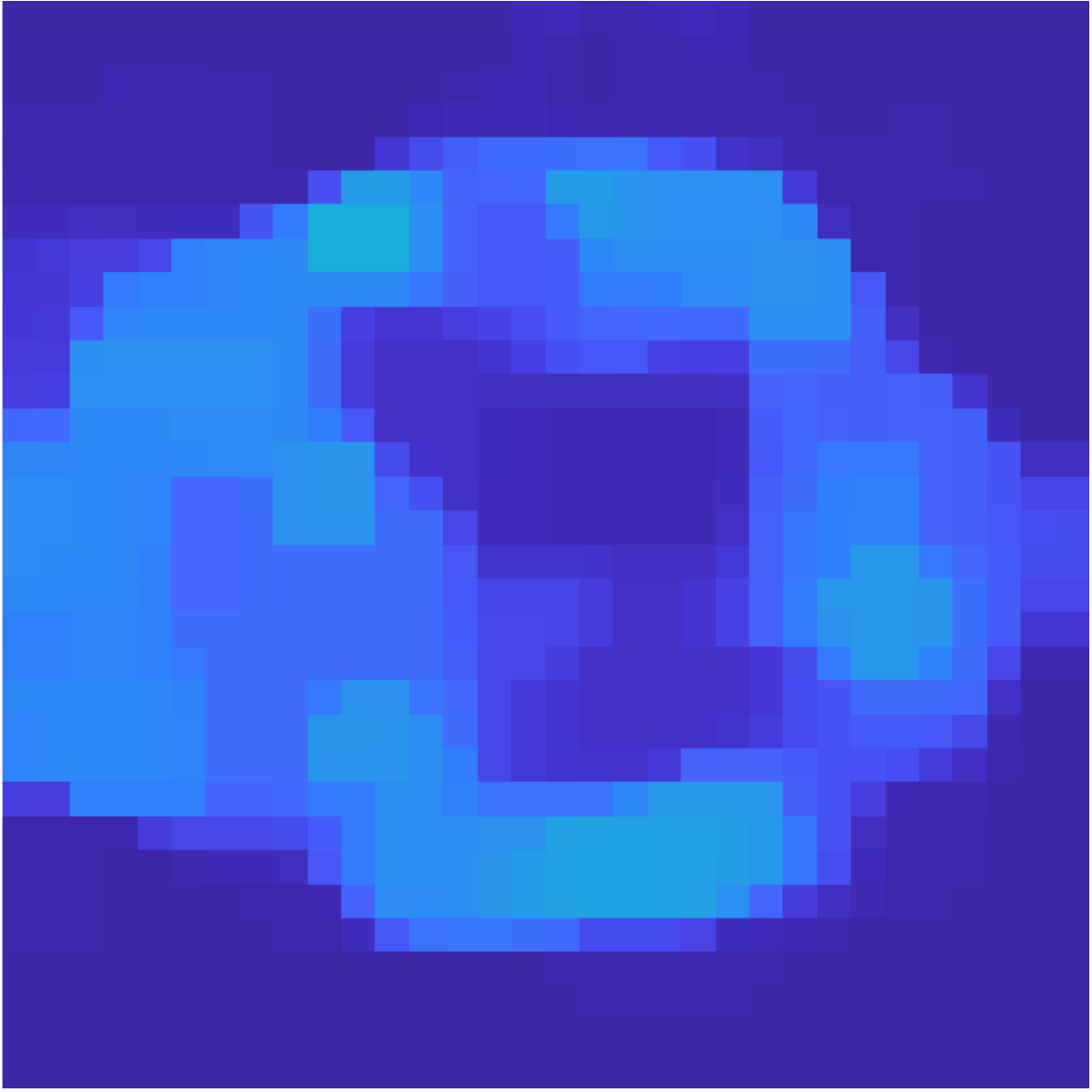} & \includegraphics[width=0.8in]{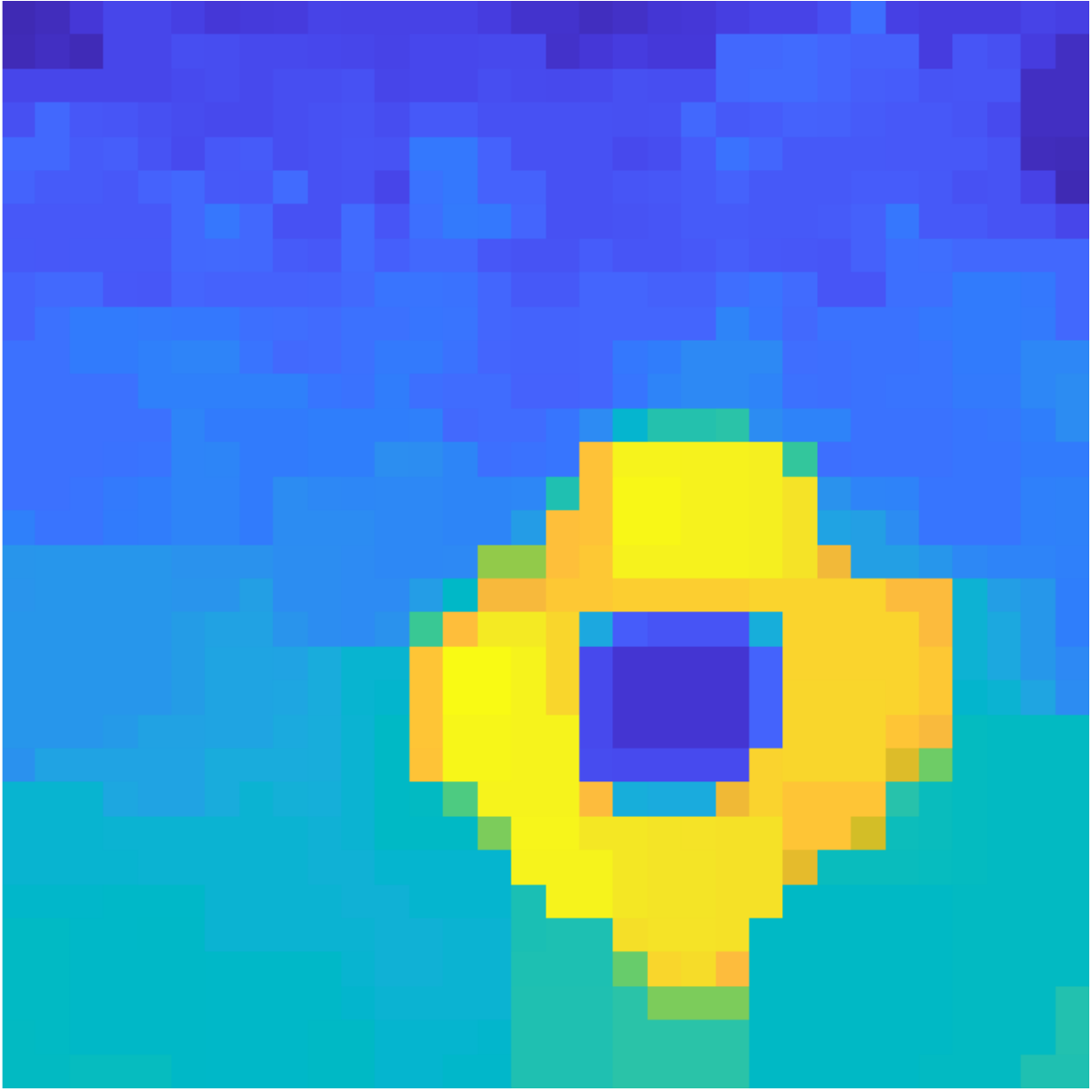} & \includegraphics[width=0.8in]{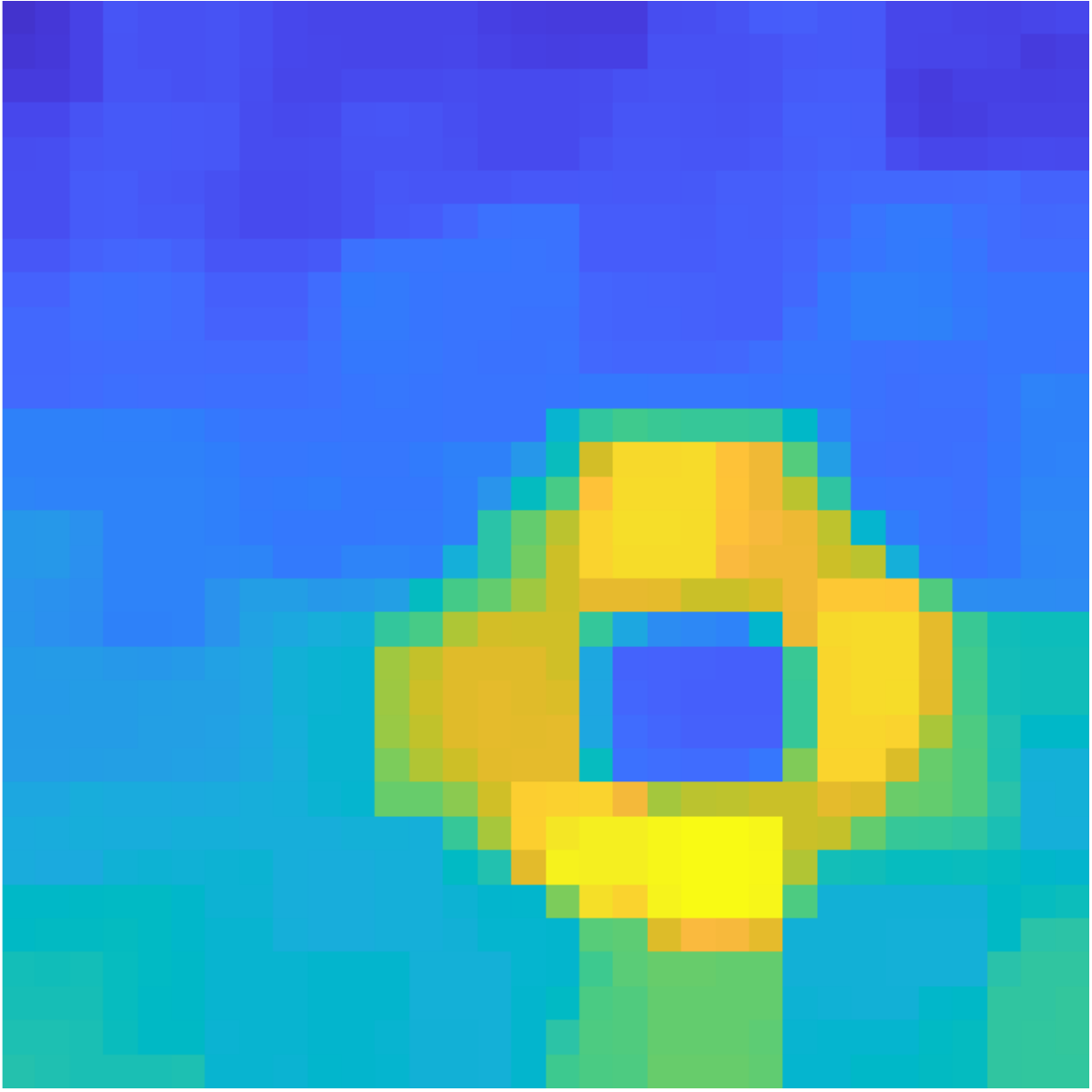}
	\end{tabular}
	\caption{ Noise robustness of proposed methods on Phantom 1 and Phantom 2 with $20\%$ noise. Left and right columns show reconstructions of Phantom 1 and 2 respectively.}
	\label{fig:Phantom1:noise}
\end{figure}

\begin{table}[!b]
	\caption{\small Comparison of methods on Phantom 1 and Phantom 2}
	\centering
	\small
	{\setlength{\tabcolsep}{0.2em}
	\begin{tabular}{r | r | c c c c | c c c c }
		\toprule
		$\mathbf{f}_{\textbf{max}}$ & & \multicolumn{4}{c|}{\textbf{Phantom 1}} & \multicolumn{4}{c}{\textbf{Phantom 2}}  \\
		& & \textbf{CISOR} & \textbf{RL} & \textbf{SF-$\tau$} & \textbf{SF-$\sigma$} & \textbf{CISOR} & \textbf{RL} & \textbf{SF-$\tau$} & \textbf{SF-$\sigma$} \\[1ex] \midrule
		\multirow{2}{*}{\textbf{1}} & DR & 0.87 & 45.89  & 0.74 & 2.36 & 0.32 & 29.16  & 0.05 & 0.06 \\
							& SNR & 14.73 & 3.87 & 15.12 & 9.19 & 27.63 & 8.84 & 42.79 & 18.22 \\ \midrule
		\multirow{2}{*}{\textbf{10}}	& DR & 28.08 & 75.27 & 8.78 & 11.13 & 945.16 & 260.43 & 3.77 & 24.75 \\
							& SNR & 2.17 & 1.94 & 3.83 & 4.47 & 0.16 & 11.08 & 47.07 & 18.00  \\ \midrule
		\multirow{2}{*}{\textbf{100}} & DR & 295.41 & 97.12 & 2.72 & 4.95 & 344.59 & 69.02 & 5.76 & 1.52  \\
							& SNR & 0.27 & 1.15 & 2.60 & 3.08 & 0.11 & 10.72 & 17.18 & 14.42  \\ \bottomrule
	\end{tabular}
	}
	\label{table:phantom1:comp}
\end{table}


\begin{table}[!b]
	\caption{\small   Noise-Robustness of SF-$\tau$ and SF-$\sigma$ }
	{
	\centering
	\small
	{\setlength{\tabcolsep}{0.3em}
	\renewcommand{\arraystretch}{1.2}
	\begin{tabular}{r|r|rr|rr|rr|rr}
		\toprule
		$\mathbf{f}_{\textbf{max}}$ & & \multicolumn{4}{c|}{\textbf{Phantom 1}} & \multicolumn{4}{c}{\textbf{Phantom 2}} \\
		& &  \multicolumn{2}{c|}{\textbf{SF-$\tau$}}  & \multicolumn{2}{c|}{\textbf{SF-$\sigma$}} & \multicolumn{2}{c|}{\textbf{SF-$\tau$}}  & \multicolumn{2}{c}{\textbf{SF-$\sigma$}} \\
		& & $\mathbf{10 \%}$ &  $\mathbf{20 \%}$ & $\mathbf{10 \%}$ & $\mathbf{20 \%}$ & $\mathbf{10 \%}$ &  $\mathbf{20 \%}$ & $\mathbf{10 \%}$ & $\mathbf{20 \%}$ \\ \midrule
		\multirow{2}{*}{\textbf{1}} & DR & 7.00 & 15.10 & 13.20 & 25.79 & 9.53 & 19.77 & 11.50 & 22.73 \\
							& SNR & 10.59 & 8.64 & 6.97 & 5.35 & 19.74 & 15.77 & 14.06 & 12.04 \\ \midrule
		\multirow{2}{*}{\textbf{10}}	& DR & 19.91 & 35.38 & 21.77 & 51.59 & 41.49 & 52.53 & 42.24 & 62.78 \\
							& SNR & 3.23 & 3.23 & 4.37 & 3.43 & 18.98 & 15.15 & 14.68 & 12.34 \\ \midrule
		\multirow{2}{*}{\textbf{100}} & DR & 42.42 & 74.96 & 49.12 & 87.96 & 27.85 & 49.16 & 35.49 & 73.38 \\
							& SNR & 2.10 & 1.27 & 3.15 & 2.54 & 13.69 & 14.02 & 12.43 & 11.46 \\ \bottomrule
	\end{tabular}
	}
	\label{table:phantom1:noise}
	}
\end{table}

\subsection{Performance Measures}
We use the following measures to evaluate the performance of the proposed methods and to compare with other methods.
\begin{description}
	\item[DR:] The data residual (DR) measures the distance of the modeled data for reconstructed model with the actual data in the euclidean sense. For multi-frequency data the DR takes the following form
	\[
		\mbox{DR} \triangleq 100 \times \frac{\sum_{j \in \mathcal{J}} \mathcal{F}_j (\mathbf{f}^\star)}{\sum_{j \in \mathcal{J}} \|\mathbf{Y}_j \|^2},
	\]
	where $\mathbf{f}^\star$ is the reconstructed solution. Here, $\| \mathbf{Y} \|$ denotes the Frobenius norm for the matrix $\mathbf{Y}$. DR must be close to the noise-level for a method to be considered good.
	\item[SNR:] The signal-to-noise ratio (SNR) for the reconstructed model $\mathbf{f}^\star$ with respect to the ground truth $\mathbf{f}^\text{true}$ is
	\[
		\mbox{SNR} \triangleq -20 \log_{10} \left( \frac{\| \mathbf{f}^\star - \mathbf{f}^{\text{true}} \|}{\| \mathbf{f}^{\text{true}} \|}\right).
	\]
	A reconstruction is considered good if it has high SNR. This measure is only available if we know the ground truth. 
\end{description}

\begin{figure*}[!htbp]
	\centering
	\begin{tabular}{cccc}
	\includegraphics[height=0.13\textheight]{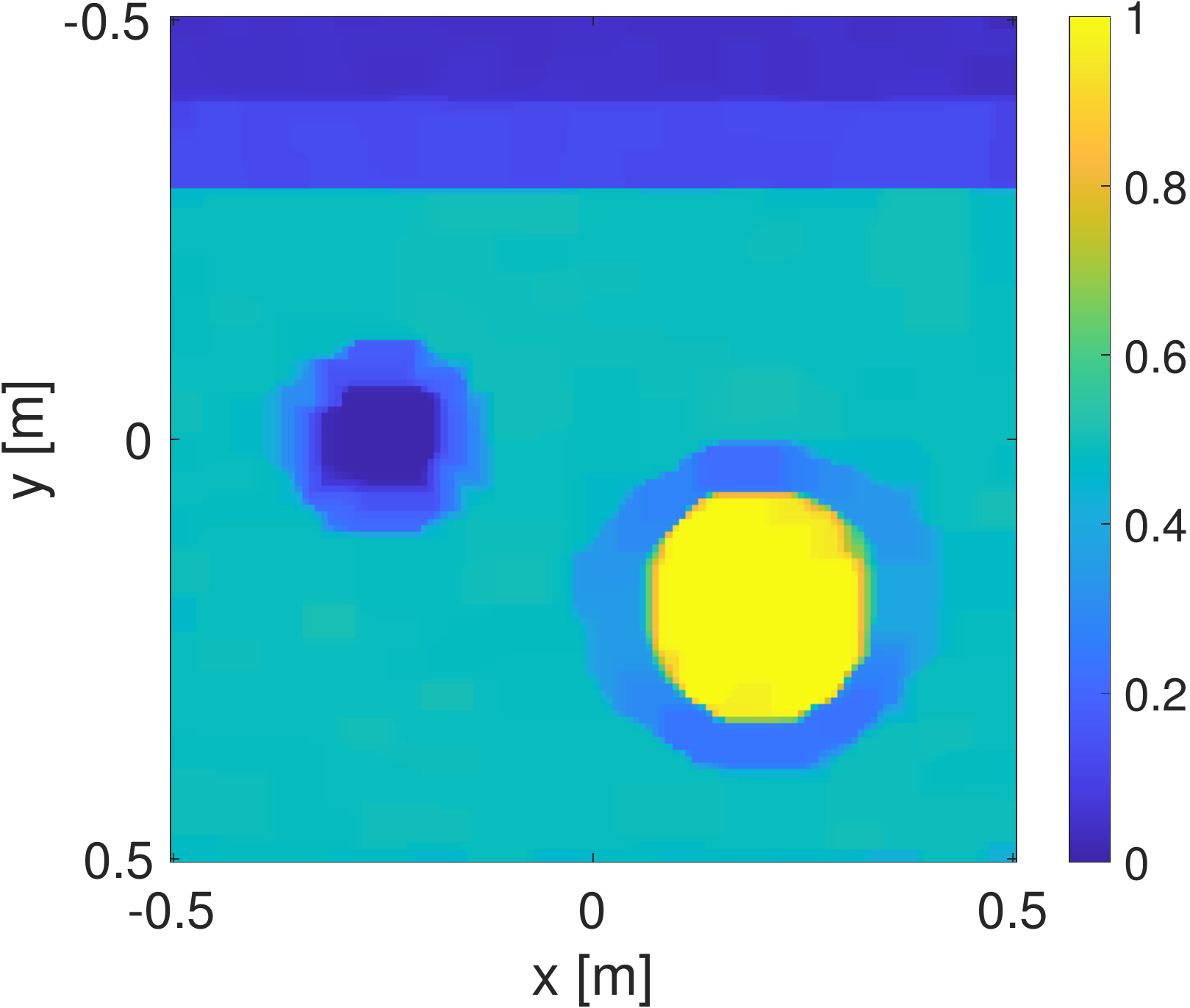} & \includegraphics[height=0.13\textheight]{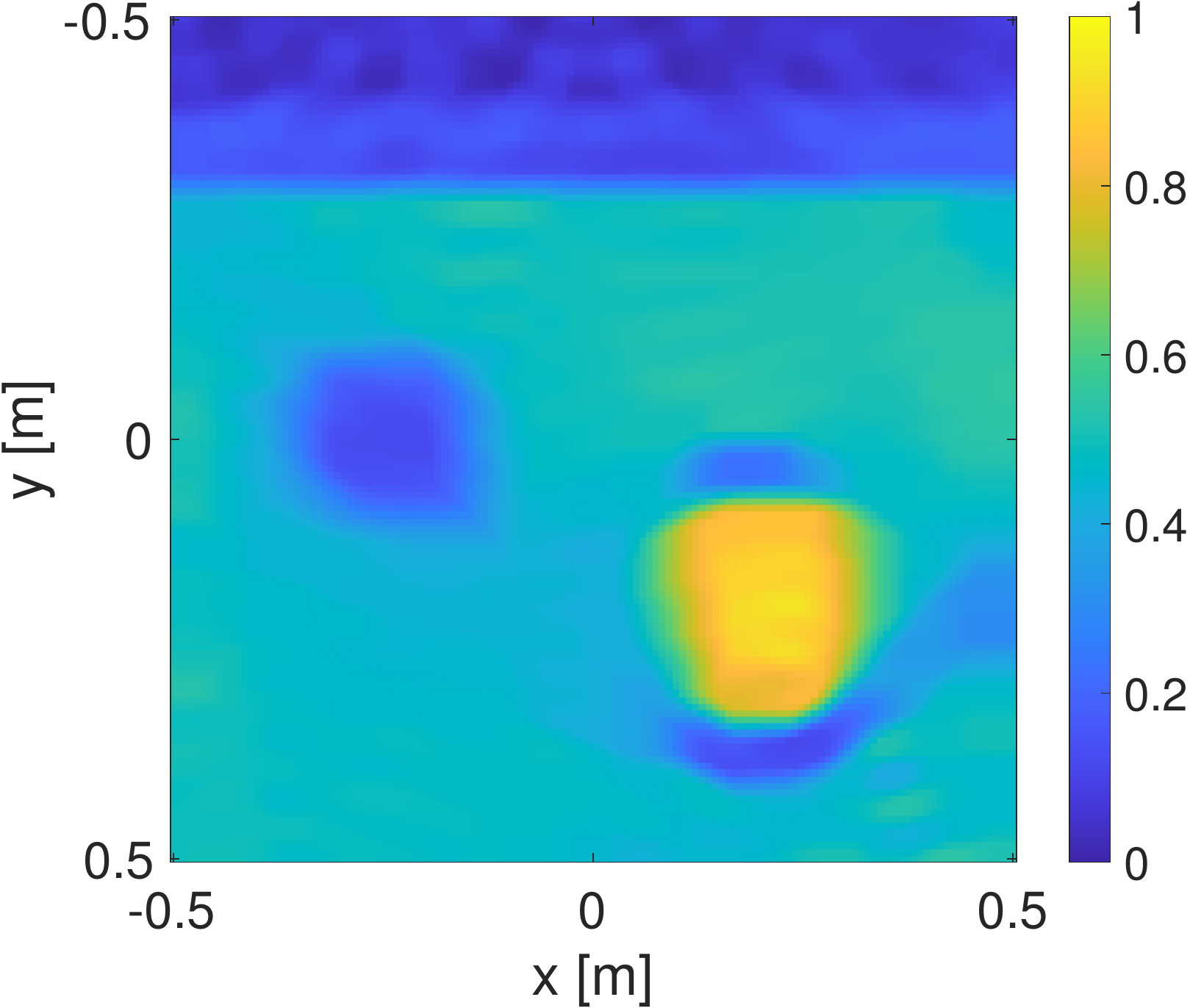} & \includegraphics[height=0.13\textheight]{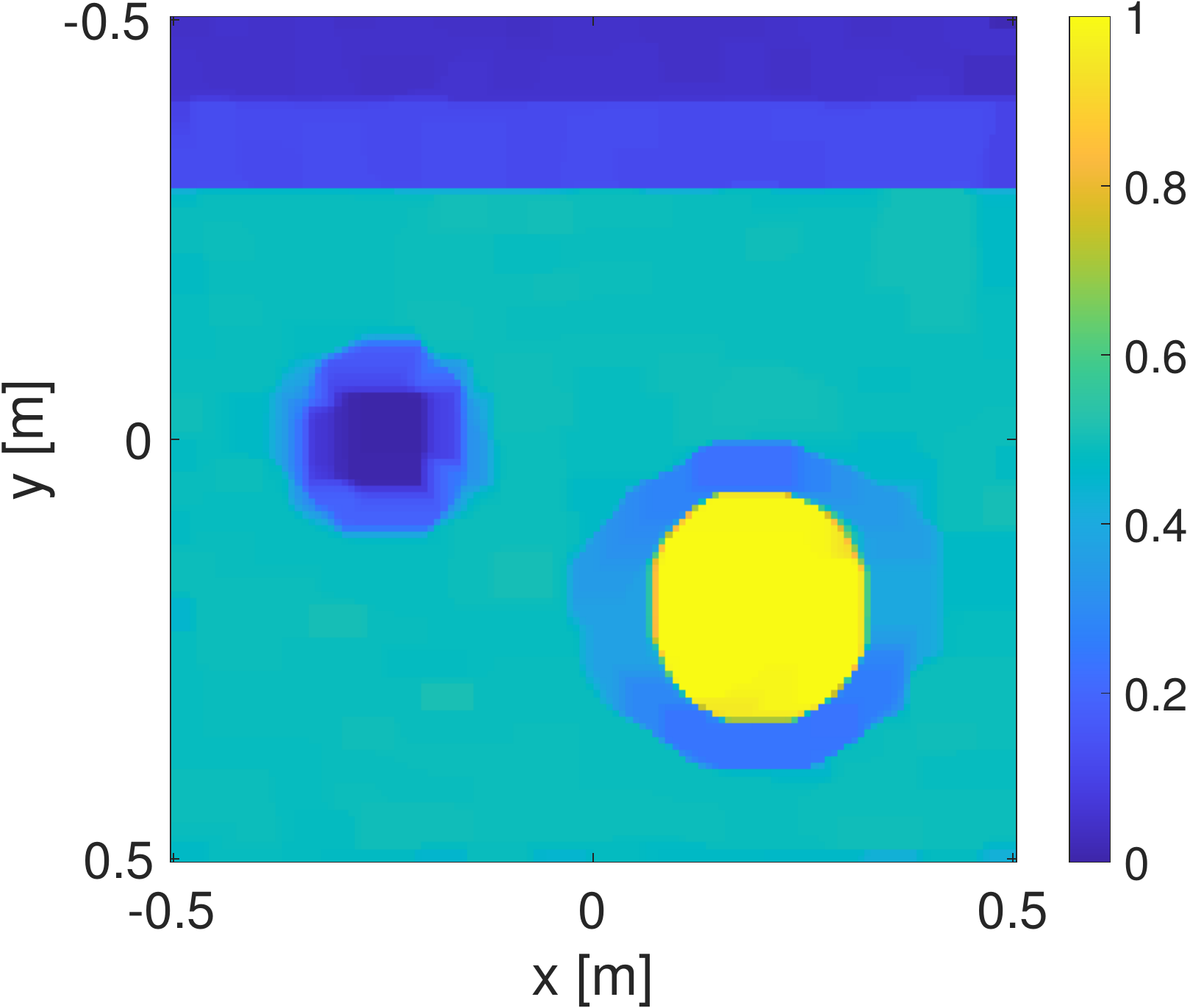}  & \includegraphics[height=0.13\textheight]{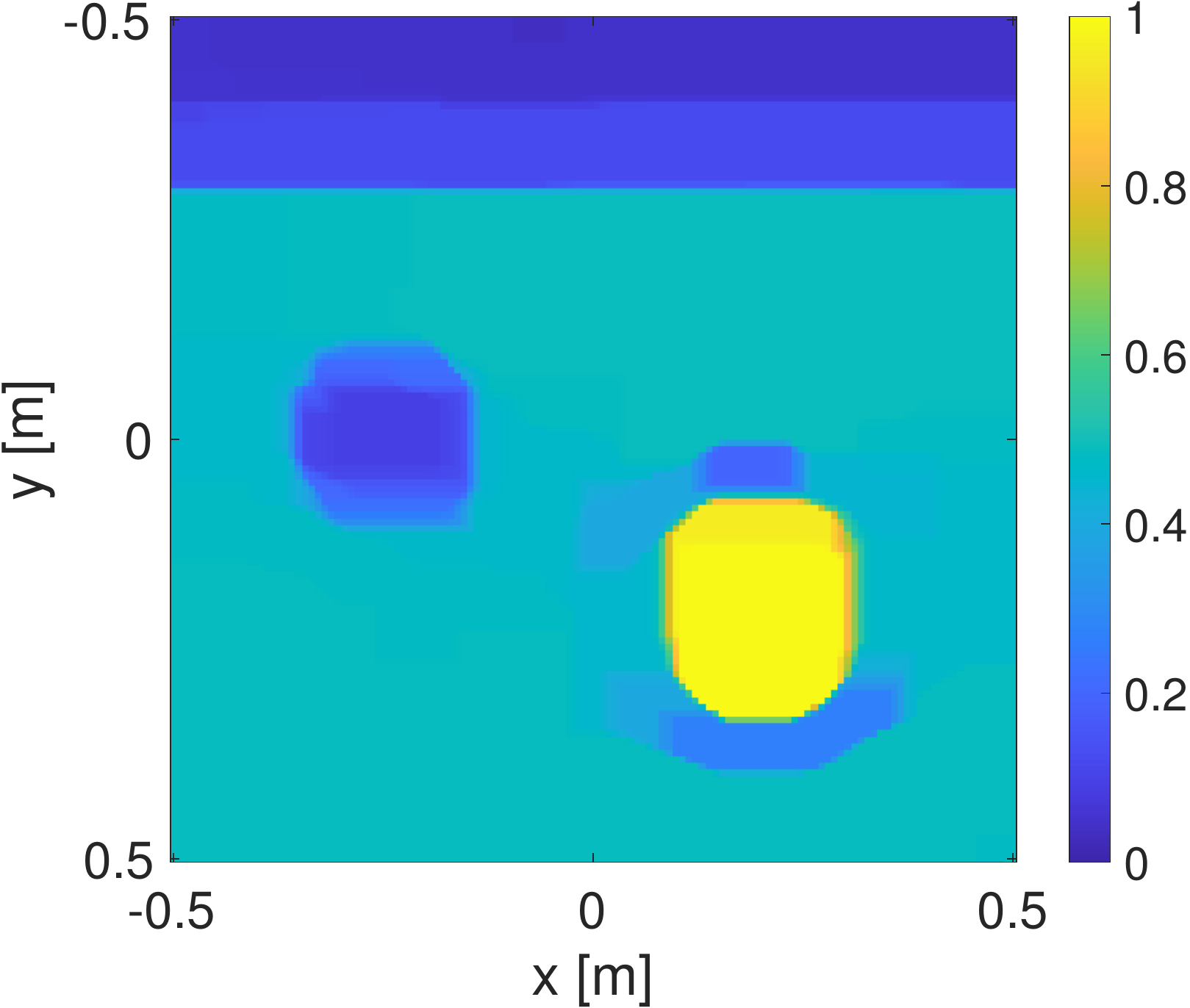}  \\
	(a) CISOR & (b) RL & (c) SF-$\tau$ & (d) SF-$\sigma$ \\
	\end{tabular}
	
	\vspace{0.2cm}
	
	\begin{tabular}{ccc}
	\includegraphics[height=0.13\textheight]{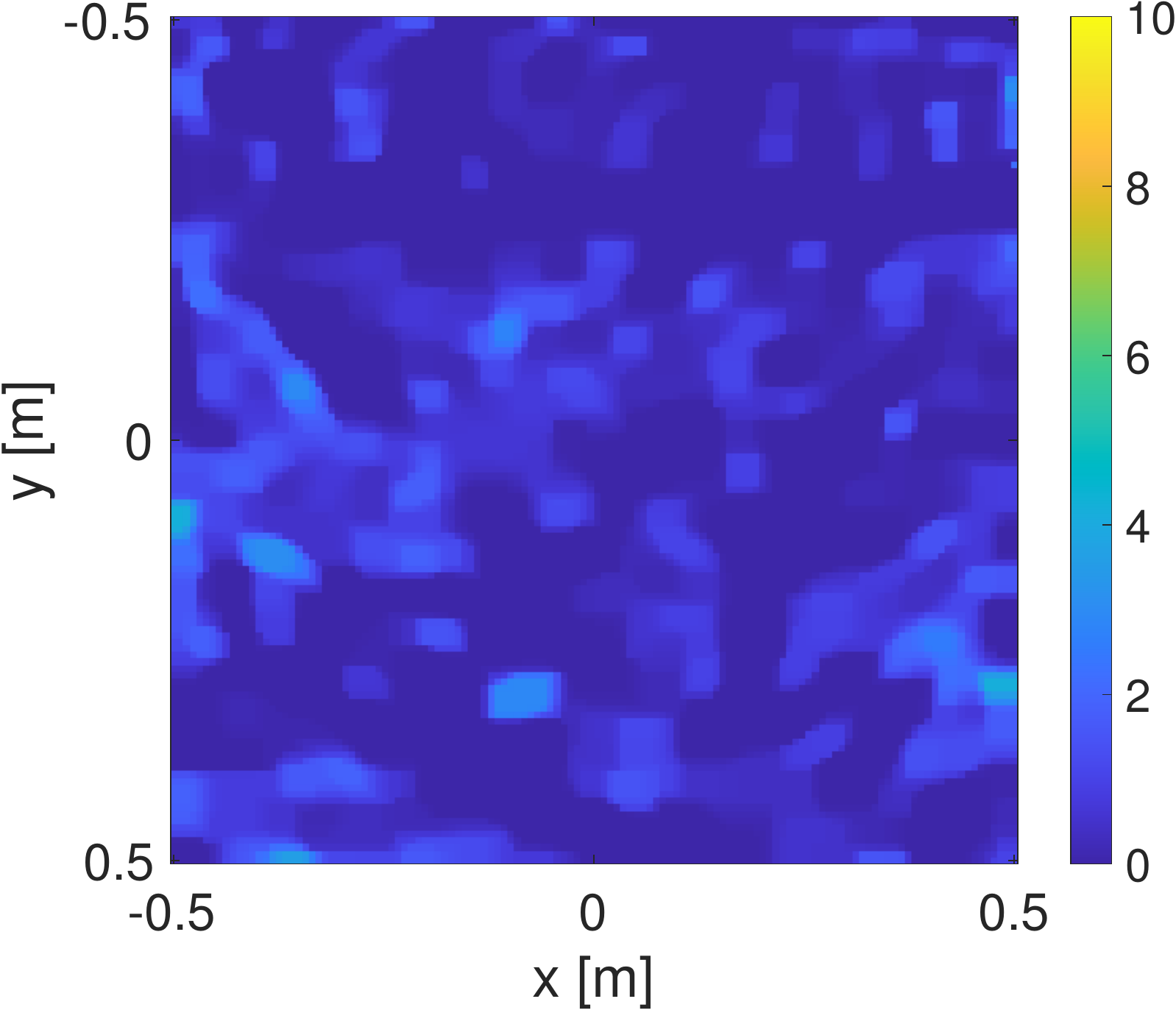} & \includegraphics[height=0.13\textheight]{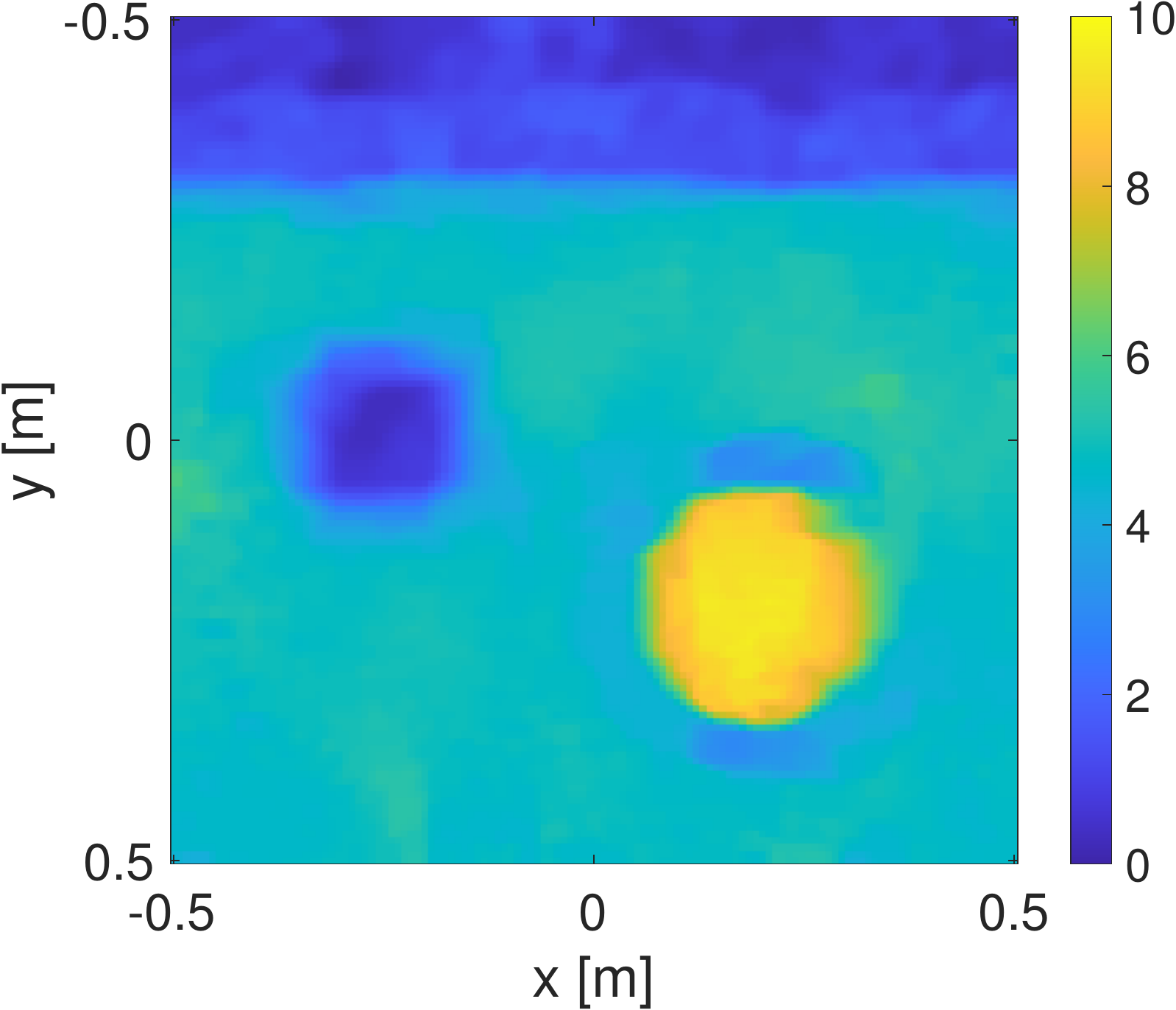} &  \includegraphics[height=0.13\textheight]{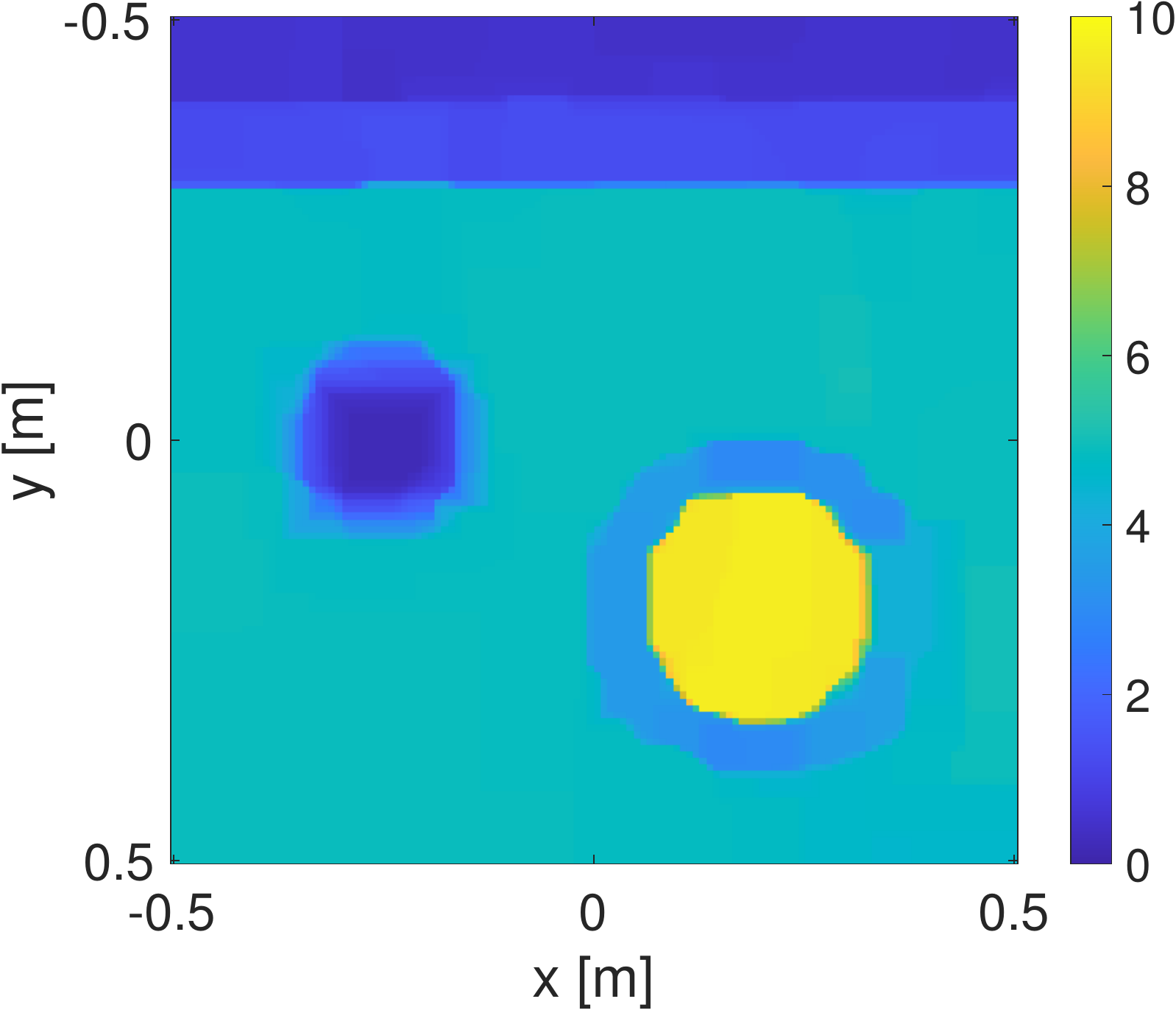}  \\
	(e) CISOR & (f) RL & (g) SF-$\sigma$ 
	\end{tabular}
	
	\caption{Inexact model experiments on Phantom 3. The top row shows the reconstructions on low-contrast phantom, while the bottom row shows the reconstructions on medium-contrast phantom.} \vspace{-3mm}
	\label{fig:NICSF}
\end{figure*}

\subsection{Exact-model experiments}
To evaluate our approach, we perform both noise-free and noisy experiments, in which the exact model is known. In the noise-free experiments, we compare our methods with the other two methods (CISOR and RL). In noisy ones, we only examine the robustness of our methods against various levels of noise.

\subsubsection{Noise-free experiment}
We consider Phantom 1 and 2 for this experiment. We produce three types of phantoms by scaling these phantoms with a maximum contrast ($\mathbf{f_\text{max}}$) of $\{ 1, 10, 100\}$, which we consider to be low, medium, and high-contrast phantoms, respectively. For the simulations we use the reflection tomography setup illustrated in Figure~\ref{fig:setupPh}(a) with noiseless data. We examine the performance of the methods SF-$\tau$ and SF-$\sigma$, and compare it with the CISOR and RL method. Figures~\ref{fig:Phantom1:comp} and~\ref{fig:Phantom2:comp} show the reconstructions for various contrast levels for Phantom 1 and 2,respectively. We see that SF-$\tau$ consistently performs well except in the case of $\mathbf{f_\text{max}}$ = 100 for Phantom 1, where all the methods fail. The reason for the failure is that Phantom 1 is ideal for transmission or full-view tomography and not for reflection tomography. For an underground scene (depicted by Phantom 2) we see that the proposed methods performs well with the reflection tomography.
We tabulate the values for the performance measures in Table~\ref{table:phantom1:comp}. We conclude that the SF-$\tau$ and SF-$\sigma$ perform superior to the existing methods (CISOR and RL).

\subsubsection{Noisy experiment}
We consider Phantom 1 and 2, with the scaling $\{1, 10, 100\}$. We add a Gaussian noise of relative energy $10\%$ (20dB measurement SNR), and $20\%$ (14dB measurement SNR), and examine the performance of SF-$\tau$ and SF-$\sigma$ on these noise levels. Figure~\ref{fig:Phantom1:noise} shows the reconstructions using these methods for $20\%$ relative noise energy and various levels of contrast values. The performance measures are tabulated in Table~\ref{table:phantom1:noise}. We observe that SF-$\tau$ and SF-$\sigma$ are robust against high noise in the low-contrast phantoms. SF-$\tau$ is also stable for moderate level of noise in high-contrast regime.

\begin{figure}
    \centering
    \setlength\tabcolsep{1.5pt}
	\begin{tabular}{c c c c}
	\includegraphics[width=0.12\textwidth]{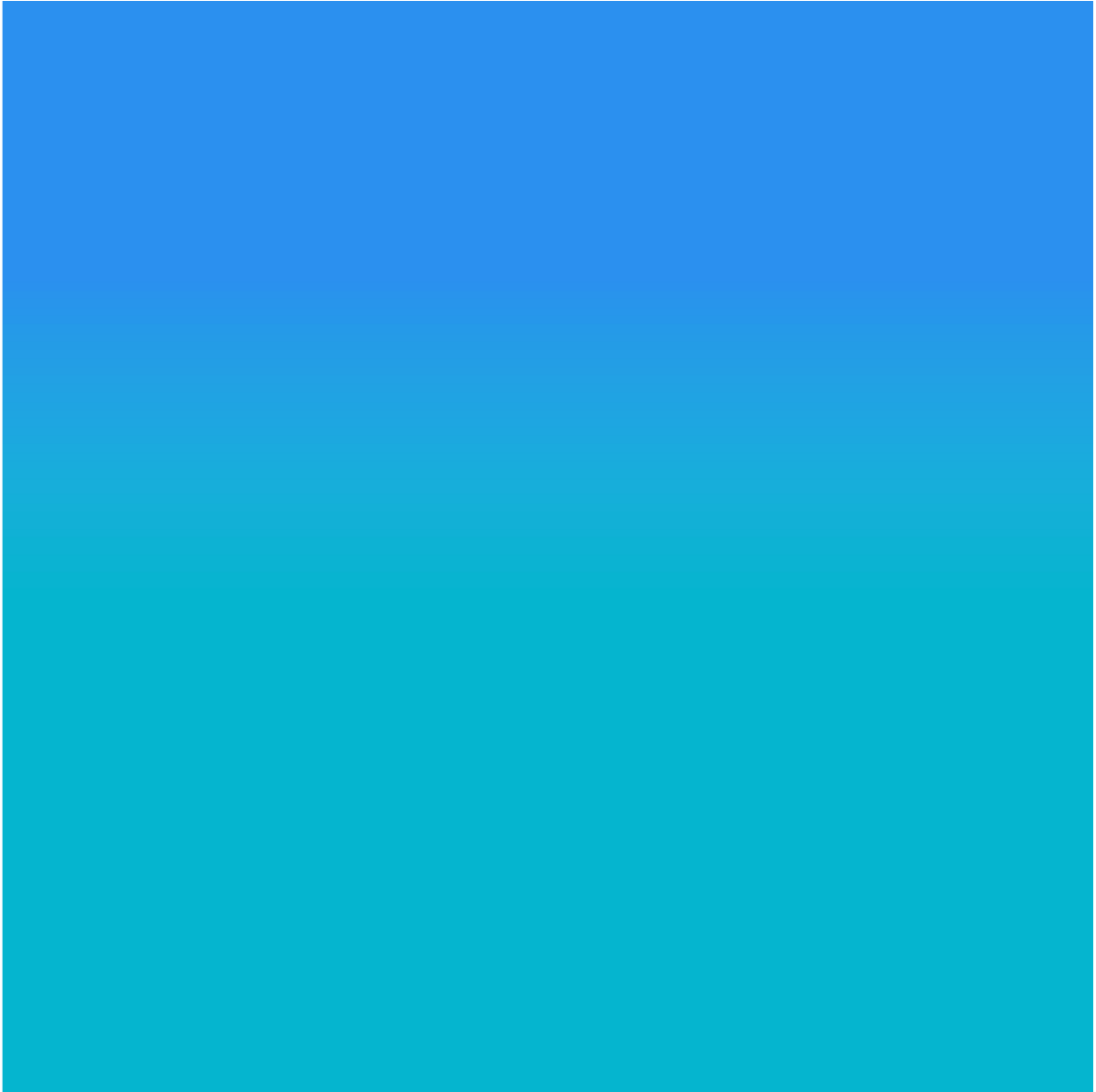} & \includegraphics[width=0.12\textwidth]{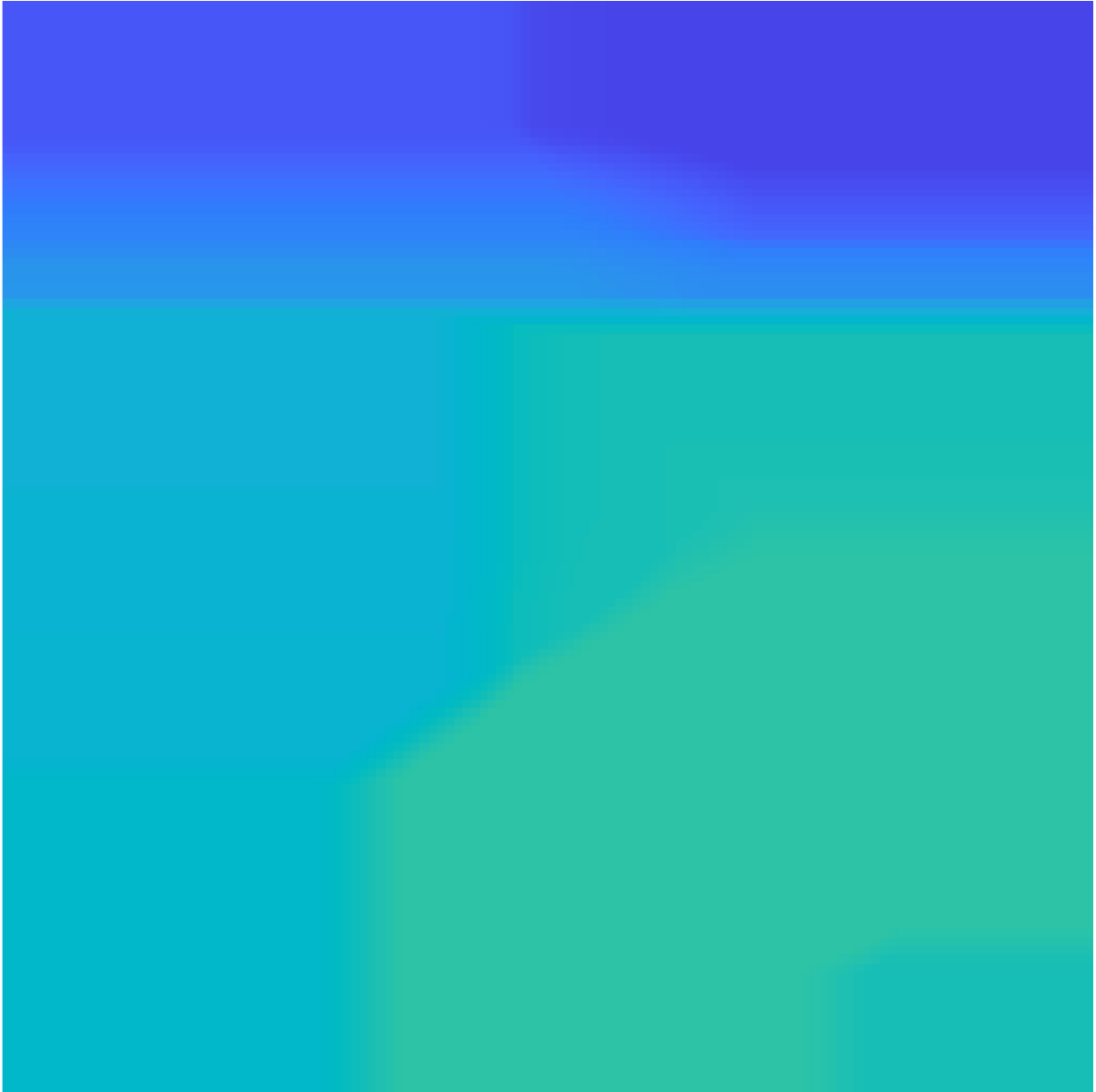} & \includegraphics[width=0.12\textwidth]{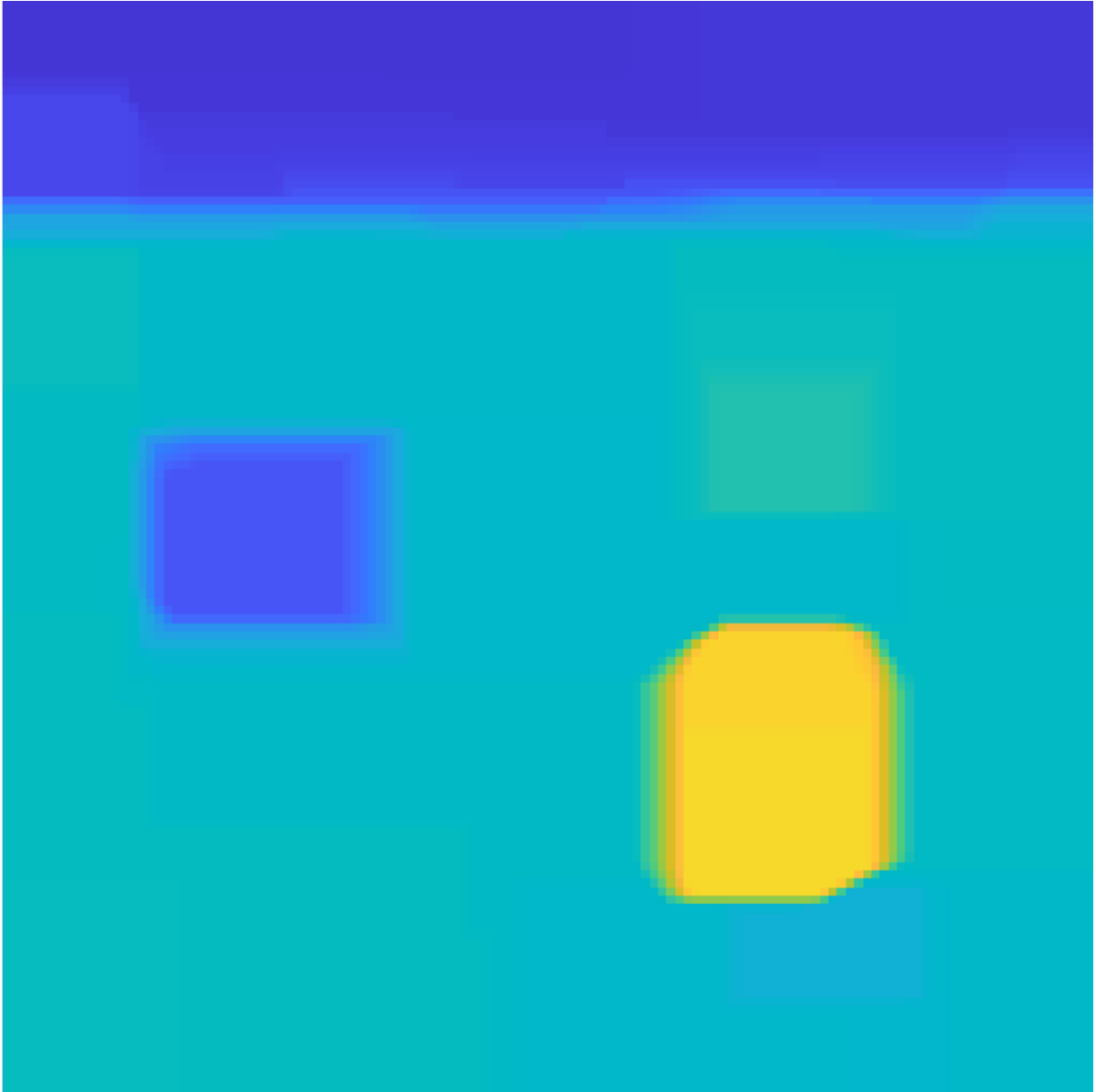} & \includegraphics[width=0.12\textwidth]{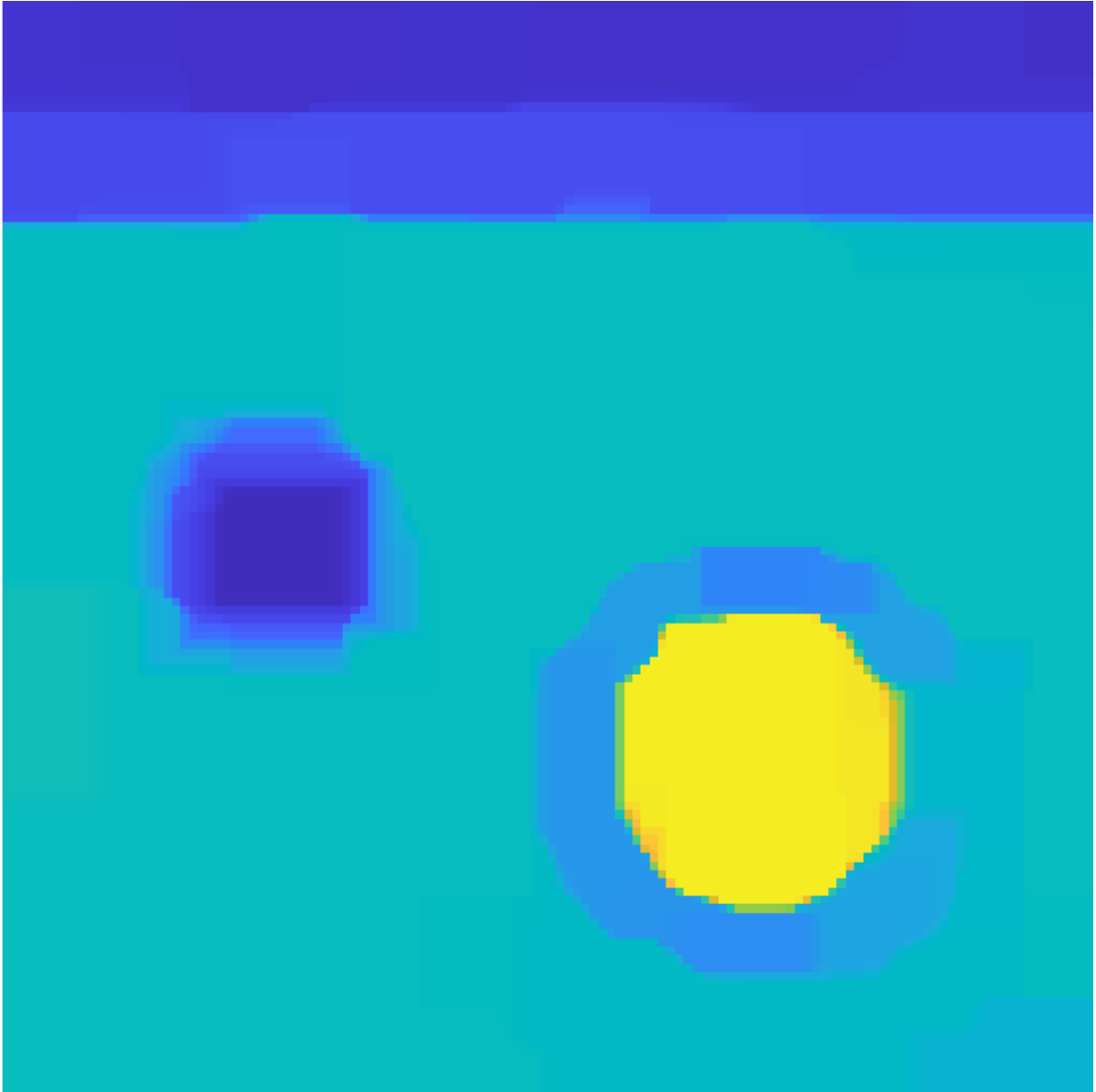} \\
	(a) 10~Hz & (b) 100~Hz & (c) 500~Hz & (d) 1000~Hz 
	\end{tabular}
    \caption{Intermediate reflectivity maps from SF-$\sigma$ for the inexact model for Phantom 3 on medium-contrast phantom highlighting the low-to-high spatial resolution of the reconstruction.} \vspace{-5mm}
    \label{fig:evolution}
\end{figure}

\subsection{Inexact-model experiment}
\label{sec:nonInvCrime}
To verify the robustness of our approach, we consider Phantom 3 for this test that has a resolution of $128 \times 128$, using an inexact model for the reconstruction. { To avoid \textit{inverse-crime}, we generate the measurements with a high-resolution modeling grid with additional Gaussian noise, and use a low-resolution grid as a forward solver~\cite{kaipio2007statistical}}. In particular, we first discretize the model on a high-resolution grid of size $192 \times 192$. We use the nearest-neighbor algorithm for the rescaling to a high-resolution grid. We generate the data on the high-resolution grid, and add $10\%$ (20dB measurement SNR) white Gaussian noise relative to the signal power. As a sanity check, we look at the difference between the data for high-resolution and low-resolution model, and found the relative difference is less than $20\%$. { We test CISOR, RL, SF-$\tau$ and SF-$\sigma$ on low-constrast phantom ($\mathbf{f}_\text{max} = 1$) with this high-resolution dataset. We assume a noise level of $20\%$ for SF-$\sigma$, while we set $\tau$ to be the TV-value of the ground truth (low-resolution model) for CISOR, RL and SF-$\tau$. The reconstruction results for these methods are presented in Figure~\ref{fig:NICSF}(a)-(d). CISOR has DR of 1.47 and SNR of 20.32dB, while RL has DR of 12.52 and SNR of 14.23dB. Similarly,  SF-$\tau$ has DR of 1.46, and SNR of 19.95dB, while SF-$\sigma$ has a DR of 4.74, and SNR of 16.71dB. We observe that CISOR and SF-$\tau$ are able to reconstruct the ground scene accurately: the top and the bottom regions of the pipes are retrieved to high precision. SF-$\sigma$ is able to locate the high-contrast and the low-contrast objects in the pipes but fails to get the boundary of the pipes accurately. Moreover, RL struggles to predict the accurate geometries of pipes as well as layers. As we have seen in Figures~\ref{fig:Phantom1:comp} and \ref{fig:Phantom2:comp}, we conclude that the low-contrast phantom can be reconstructed with CISOR, SF-$\sigma$ and SF-$\tau$.

Next, we run a similar test for medium-contrast phantom, \ie, Phantom 3 scaled to the maximum contrast of 10. For this test, we only show the reconstructions from SF-$\sigma$ with CISOR and RL, since SF-$\tau$ consistently performs better. Figure~\ref{fig:NICSF}(e)-(g) provide reconstruction results for these methods. The CISOR has DR of 342.82, SNR of 0.64dB, while RL achieves DR of 24.24 and SNR of 15.19dB. Compared to these methods, SF-$\sigma$ has a DR of 13.78 and SNR of 18.73dB. Hence, CISOR fails on medium-contrast phantoms, while RL still struggles to provide an accurate picture of target image. On the other hand, the reconstruction from SF-$\sigma$ quite accurately recovers the layers and the pipes.

Finally, we plot in Figure~\ref{fig:evolution} the intermediate solutions from SF-$\sigma$ on medium-contrast phantoms. In particular, we plot the solutions for batches 10~Hz, 10-100~Hz, 10-500~Hz, and 10-1000~Hz. We observe that the solution at 10~Hz obtains an almost constant image due to the low TV constraint. As we move towards higher frequencies, the reconstruction method starts to fill in the details in the image by allowing for higher values of TV and higher frequencies in the measurements. This behavior is reminiscent of ``multi-grid methods" while avoiding their complex bookkeeping requirements.}

\section{Conclusions}
\label{sec:conc}
We consider limited-angle reflection tomography of high-contrast objects and show that the tomography problem is severely ill-posed due to the absence of low-frequency content and multiple scattering of waves. To find a feasible solution to this ill-posed problem, we develop a regularized multiscale approach. We pose the imaging problem as a nonlinear least-squares problem with constraints. The cost function includes the wave-based modeling that accounts for multiple scattering and a regularization term that includes non-negativity and total variation constraints. The total cost function is decomposed according to the frequency, and we observe that the low-frequencies promote smoothness while higher frequencies add details in the reconstruction.  Hence, we solve a sequence of subproblems, where the $k^{\text{th}}$ subproblem has a constrained cost function measured over the first $k$ frequencies. We propose a proximal-Quasi-Newton method to solve the resulting constrained problem. The underlying proximal operations are performed using a primal-dual approach. We also propose an automatic strategy to update the TV-constraint parameter based on the noise-level in the data. Through numerical experiments, we demonstrate that our methodologies outperform the existing methods and is robust against moderate noise. The proposed techniques can retrieve high-contrast object (contrast up to 100) for scenes similar to the underground.

\appendices

\numberwithin{equation}{section}

\section{Scattering Formalism}
Consider a scattering setup illustrated in Figure~\ref{fig:acq-setup}. The scene (free-space with permittivity $\epsilon_b$) has a dimension $d$. The transmitter domain $\Gamma_t \subset \mathbb{R}^d$ emits a source function $q: \Gamma_t \mapsto \mathbb{C}$, which generates an incident wavefield $u_\text{in}: \mathbb{R}^d \mapsto \mathbb{C}$ everywhere. This incident wavefield interacts with an object in domain $\Omega \subset \mathbb{R}^d$ and generates a total wavefield $u: \mathbb{R}^d \mapsto \mathbb{C}$. The scattered wavefield $u_\text{sc} := u - u_\text{in}$ is then measured in the receiver domain $\Gamma_r \subset \mathbb{R}^d$.

The total wavefield is a superposition of an incident field $u_\text{in}(\mathbf{r})$ and a scattered field $u_\text{sc}(\mathbf{r})$,
\begin{equation}
	u(\mathbf{r}) = u_\text{in}(\mathbf{r}) + u_\text{sc}(\mathbf{r}), \qquad \mathbf{r} \in \mathbb{R}^d.
	\label{eq:wavfieldRelation}
\end{equation}
The incident wavefield is the field in the absence of the scatterer, while the scattered field takes the presence of object into account. The incident wavefield satisfies the Helmholtz equation
\[
	\nabla^2 u_\text{in}(\mathbf{r}) - k^2 \epsilon_b u_\text{in}(\mathbf{r}) = -q(\mathbf{r}) \qquad \forall \mathbf{r} \in \mathbb{R}^d,
\]
where $k$ denotes the wavenumber. It is convenient to consider the above equation for inside and outside the object domain $\Omega$:
\begin{equation}
	\begin{split}
	 \nabla^2 u_\text{in}(\mathbf{r}) - k^2 \epsilon_b u_\text{in}(\mathbf{r}) = 0 \qquad &  \forall \mathbf{r} \in \Omega, \\
	  \nabla^2 u_\text{in}(\mathbf{r}) - k^2 \epsilon_b u_\text{in}(\mathbf{r}) = -q(\mathbf{r}) \qquad & \forall \mathbf{r} \notin \Omega,
	  \end{split}
	  \label{eq:incWavefield}
\end{equation}
Similarly, the total wavefield satisfies the Helmholtz equation, and we can express it inside and outside the domain as follows,
\begin{equation}
	\begin{split}
	 \nabla^2 u(\mathbf{r}) - k^2 \epsilon(\mathbf{r}) u(\mathbf{r}) = 0 \qquad & \forall \mathbf{r} \in \Omega, \\
	  \nabla^2 u(\mathbf{r}) - k^2 \epsilon_b u(\mathbf{r}) = -q(\mathbf{r}) \qquad & \forall \mathbf{r} \notin \Omega,
	  \end{split}
	  \label{eq:totWavefield}
\end{equation}
where $\epsilon(r)$ is the permittivity of the object. Now, from the equations \eqref{eq:wavfieldRelation}, \eqref{eq:incWavefield} and \eqref{eq:totWavefield}, the governing equation for the scattered wavefield reads
\begin{equation*}
	\begin{split}
	 \nabla^2 u_\text{sc}(\mathbf{r}) - k^2 \epsilon_b u_\text{sc}(\mathbf{r}) = -k^2\left( \epsilon_b - \epsilon(\mathbf{r}) \right) u(\mathbf{r}) \qquad & \forall \mathbf{r} \in \Omega, \\
	  \nabla^2 u_\text{sc}(\mathbf{r}) - k^2 \epsilon_b u_\text{sc}(\mathbf{r}) = 0 \qquad & \forall \mathbf{r} \notin \Omega,
	  \end{split}
\end{equation*}
These equations can be compactly written as
\begin{align}
	 \nabla^2 u_\text{sc}(\mathbf{r}) - k^2 \epsilon_b u_\text{sc}(\mathbf{r}) = -k^2 f(\mathbf{r}) u(\mathbf{r}) \qquad \forall \mathbf{r} \in \mathbb{R}^d
	 \label{eq:scatWavefield}
\end{align}
where $f(\mathbf{r})$ is a \textit{contrast function} that is equal to the difference between the permittivity, $\epsilon(r) - \epsilon_b$, inside the object domain $\Omega$ and $0$ outside. We supplement the scattered wavefield equation \eqref{eq:scatWavefield} with the Sommerfeld radiation condition
\[
	\lim_{r \mapsto \infty} r \left( \frac{\partial u_\text{sc}}{\partial r} - i k u_\text{sc} \right) = 0
\]
where $r = \| \mathbf{r} \|$. Equation \eqref{eq:scatWavefield} can be converted to an equivalent integral equation by introducing the free space Green function. The free space Green function $g: \mathbb{R}^d \mapsto \mathbb{R}^d$ satisfies
\[
	\nabla^2 g(\mathbf{r}) + k^2 \epsilon_b g(\mathbf{r}) = -\delta (\mathbf{r}), \qquad \forall \; \mathbf{r} \in \mathbb{R}^d
\]
together with the Sommerfeld radiation conditions. Here, $\delta$ is a dirac-delta function. The explicit representation for the Green function reads
\begin{align*}
	g(\mathbf{r}) \triangleq \begin{cases}
		-\frac{i}{2k}e^{-ikr} & \quad d = 1 \\[1ex]
		-\frac{i}{4}H_0^{(2)}(kr) & \quad d = 2 \\[1ex]
		\frac{1}{4 \pi r} e^{-ikr} & \quad d = 3
	\end{cases},
\end{align*}
where $r = \|\mathbf{r} \|$, and $H_0^{(2)}$ is the zero-order Hankel function of second kind.  Hence, the integral representation for the input wavefield is
\begin{align*}
	u_\text{in}(\mathbf{r}) = k^2 \int_{\mathbf{r}' \in \Gamma_t} \, g(\mathbf{r} - \mathbf{r}')q(\mathbf{r}') \, \mathrm{d} \mathbf{r}' \qquad \forall \, \mathbf{r} \in \mathbb{R}^d,
\end{align*}
and similarly, for the scattered wavefield is
\begin{align*}
	u_\text{sc}(\mathbf{r}) = k^2 \int_{\mathbf{r}' \in \Omega} \, g(\mathbf{r} - \mathbf{r}') u(\mathbf{r}') f(\mathbf{r}')\, \mathrm{d} \mathbf{r}' \qquad \forall \, \mathbf{r} \in \mathbb{R}^d.
\end{align*}
Noting that the scattered wavefield is the difference of the total wavefield and the input wavefield (see \eqref{eq:wavfieldRelation}) and restricting our observations to the object domain $\Omega$, we arrive at the well-known Lippmann-Schwinger equation
\begin{equation*}
	u(\mathbf{r}) = u_\text{in}(\mathbf{r}) +  k^2 \int_{\mathbf{r}' \in \Omega} \, g(\mathbf{r} - \mathbf{r}') u(\mathbf{r}') f(\mathbf{r}') \, \mathrm{d} \mathbf{r}' \qquad \forall \, \mathbf{r} \in \Omega
\end{equation*}
The equation above describes the relation between the total-wavefield and the contrast function inside the object domain $\Omega$.  The scattered wavefield is then measured in the receiver domain $\Gamma_r$ resulting in the following data equation:
\begin{equation*}
	y(\mathbf{x}) = \int_{\Omega}  g(\mathbf{x} - \mathbf{r}) f(\mathbf{r}) u(\mathbf{r}) \;\mathrm{d}\mathbf{r} , \qquad \forall \, \mathbf{x} \in \Gamma_r.
\end{equation*}

\section{Gradient Computation}
\label{sec:gradCompute}
In this section, we derive a gradient for an equality constrained cost function
\begin{equation}
	\mathcal{F}(\mathbf{f}) = \Big\lbrace h(\mathbf{f},\mathbf{u}) \quad \mbox{subject to} \quad \mathbf{k}(\mathbf{f}, \mathbf{u}) = \mathbf{0} \Big\rbrace
	\label{eq:grad:function}
\end{equation}
where $h: \mathbb{R}^n \times \mathbb{C}^n \mapsto \mathbb{R}$ is a real-valued function and $\mathbf{k}: \mathbb{R}^n \times \mathbb{C}^n \mapsto \mathbb{C}^n$ is a set valued function. We assume that both the functions $h$ and $\mathbf{k}$ are smooth and hence, differentiable. For the constrained problem \eqref{eq:grad:function}, the Lagrangian reads
\begin{align}
	\mathcal{L} \left( \mathbf{f}, \mathbf{u}, \boldsymbol{\lambda} \right) = h(\mathbf{f},\mathbf{u}) + \boldsymbol{\lambda}^H \mathbf{k}(\mathbf{f},\mathbf{u}),
	\label{eq:grad:lag}
\end{align}
where $\boldsymbol{\lambda} \in \mathbb{C}^n$ is a Lagrange multiplier corresponding to the constraints, and $\mathbf{x}^{H}$ represents the conjugate transpose of the vector $\mathbf{x}$ with complex entries. The stationary point of the Lagrangian $\mathcal{L}$, denoted by $\left(\mathbf{f}, \mathbf{u}^\star , \boldsymbol{\lambda}^\star \right)$, satisfies
\begin{align*}
	\frac{\partial \mathcal{L}}{\partial \mathbf{u}} = \mathbf{0}, \qquad \frac{\partial \mathcal{L}}{\partial \boldsymbol{\lambda}} = \mathbf{0}.
\end{align*}
The first condition gives rise to an adjoint equation
\begin{equation}
	\frac{\partial h}{\partial \mathbf{u}}\left( \mathbf{f}, \mathbf{u}^\star\right) + \left( \frac{\partial \mathbf{k}}{\partial \mathbf{u}}\left( \mathbf{f}, \mathbf{u}^\star \right) \right)^H \boldsymbol{\lambda}^\star = 0,
\end{equation}
while the second condition is the states equation
\begin{equation}
	\mathbf{k}\left(\mathbf{f}, \mathbf{u}^\star \right) = \mathbf{0}.
\end{equation}
The states equation generates a wavefield $\mathbf{u}^\star$ for a given parameter value $\mathbf{f}$. The adjoint equation calculates the Lagrange multiplier (also called adjoint wavefield) correspoding to wavefield $\mathbf{u}^\star$ for given $\mathbf{f}$. Tthe gradient of $\mathcal{F}$ is now retrieved from the partial derivative of the Lagrangian with respect to $\mathbf{f}$,
\begin{align}
	\nabla \mathcal{F}(\mathbf{f}) = \frac{\partial \mathcal{L}}{\partial \mathbf{u}} = \frac{\partial h}{\partial \mathbf{f}}\left(\mathbf{f}, \mathbf{u}^\star \right)  + \left( \frac{\partial \mathbf{k}}{\partial \mathbf{f}} \left( \mathbf{f}, \mathbf{u}^\star \right) \right)^H \! \boldsymbol{\lambda}^\star.
\end{align}
 This method is known as the \textit{adjoint-state} method \cite{plessix2006review}.

\subsection*{Inverse scattering Example}
For an inverse scattering problem, $h$ represents the misfit function between the simulated and the measured wavefields and $\mathbf{k} = \mathbf{0}$ is a Lippmann-Schwinger equation,
\begin{align*}
	\quad h \triangleq \tfrac{1}{2} \| \mathbf{y} - \mathbf{H} \diag (\mathbf{u}) \mathbf{f} \|^2, \quad \mathbf{k} \triangleq \left( \mathbf{I} - \mathbf{G} \diag(\mathbf{f}) \right) \mathbf{u} - \mathbf{v}.
\end{align*}
At a given value of $\mathbf{f}$,  the adjoint system for the Lippmann-schwinger equation is
\begin{align}
	\left( \mathbf{I} - \mathbf{G}^{H} \! \diag( \mathbf{f}) \right) \boldsymbol{\lambda}^\star = \diag(\mathbf{f}) \mathbf{H}^H \! \left( \mathbf{y} - \mathbf{H} \diag(\mathbf{f}) \mathbf{u}^\star\right).
	\label{eq:grad:adjointEqnIS}
\end{align}
Here, $\boldsymbol{\lambda}^\star$ is the adjoint wavefield and the $\mathbf{u}^\star$ is obtained satisfying the constraints at given value of $\mathbf{f}$:
\begin{equation}
	\left( \mathbf{I} - \mathbf{G} \diag (\mathbf{f}) \right) \mathbf{u}^\star = \mathbf{v}.
	\label{eq:grad:forward}
\end{equation}
Once the forward wavefield $\mathbf{u}^\star$ and the adjoint wavefield $\boldsymbol{\lambda}^\star$ are computed, the gradient is
\begin{equation}
	\begin{split}
	\nabla \mathcal{F}(\mathbf{f}) &= \diag(\mathbf{u}^\star)^H \mathbf{H}^H \left( \mathbf{H} \diag (\mathbf{u}^\star) \mathbf{f} - \mathbf{y}\right)  \\
	& \qquad - \diag(\mathbf{u}^\star)^H \mathbf{G}^H \boldsymbol{\lambda}^\star.
	\end{split}
	\label{eq:grad:final}
\end{equation}
Computing the gradient requires solving the forward~\eqref{eq:grad:forward} and the adjoint~\eqref{eq:grad:adjointEqnIS} systems only once each.

\section{Primal-Dual Method}
\label{sec:primal-dual}
We consider a class of optimization problems
\begin{align}
	\min_{\mathbf{x}} &\quad h(\mathbf{x}) + g(\mathbf{L} \mathbf{x}) +  k(\mathbf{x}) ,
	\label{eq:3PtProblem:main}
\end{align}
where $h: \mathbb{R}^n \mapsto \mathbb{R}$ is a differentiable closed convex function. $g: \mathbb{R}^m \mapsto \mathbb{R}$ and $k: \mathbb{R}^n \mapsto \mathbb{R}$ are closed non-differentiable convex functions. We assume that the proximal operators for the functions $h, g$ and $k$ are inexpensive. $\mathbf{L} \in \mathbb{R}^{m \times n}$ denotes a structured matrix. For example, in TV regularization, $\mathbf{L}$ represents a discrete gradient operator. We assume that the matrix $\mathbf{L}$ may be potentially non-invertible, such is the case in TV regularization. In this section, we derive a primal-dual algorithm to find an optimal solution to problem \eqref{eq:3PtProblem:main}.

\subsection{Preliminaries}
 A set-valued operator $\mathcal{H}: \mathbb{R}^n \mapsto \mathbb{R}^n$, that maps a point $\mathbf{x} \in \mathbb{R}^n$ to sets $\mathcal{H}(\mathbf{x}) \in \mathbb{R}^n$ is \textit{monotone} if
 \[
 	\left( \mathcal{H}(\mathbf{x}) - \mathcal{H}(\hat{\mathbf{x}}) \right)^T \left(\mathbf{x} - \hat{\mathbf{x}} \right) \geq 0 \quad \forall \; \mathbf{x}, \hat{\mathbf{x}}.
 \]
The operator $(\mathcal{I} + \gamma \mathcal{H})^{-1}$, with $\gamma > 0$ is called as the resolvent of the operator $\mathcal{H}$, where $\mathcal{I}$ is an identity operator.  The value $\mathbf{x} = (\mathcal{I} + \gamma \mathcal{H})^{-1}(\mathbf{y})$ of the resolvent is the unique solution of the monotone inclusion $\mathbf{y} \in \mathbf{x} + \gamma \mathcal{H} (\mathbf{x})$. A resolvent of a monotone operator is a non-expansive operator. An operator $\mathcal{H}$ is non-expansive if $
 	\| \mathcal{H}(\mathbf{x}) - \mathcal{H}(\mathbf{y}) \| \leq \| \mathbf{x} - \mathbf{y} \|,  \forall \; \mathbf{x}, \mathbf{y}.$

A \textit{proximal operator} (also known as\textit{ prox-operator}) of a closed convex function $h$ is the resolvent with $\mathcal{H} = \partial h$, a sub-differential of a function $h$. The \textit{prox-operator} reads as
\[
	\prox_{\alpha h} = (\mathcal{I} + \gamma \partial h)^{-1},
\]
and it maps $\mathbf{x} \in \mathbb{R}^n$ to the unique solution of the optimization problem
\[
	\mathbf{x}^\star = \argmin_{\mathbf{y}} \left\lbrace h(\mathbf{y}) + \frac{1}{2\gamma} \|\mathbf{y} - \mathbf{x} \|^2 \right\rbrace.
\]
A (convex) conjugate of a general function $h: \mathbb{R}^n \mapsto \mathbb{R}$ is
\[
	h^\star(\mathbf{y}) =  \sup_{\mathbf{x} \in \mathbb{R}^n} \Big\lbrace \, \mathbf{y}^T \mathbf{x} - h(\mathbf{x}) \, \Big\rbrace.
\]
A conjugate of a function is always \textit{convex}. The prox-operator of a function and its conjugate is related by the \textit{Moreau identity},
\begin{equation}
	\prox_{\alpha h^\star} (\mathbf{x}) + \alpha \prox_{h/\alpha}\left( \frac{\mathbf{x}}{\alpha}\right) = \mathbf{x} .
	\label{eq:PD:MoreauId}
\end{equation}

\subsection{Fixed point method}
\label{sec:primal-dual:FP}
A fixed point of an operator $\mathcal{T}: \mathbb{R}^n \mapsto \mathbb{R}^n$ is defined as the set of points $\mathbf{x} \in \mathbb{R}^n$ such that $\mathcal{T} (\mathbf{x}) = \mathbf{x}$. A fixed point method finds one such point by generating a sequence of iterates $\mathbf{x}^{(k)}$ with $k=1, \dots, n$ of form
\[
	\mathbf{x}^{(k+1)} = \mathcal{T}\left( \mathbf{x}^{(k)} \right)
\]
for a given initial point $\mathbf{x}^{(0)}$. The iterates converge to one of the fixed point if $\mathcal{T}$ is a non-expansive operator.

Now recall that the resolvent of a monotone operator $\mathcal{H}$ is a non-expansive operator, \ie,  $\mathcal{T} = \left( \mathcal{I} + \alpha \mathcal{H} \right)^{-1} $. Also, it can be easily seen that the zeros of the monotone operator $\mathcal{H}$ are the fixed points of its resolvent. Hence, the fixed point iterations takes the following form to find the zeros of a monotone operator $\mathcal{H}$:
\[
	\mathbf{x}^{(k+1)} = \left( \mathcal{I} + \alpha \mathcal{H} \right)^{-1} \mathbf{x}^{(k)} .
\]
A more efficient scheme to find the zero of $\mathcal{H}$ is a preconditioned fixed-point method. This iteration scheme generates a sequence
\[
	\mathbf{x}^{(k+1)} = \left( \mathcal{I} +  \mathcal{P}^{-1} \mathcal{H} \right)^{-1} \mathbf{x}^{(k)},
\]
with $\mathcal{P}$ as a symmetric positive-definite linear operator. This sequence can be simplified to
\begin{equation}
	 \left(  \mathcal{P} + \mathcal{H} \right) \mathbf{x}^{(k+1)}  =  \mathcal{P} \mathbf{x}^{(k)}
	 \label{eq:PD:precFP}
\end{equation}

\subsection{Primal-Dual algorithm}
To compute monotone operator for \eqref{eq:3PtProblem:main}, we look at its first-order optimality condition. It states that a zero-vector must be in the subdifferential of the cost function, \ie,
\begin{align}
	\mathbf{0} \in \nabla h (\mathbf{x}) + \mathbf{L}^T \partial g (\mathbf{L} \mathbf{x}) + \partial k (\mathbf{x}),
	\label{eq:3PtProblem:optCond}
\end{align}
where $\partial g: \mathbb{R}^m \mapsto \mathbb{R}^m$ and $\partial k: \mathbb{R}^n \mapsto \mathbb{R}^n$ are the respective subdifferentials of functions $g$ and $k$. Let's consider variables $\mathbf{u} \in \mathbb{R}^m$ in the subdifferential of $g$ and $\mathbf{v} \in \mathbb{R}^n$ in the subdifferential of $k$,
\begin{align}
	\mathbf{u} \in \partial g (\mathbf{L} \mathbf{x}), \qquad \mathbf{v} \in \partial k (\mathbf{x}) .
	\label{eq:3PtProblem:var}
\end{align}
The equations in \eqref{eq:3PtProblem:var} can be restated as follows.
\begin{align}
	\mathbf{0} \in \partial g^\star (\mathbf{u}) - \mathbf{L} \mathbf{x}, \qquad \mathbf{0} \in \partial k^\star (\mathbf{v}) - \mathbf{x}
	\label{eq:3PtProblem:varCond}
\end{align}
where $g^\star$ and $k^\star$ are the convex conjugate of the functions $g$ and $k$ respectively. From equations \eqref{eq:3PtProblem:optCond} and \eqref{eq:3PtProblem:varCond}, we can write the optimality conditions in the form of the following system
\begin{align}
	\begin{bmatrix}
	\mathbf{0} \\ \mathbf{0} \\ \mathbf{0}
	\end{bmatrix}
	\in
	\underbrace{\begin{bmatrix}
	\nabla h & \mathbf{L}^T & \mathbf{I}_n \\
	-\mathbf{L} & \partial g^\star & \mathbf{0} \\
	-\mathbf{I}_n & \mathbf{0} & \partial k^\star
	\end{bmatrix}}_{\mathcal{H}}
	\underbrace{\begin{bmatrix}
	\mathbf{x} \\ \mathbf{u} \\ \mathbf{v}
	\end{bmatrix}}_{\mathbf{z}},
	\label{eq:3PtProblem:opH}
\end{align}
where $\mathbf{I}_n \in \mathbb{R}^{n \times n}$ is the identity matrix.
It is easy to show that the operator $\mathcal{H}$ in~\eqref{eq:3PtProblem:opH} is a monotone operator.
Consider a preconditioner operator
\[
	\mathcal{P} = \begin{bmatrix}
	\frac{1}{\gamma}\mathcal{I} & -\mathbf{L}^T & -\mathbf{I}_n \\
	-\mathbf{L} & \frac{1}{\gamma}\mathcal{I} & \mathbf{0} \\
	-\mathbf{I}_n & \mathbf{0} & \frac{1}{\gamma}\mathcal{I}
	\end{bmatrix},
\]
with $\gamma > 0$, the preconditioned fixed-point iteration scheme in \eqref{eq:PD:precFP} results in the following primal-dual algorithm:
\begin{equation*}
	\begin{split}
	\mathbf{x}^{(t+1)} &= \left( \mathcal{I} + \gamma \nabla h \right)^{-1} \! \left( \mathbf{x}^{(t)} - \gamma \mathbf{L}^T \mathbf{u}^{(t)} - \gamma \mathbf{v}^{(t)}\right) \\
	\mathbf{u}^{(t+1)} &= \left( \mathcal{I} + \gamma \partial g^\star \right)^{-1} \! \left( \mathbf{u}^{(t)} - \gamma \mathbf{L} \left( \mathbf{x}^{(t)} - 2\mathbf{x}^{(t+1)} \right) \right) \\
	\mathbf{v}^{(t+1)} &= \left( \mathcal{I} + \gamma \partial k^\star \right)^{-1} \! \left( \mathbf{v}^{(t)} - \gamma \left( \mathbf{x}^{(t)} - 2\mathbf{x}^{(t+1)} \right) \right)
	\end{split}
\end{equation*}
If the proximal operators of functions $h, g$ and $k$ are simple, then the each iteration can be computed efficiently.

\section{Proximal Quasi-Newton method}
\label{sec:prox-QN}
In this section, we discuss the Quasi-Newton (QN) method and its proximal version (prox-QN). Assuming the cost function $f$ is twice differentiable, QN aims to solve
\begin{align}
	\mathbf{x}^\star = \argmin_{\mathbf{x}} \quad f(\mathbf{x})
\end{align}
by generating a sequence based on the quadratic approximation to the fuction $f$ at every iterate of the sequence. The procedure is as follows:
\begin{equation}
	\begin{split}
	\mathbf{s}^{(k)} &= - \mathbf{H}_k^{-1} \nabla f\left( \mathbf{x}^{(k)}\right), \\
	\alpha_k &= \text{linesearch} \left( f(\mathbf{x}^{k} + \alpha \mathbf{s}^{(k)}) \right), \\
		\mathbf{x}^{(k+1)} &= \mathbf{x}^{(k)} + \alpha_k \mathbf{s}^{(k)}.
		\label{eq:Quasi-Newton}
	\end{split}
\end{equation}
Here, $\mathbf{H}_k$ is 
(an approximation of) the Hessian of function $f$ at $\mathbf{x}^{(k)}$. This method differs from the Newton method, as the former relies on an approximation, while the latter computes the exact Hessian. If $f$ is a convex function, the QN method converges to a global minimum. If $f$ is non-convex, the QN can only guarantee the convergence to a local optimum.

We are interested in adapting the QN method to solve problems of form
\begin{align}
	\mathbf{x}^\star = \argmin_{\mathbf{x}} \bigg\lbrace f(\mathbf{x}) \mbox{  subject to  } g(\mathbf{L}\mathbf{x})\leq \tau, \mathbf{x} \geq 0 \bigg\rbrace.
	\label{eq:proxQN-problem}
\end{align}
Here, $f: \mathbb{R}^n \mapsto \mathbb{R}$ is a twice-differentiable function, and $g: \mathbb{R}^n \mapsto \mathbb{R}$ is a convex but potentially non-differentiable function. For convenience, we rewrite the problem \eqref{eq:proxQN-problem} as
\begin{align}
	\mathbf{x}^\star = \argmin_{\mathbf{x}} \Big\lbrace f(\mathbf{x}) + \delta_{g} (\mathbf{L}\mathbf{x}) + \delta_{k}(\mathbf{x}) \Big\rbrace,
\end{align}
where, $\delta_g$ is an indicator to the set $\{ \mathbf{x} : g(\mathbf{x}) \leq \tau \}$, and  $\delta_k$ is an indicator to the set $\{ \mathbf{x} : \mathbf{x} > \mathbf{0} \}$. We propose a following modification to the Quasi-Newton method, and call it  Proximal Quasi-Newton (prox-QN) method:
\begin{equation}
	\begin{split}
	(a) \quad & \mathbf{s}^{(k)} = \argmin_{\mathbf{s}} \bigg\lbrace \mathbf{s}^T \nabla f \left( \mathbf{x}^{(k)} \right)  + \frac{1}{2}\mathbf{s}^T \mathbf{H}_k \mathbf{s}   \\
	& \qquad + \delta_g \left( \mathbf{L} (\mathbf{x}^{(k)} + \mathbf{s} )\right) + \delta_k \left( \mathbf{x}^{(k)} + \mathbf{s}\right)\bigg\rbrace \\
	 (b) \quad  & \mbox{define} \quad \hat{\mathbf{x}}(\alpha) = \prox_{\alpha (\delta_g+\delta_k)} \left( \mathbf{x}^{k} + \alpha \mathbf{s}^{(k)} \right) \\
	(c) \quad & \alpha_{k} = \argmin_{\alpha} \left\lbrace f\left( \hat{\mathbf{x}}(\alpha) \right) \right\rbrace \\
	(d) \quad & \mathbf{x}^{(k+1)} = \hat{\mathbf{x}} \left( \alpha_k \right)
	\end{split}
	\label{eq:prox-QN:iterates}
\end{equation}
The steps in \eqref{eq:prox-QN:iterates} can be summarized as follows: Step (a) finds a search direction $\mathbf{s}^{k}$. It minimizes the quadratic approximation of $f$ at $\mathbf{x}^{(k)}$, ensuring that it satisfies the constraints. In step (b), we define a function $\hat{\mathbf{x}}: \mathbb{R} \mapsto \mathbb{R}^n$ which is a proximal of the iterate $\mathbf{x}^{k} + \alpha \mathbf{s}^{(k)}$ with respect to indicators to functions $g$ and $k$. The function $\hat{\mathbf{x}}$ ensures that the step length, $\alpha$, must satisfy the constraints. Step (c) does a linesearch with respect to the feasible $\alpha$. Once we obtain the correct $\alpha$, we update our variable of interest $\mathbf{x}$ in step (d).

The minimization problem in step (a) of \eqref{eq:prox-QN:iterates}, is a convex minimization problem. The cost function is the sum of three functions $h$, $\delta_g$, $\delta_k$. The function
\[
	h(\mathbf{s}) = \mathbf{s}^T \nabla f \left( \mathbf{x}^{(k)} \right)  + \frac{1}{2}\mathbf{s}^T \mathbf{H}_k \mathbf{s}
\]
is a convex quadratic function, while the remaining two, $\delta_g$ and $\delta_k$, are non-differentiable convex functions. To solve this minimization problem, we use first-order primal-dual method described in Appendix~C. The iterates for $t = 0, \dots, T$ are
\begin{equation}
	\begin{split}
	\mathbf{s}^{(t+1)} &= \prox_{\gamma h} \! \left( \mathbf{s}^{(t)} - \gamma \mathbf{L}^T \mathbf{u}^{(t)} - \gamma \mathbf{v}^{(t)} \right), \\
	\mathbf{u}^{(t+1)} &= \prox_{\gamma \delta_g^\star} \! \left( \mathbf{u}^{(t)} + \gamma \mathbf{L} \left(2 \mathbf{s}^{(t+1)} - \mathbf{s}^{(t)} \right)\right), \\
	\mathbf{v}^{(t+1)} &= \prox_{\gamma \delta_k^\star} \! \left( \mathbf{v}^{(t)} + \gamma \left( 2 \mathbf{s}^{(t+1)} - \mathbf{s}^{(t)} \right)\right),
	\label{eq:prox-QN:primalDual}
	\end{split}
\end{equation}
with $\gamma > 0$ controlling the speed of convergence. The proximal operations for $h, \delta_g$ and $\delta_k$ are expressed as follows:
\begin{align*}
	\prox_{\gamma h} (\mathbf{y}) &= \left( \mathbf{I} + \gamma \mathbf{H}_k \right)^{-1} \left( \mathbf{y} - \gamma \nabla f \left( \mathbf{x}^{(k)}\right)\right) \\
	\prox_{\gamma\delta_g}(\mathbf{y}) &=  \mbox{\textbf{proj}}_{\|\cdot \|_1 \leq \tau} (\mathbf{y}) \\
	\prox_{\gamma\delta_k}(y) &= \begin{cases}
	y & y > 0 \\
	0 & y \leq 0
	\end{cases}
\end{align*}
The proposed method (prox-QN) differs from \cite{lee2014proximal} in two aspects: \textit{(i)} The function $g$ can be potentially be more than $\ell_1$ type penalty. For example, we can work with total-variation-type regularization. \textit{(ii)} The linesearch ensures that the chosen $\alpha$ is strictly feasible.




\bibliographystyle{IEEEtran}
\bibliography{refs}


\begin{IEEEbiography}[{\includegraphics[width=1in,height=1.25in,clip,keepaspectratio]{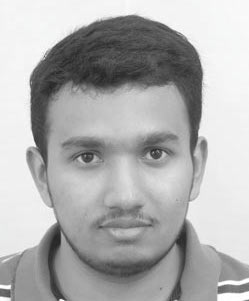}}]{Ajinkya Kadu} received the B.Tech. and M.Tech. degree in Aerospace Engineering from the Indian Institute of Technology Bombay, India in 2015, and the Ph.D. degree in Mathematics from Utrecht University, The Netherlands, in 2019. From July 2020, he is a postdoctoral researcher with the Computational Imaging group at the National Research Institute for Mathematics and Computer Science (CWI) in Amsterdam, The Netherlands. His research interests are in inverse problems, wavefield imaging, convex optimization and electron tomography.
\end{IEEEbiography}

\begin{IEEEbiography}[{\includegraphics[width=1in,height=1.25in,clip,keepaspectratio]{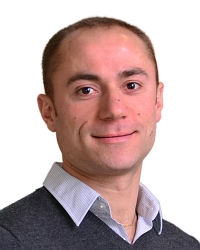}}]{Hassan Mansour} (S'99, M'09, SM'17) received the B.E. degree in computer and communications engineering from the American University of Beirut, Beirut, Lebanon, in 2003, and the M.A.Sc. degree in electrical and computer engineering and the Ph.D. degree in electrical and computer engineering from The University of British Columbia, Vancouver, BC, Canada, in 2005 and 2009, respectively. Between January 2010 and January 2013, he was a Postdoctoral Research Fellow with the Department of Computer Science, the Mathematics Department, and the Department of Earth, Ocean, and Atmospheric Sciences, The University of British Columbia. He is currently a Senior Principal Research Scientist with Mitsubishi Electric Research Laboratories, Cambridge, MA, USA. His research interests are in inverse problems, compressed sensing, sparse signal reconstruction, image enhancement, and scalable video compression and transmission. His current research is focused on the design of efficient acquisition schemes and reconstruction algorithms for natural images, radar sensing, video analytics, and inverse scattering problems. Dr. Mansour is a member of the IEEE Computational Imaging Technical Committee and the IEEE Sensor Array and Multichannel Technical Committee. He is also an Associate Editor for the IEEE Transactions on Signal Processing.
\end{IEEEbiography}

\begin{IEEEbiography}[{\includegraphics[width=1in,height=1.25in,clip,keepaspectratio]{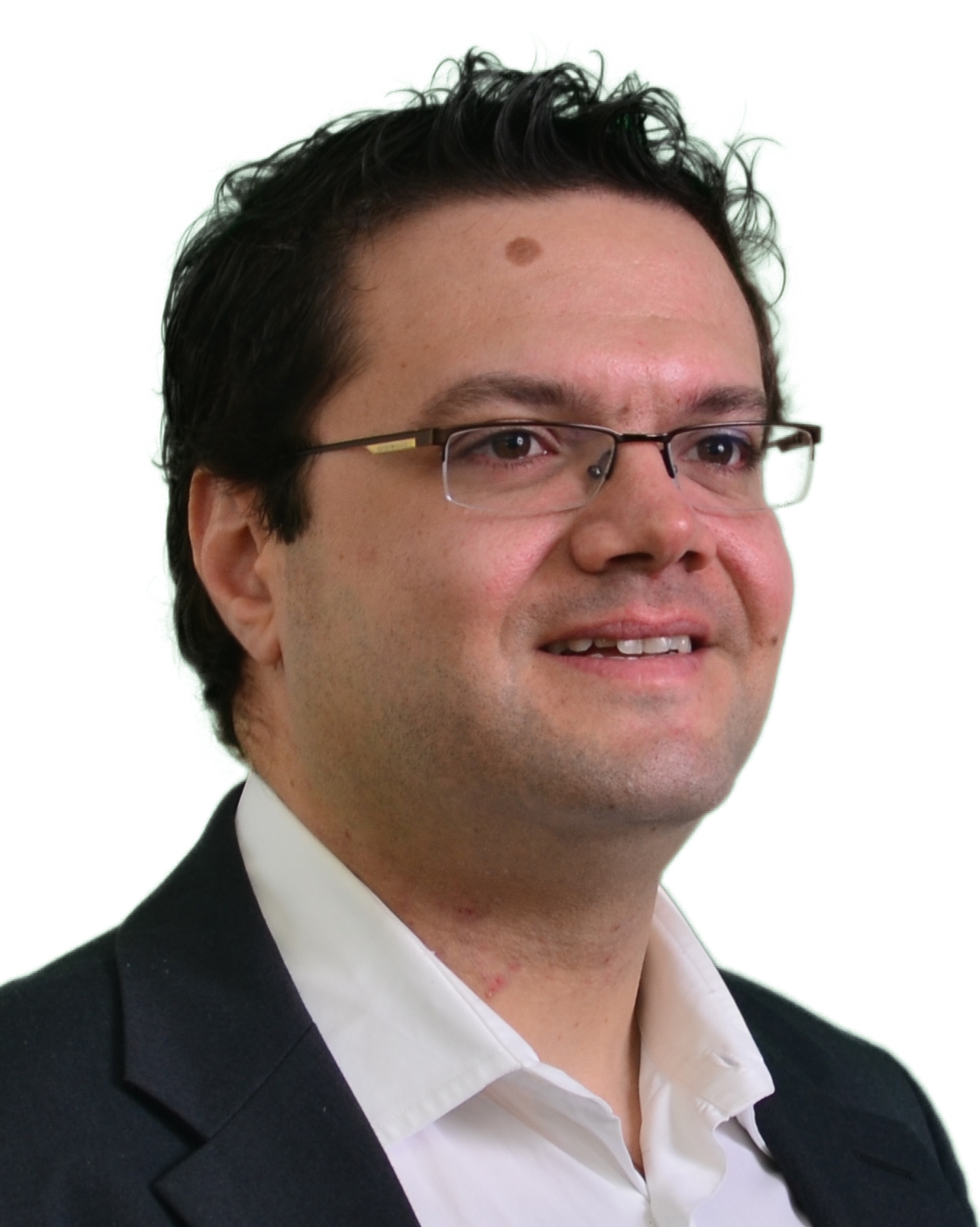}}]{Petros T.\ Boufounos} (S'02, M'06, SM'13) is a Senior Principal Research Scientist and the Computational Sensing Team Leader at Mitsubishi Electric Research Laboratories (MERL). Dr. Boufounos completed his undergraduate and graduate studies at MIT. He received the S.B. degree in Economics in 2000, the S.B. and M.Eng. degrees in Electrical Engineering and Computer Science (EECS) in 2002, and the Sc.D. degree in EECS in 2006. Between September 2006 and December 2008, he was a postdoctoral associate with the Digital Signal Processing Group at Rice University. Dr. Boufounos joined MERL in January 2009, where he has been heading the Computational Sensing Team since 2016.

Dr. Boufounos' immediate research focus includes signal acquisition and processing, computational sensing, inverse problems, frame theory, quantization, and data representations. He is also interested in how signal acquisition interacts with other fields that use sensing extensively, such as machine learning, robotics, and dynamical system theory. Dr. Boufounos has served as an Area Editor and a Senior Area Editor for the IEEE signal processing letters. He has been a part of the SigPort editorial board and is currently a member of the IEEE Signal Processing Society Theory and Methods technical committee and and an Associate Editor at IEEE Transactions on Computational Imaging. He was also named IEEE SPS Distinguished Lecturer for 2019-2020. 
\end{IEEEbiography}

\end{document}

%% file: defs.tex
\newcommand{\diag}{\mathop{\bf diag}}

\newcommand{\prox}{\mathbf{prox}}

\newcommand{\argmin}{\mathop{\rm argmin}}
\newcommand{\argmax}{\mathop{\rm argmax}}

\newcommand{\eg}{{\it e.g.}}
\newcommand{\ie}{{\it i.e.}}